\documentclass[twocolumn]{aa}

     \usepackage[authoryear]{natbib}
     \usepackage{graphicx}
     \usepackage{amssymb}
\begin{document}

   \title{A Radio and Mid-Infrared Survey of Northern Bright-Rimmed Clouds}

   \subtitle{}

   \author{L.K.~Morgan, M.A.~Thompson, J.S.~Urquhart, Glenn.J.~White \& J.~Miao }
   
   \offprints{L.K.~Morgan}

   \institute{Centre for Astrophysics \& Planetary Science,
              School of Physical Sciences,
              University of Kent,
              Canterbury,
              Kent CT2 7NR,
              UK\\
              \email{lkm8@kent.ac.uk}
	       }

   \authorrunning{L.Morgan et al.}
   \titlerunning{A Radio and Mid-IR Survey of Northern BRCs}
   
   \date{}

   \abstract{We have carried out an archival radio, optical and infrared wavelength imaging survey of 44
  Bright-Rimmed Clouds (BRCs) using the NRAO/VLA Sky Survey (NVSS) archive, images from the Digitised Sky Survey (DSS) and the Midcourse Space eXperiment (MSX). The data characterise the physical properties of the Ionised Boundary Layer (IBL) of the BRCs. We have classified the radio detections as: that associated with the ionised cloud rims; that associated with possible embedded Young Stellar Objects (YSOs); and that unlikely to be associated with the clouds at all. The stars responsible for ionising each cloud are identified and a comparison of the expected ionising flux to that measured at the cloud rims is presented. A total of 25 clouds display 20 cm radio continuum  emission that is associated with their bright optical  rims. The ionising photon flux illuminating these clouds, the ionised gas pressure and the electron density of the IBL are determined. We derive internal molecular pressures for 9 clouds using molecular line data from the literature and compare these pressures to the IBL pressures to determine the pressure balance of the clouds. We find three clouds in which the pressure exerted by their IBLs is much greater than that measured in the internal molecular material. A comparison of external pressures around the remaining clouds to a global mean internal pressure shows that the majority of clouds can be expected to be in pressure equilibrium with their IBLs and hence are likely to be currently shocked by photoionisation shocks. We identify one source which shows 20 cm emission consistent with that of an embedded high-mass YSO and confirm its association with a known infrared stellar cluster. This embedded cluster is shown to contain early-type B stars, implying that at least some BRCs are intimately involved in intermediate to high mass star formation.}

    \maketitle

    
\section{Introduction}
  It has long been suspected that the illumination of dense, quiescent clumps within molecular clouds by nearby OB stars could be responsible for their triggered collapse and subsequent star formation. BRCs associated with old ($\tau$ $\gtrsim$ 10$^{6}$ yr) HII regions are potential examples of Radiatively Driven Implosion (RDI) in which the UV flux of the associated OB star ionises the external layers of the cloud and causes the BRC to collapse. The ionisation front moves slowly into the globule, creating a dense outer shell of ionised gas (the Ionised Boundary Layer or IBL), which streams radially away from the cloud surface. The high density in the shell causes a high recombination rate which shields the globule from quickly evaporating \citep{Reipurth}.  The increased pressure in the ionised gas causes an expansion of the IBL into the intercloud medium and an ionisation front preceded by a shock in the neutral gas propagates into the cloud \citep{White97,Lefloch94}. The UV radiation from the OB stars  often sweeps the molecular material of the cloud into a cometary morphology with a dense core located at the `head' of the cometary globule.\\   
  
  The ratio of the internal pressure of the molecular cloud (mostly due to turbulent motions, but with a small thermal contribution) to the external pressure of the IBL is predicted by RDI models to be key to the evolution of the cloud. 
  If the IBL has a pressure greater than, or equal to, the turbulent and thermal pressure of the cloud then photo-ionisation shocks and a D-critical ionisation front propagate into the cloud interior, compressing and heating the molecular gas \citep{Bertoldi,Lefloch94}. However, if the interior pressure of the cloud is greater than the pressure created by the ionised gas within the IBL then the ionisation front stalls at the surface of the cloud until the effects of mass evaporation from the cloud (due to the `boiling off' of particles by the ionising radiation) and increasing recombination within the IBL raises the ionised gas pressure to equilibrium with the interior pressure \citep{Lefloch94}. As this equilibrium is reached, the evolution of the cloud follows the same path as the initially underpressured cloud case.
  An analysis of the pressure balance present in BRCs thus allows a separation of shocked and unshocked clouds. Our aim is the identification of shocked clouds, as these clouds are important subjects for future study and modelling to enable a greater understanding of the processes involved in triggered star formation.\\
  
  The general conditions prevalent within the IBLs of BRCs are not well known as most studies have, to date, accurately determined the conditions within only a few individual clouds (e.g.\citealt{searching,White99,Megeath,Lefloch97,Lefloch1995}). Recently, \citet{tuw} measured the IBL pressures and electron densities for 18 clouds from the Sugitani and Ogura southern hemisphere BRC catalogue \citep{SO94}. As a companion paper to \citet{tuw} we have utilised archival data to extend their analysis of the Sugitani and Ogura southern catalogue to incorporate the northern hemisphere Sugitani, Fukui \& Ogura (SFO) catalogue \citep{SFO91} which identified 44 BRCs associated with IRAS point sources which are therefore candidates for RDI.  
  
  From observations of the free-free emission associated with BRCs identified in the SFO catalogue we are able to determine the ionised gas pressure, the ionising photon flux impinging upon the cloud surface and the electron density of the IBL. This can be achieved using data obtained from the NRAO VLA 20 cm Sky Survey (NVSS) \citep{Condon}. Midcourse Space eXperiment (MSX) images at 8.3 $\mu$m allow us to trace the Photon Dominated Regions (PDRs) associated with the BRCs, confirming the identification of individual sources of 20 cm emission as IBLs. 
  The identification of the primary ionising stars associated with each BRC through a literature search allows us to compare the 20 cm emission measured toward each BRC with that predicted by the incident ionising flux. Through further quantification of these properties of IBLs at the edge of BRCs we hope to clarify the role of the RDI scenario in global star formation.
  To this end we have collated information using a thorough literature search and the SIMBAD database of astronomical catalogues \texttt{(http://simbad.u-strasbg.fr)} as well as analysing optical, radio and infrared images to identify the stars responsible for ionising the clouds.

\section{Survey Procedure}
We obtained 20 cm radio images from the NVSS radio catalogue \citep{Condon} downloaded using the postage stamp server \texttt{(http://www.cv.nrao.edu/nvss/postage.shtml)}. The NVSS was a 20 cm sky survey complete north of $\delta$=-40$^{\circ}$ carried out using the VLA in its D configuration. The angular resolution of the NVSS is 45$^{\prime\prime}$ \citep{Condon} and the limiting 1$\sigma$ noise of the survey is $\sim$0.5 mJy. Our sample comprises 44 15$^{\prime}\times$15$^{\prime}$ images centred upon the coordinates of each IRAS source associated with the 44 BRCs contained in the SFO catalogue.\\ 

  Ionising radiation excites Polycyclic Aromatic Hydrocarbon (PAHs) which re-emit the absorbed energy at infrared wavelengths \citep{Leger84}.  MSX 8.3 $\mu$m images incorporate the PAH spectral features at 7.7 $\mu$m and 8.6 $\mu$m as well very small dust grain continuum emission.  Images of the cloud sample were acquired in order to trace any associated PAH emission and confirm that the 20 cm emission at the cloud rims is free-free continuum emission from the IBL. The MSX surveyed the entire galactic plane within the range $\mid$b$\mid$ $\le$ 5$^{\circ}$ in four mid-infrared spectral bands between 6 and 25 $\mu$m at a spatial resolution of $\sim$18.3$^{\prime\prime}$ \citep{msx}. The A band of the MSX corresponds to a wavelength of 8.3 $\mu$m and is the most sensitive of the MSX bands. The A band of the MSX ranges from 6.8 - 10.8 $\mu$m and includes the broad silicate band at 9.7 $\mu$m and the PAH emission features at 7.7 $\mu$m and 8.6 $\mu$m, which are of particular interest to this study as they trace the interface between molecular material and ionisation features associated with 20 cm emission. Images were downloaded from the NASA/IPAC Infrared Science Archive \texttt{(http://irsa.ipac.caltech.edu)} and were analysed using GAIA, part of the Starlink Software Collection. The MSX 8.3 $\mu$m images were smoothed to a resolution of 45$^{\prime\prime}$, to match that of the NVSS images and improve the signal to noise ratio. The r.m.s. noise of each individual image was determined from off-source sky measurements and ranged from 1.5 - 39.1 $\times$10$^{5}$ W m$^{-2}$ sr$^{-1}$.
  
\section{Results and Analysis}
\subsection{Source Identification and Classification}
 Using NVSS data we have studied the BRCs from the survey of \citet{SFO91}. Identification of bright rims was initially achieved by overlaying contours of the 20 cm emission onto the R-band images of the clouds obtained from the Digitised Sky Survey (DSS) and searching for 20 cm emission that is positionally coincident with the bright optical rims of the clouds. 
 We detected no 20 cm emission (to a level of three times the r.m.s. noise) associated with the rims of the clouds SFO 3, 8, 9, 19, 20, 22, 23, 24, 26, 33, 34 and 39. The images of SFO 16, 17 \& 18 were disregarded due to their relatively low quality, which is likely to be due to the sidelobes of nearby confusing sources combined with the low surface brightness of the 20 cm rim emission from these clouds. For the clouds SFO 2 \& 44 there are no MSX data to assist with the identification, however, as the emission is clearly extended the inclusion of these sources was felt to be justified. A total of 26 radio sources were determined to be positionally associated with the BRCs from the SFO catalogue and the coordinates of the peak emission, peak fluxes and integrated flux densities of these sources are presented in Table \ref{tbl:rem}. DSS images overlaid with NVSS and MSX 8.3 $\mu$m emission contours are presented in Fig. \ref{fig:images}, centred on the IRAS source position given in \citet{SFO91} with a field of view of 2$^{\prime}$ unless otherwise stated. Images of the clouds SFO 2 and 44 are presented separately in Fig. \ref{fig:2and44} as they consist of NVSS emission contours only.
  \begin{center}
 \begin{table*}[th!]
 \begin{center}
 \caption{Properties of Radio Emission associated with the SFO Objects.}
   \begin{tabular}{|cccccc|}
 
    \hline
 
      \multicolumn{1}{|c}{SFO object}
    & \multicolumn{1}{c}{peak emission}
    & \multicolumn{1}{c}{peak emission}
    & \multicolumn{1}{c}{Peak Flux}
    & \multicolumn{1}{c}{Integrated Flux}
    & \multicolumn{1}{c|}{Source Classification}\\
     &$\alpha$(2000)&$\delta$(2000)&(mJy/beam)&(mJy)&\\
    \hline
 
1	&	23 59 33.7 	 & +67 23 54     &	    69.2   &		728.4 &       1       \\

2	&	00 03 50.3 	 & +68 32 09     &	    4.4    &		13.2  &       1       \\

4	&	00 58 60.0 	 & +60 53 16     &	    5.3    &		26.6  &       1       \\

5	&	02 29 01.8 	 & +61 33 17     &	    11.5   &		21.4  &       1       \\

6	&	02 34 45.0 	 & +60 48 18     &	    4.6    &		42.9  &       1$^{T}$ \\

7	&	02 34 47.8 	 & +61 49 14     &	    11.7   &		86.3  &       1       \\

10	&	02 48 07.3 	 & +60 25 35     &	    30.7   &		519.6 &       1       \\

11	&	02 51 33.0 	 & +60 03 43     &	    4.0    &		10.1  &       1       \\

12	&	02 55 00.9 	 & +60 35 44     &	    12.0   &		127.0 &       1       \\

13	&	03 00 52.5 	 & +60 40 19     &	     3.7   &		6.3   &       1       \\
			    	   	        			 
14	&	03 01 24.2 	 & +60 29 12     &	     8.1   &		41.9  &       3       \\
			    	   	        		 
15	&	05 23 26.5 	 & +33 11 54     &	     3.3   &		3.0   &       1       \\

16	&	05 20 00.9 	 & -05 50 22     &	     1.1   &		0.6   &       4       \\

17	&	05 31 27.1	 & +12 04 54	 &	     1.9   &	        7.7   &       4 	\\

18	&	05 44 28.7	 & +09 08 40	 &	     8.1   &	        9.1   &       4 	\\

21	&	05 39 38.3 	 & -03 36 25     &	     1.3   &		1.3   &       4       \\

25	&	06 41 06.3 	 & +10 14 30     &	     3.0   &		15.5  &       1       \\

27	&	07 04 03.8 	 & -11 23 19     &	     3.6   &		39.0  &       1$^{T}$ \\

28	&	07 04 41.4 	 & -10 22 15     &	     1.3   &		0.6   &       4       \\

29	&	07 04 54.5 	 & -12 09 42     &	     2.2   &		0.6   &       4       \\

30	&	18 18 46.1 	 & -13 44 39     &	    57.1   &		125.6 &       1       \\
  		    	   	        					
31	&	20 50 48.8 	 & +44 21 29     &	    12.8   &		345.5 &       1       \\
			    	   	        				
32	&	21 32 34.9 	 & +57 24 08     &	     2.9   &		5.0   &       1       \\

35	&	21 36 09.3 	 & +58 31 53     &	     1.8   &		12.1  &       1       \\

36	&	21 36 10.6 	 & +57 26 34     &	     3.8   &		11.5  &       1       \\

37	&	21 40 29.0 	 & +56 36 13     &	     3.5   &		11.0  &       1       \\
  		    	   	        					
38	&	21 40 44.3  	 & +58 15 01     &	     6.0   &		26.6  &       1       \\
			    	   	        				
40	&	21 46 09.1 	 & +57 09 59     &	     4.9   &		14.6  &       1       \\

41	&	21 46 29.0 	 & +57 18 10     &	     2.7   &		28.5  &       1       \\

42	&	21 46 38.6 	 & +57 11 39     &	     4.9   &		1.2   &       1       \\

43	&	22 47 50.3	 & +58 02 51	 &	     35.3  &		204.5 &       1       \\

44	&	22 28 59.1       & +64 12 11	 &	     37.0  &		95.8  &       1       \\
\hline

\end{tabular}
      \label{tbl:rem}\\
\scriptsize
$^{T}$Tentative classification, see Section \ref{sec:confusion} for details
      \end{center}
      \end{table*}
    \end{center}

  The radii of the optical bright rims (taken from \citealt{SFO91}) of the clouds toward which 20 cm emission has been detected range from 13 - 133$^{\prime\prime}$, corresponding to a physical range of 0.04 - 0.83 pc. From the source counts of \citet{Condon} we determine that there is a probability of P($<$r) $\sim$0.2 of finding a background radio source within 133$^{\prime\prime}$ of the IRAS point. This implies that up to $\sim$5 sources may potentially be due to confused background radio galaxies. In order to rule out any possible confusion the NVSS archival data were compared with optical Digitised Sky Survey (DSS) images and mid-infrared MSX 8.3 $\mu$m images. \\
  
  PAH emission is a tracer of UV-dominated PDRs \citep{Leger84} as the PAHs are transiently heated by the absorption of UV photons, therefore MSX images of the northern SFO catalogue can help to identify PDRs that are associated with the detected radio emission. Comparing emission in the optical, radio and infrared regimes helps to eliminate chance associations and can identify true emission from the bright rim.
  The radio sources detected in the NVSS survey images were classified according to the scheme of \citet{tuw}:
 
\begin{enumerate}

\item Bright-rim emission clouds with 20 cm and 8.3 $\mu$m emission positionally coincident with their bright
optical rims. 

\item Broken-rimmed clouds, in which the 20 cm and MSX 8.3 $\mu$m emission is
positionally coincident with the rim of the cloud (as Type 1) but the rim has a
reverse curvature with respect to the normal orientation, i.e.~the rim is curved
towards the molecular cloud, rather than the ionising star (e.g. the well known broken-rimmed globule CG4 in the Gum Nebula \citep{Reipurth}. No clouds of this type were identified in this survey, the classification has been retained to maintain consistency with \citet{tuw}.)

\item Embedded objects with compact and coincident 20 cm and mid-infrared emission that is
set back from the rim toward the centre of the cloud. 

\item 20 cm emission that is uncorrelated with either the bright optical rim or MSX
8.3 $\mu$m emission. 

\end{enumerate}

\subsection{Confusing Sources}
\label{sec:confusion}
 
We have tentatively labelled the sources SFO 6 \& 27 as Type 1 sources though there is some confusion as to their identification. These objects shall now be examined in more detail and justification for their nominal identifications given.

\subsubsection{SFO 6}
SFO 6 has strong ($\sim$9$\sigma$) 20 cm emission associated with the optical rim which appears to follow the optical bright rim (Fig.\ref{fig:images}). There is a small region of 8.3 $\mu$m emission centred at the peak of the optical rim which, due to the confused nature of the emission, is not definitive enough to infer any clear association with the cloud rim but does not contradict an identification of the source as Type 1. However, the 20 cm emission near the optical rim is confused by a stronger unassociated source that has no optical or infrared counterpart and is thus likely to be an extragalactic background source. We suggest that this source may be NEK 135.2+00.3, identified in the Clark Lake 30.9 MHz galactic plane survey \citep{Kassim}. The source is unresolved in the survey and only marginally resolved in the NVSS images. 

\subsubsection{SFO 27}
SFO 27 has strong ($\sim$6$\sigma$) 20 cm emission associated with the optical rim (Fig.\ref{fig:images}). There is widespread MSX 8.3 $\mu$m emission associated with the IRAS source and extending along the optical rim which supports an identification of the source as Type 1. However, the 20 cm emission near the optical rim is confused by a stronger unassociated source that has no obvious optical or infrared counterpart. 

We have continued the analyses of SFO 6 and SFO 27 as Type 1 objects and have masked the emission from the background sources as much as it is possible to do so.

\subsection{Ionised Rims Associated With SFO Objects}
\label{sec:rims}
For those objects that have been identified as Type 1 sources and are likely true bright-rim candidates we have evaluated the ionising photon flux impinging upon the clouds, the electron densities and the pressures in the IBLs. These quantities were determined by using the general equations from \citet{Lefloch97}. Rearranging their Eq.(6), the ionising photon flux $\Phi$ arriving at the cloud rim may be written in units of cm$^{-2}$ s$^{-1}$ as:\\

\begin{equation}
\Phi = 1.24 \times 10^{10} ~S_{\nu} ~T_{e}^{0.35} ~\nu^{0.1} ~\theta^{-2}
\end{equation}

where S$_{\nu}$ is the integrated radio flux in mJy, T$_{e}$ is the effective
electron temperature of the ionised gas in K, $\nu$ is the frequency of the free-free
emission in GHz and $\theta$ is the angular diameter over which the emission is
integrated in arcseconds.

The 20 cm flux measured at the ionised rim may possibly be overestimated due to nebula emission from the HII regions in which the BRCs are embedded. This emission forms a background which may add to our actual rim brightness at 20 cm. However, due to the poor ($u$,$v$) coverage of the NVSS snapshot observations any structure larger than $\sim$5$^{\prime}$ is filtered out \citep{Condon}, therefore if any large scale nebula emission is present we may expect it to be at a low level. This is supported by comparison of the DSS images, in which nebula emission shows up as large, diffuse, red regions, with NVSS data in which we find no correlating 20 cm emission.

The electron density ($n_{e}$) of the IBL surrounding the cloud
may also be derived from the integrated radio flux S$_{\nu}$ by substituting for the
ionising photon flux in Eq.(6) of \citet{Lefloch97}. The electron density in cm$^{-3}$ is given by:\\

\begin{equation}
n_{e} = 122.41 \sqrt{\frac{S_{\nu} T_{e}^{0.35} \nu^{0.1} \theta^{-2}}{\eta R}}
\end{equation}

where those quantities common to both Eq.(1) and (2) are in
the same units, $R$ is the radius of the cloud in pc and  $\eta$ is the effective
thickness of the IBL as a fraction of the cloud radius (typically $\eta$ $\sim$$0.2$,
\citealt{Bertoldi}). \\
  As the ionised flow is sonic at the cloud surface \citep{Lefloch94} we must take into account the ram pressure of the flow, as well as the bulk pressure of the ionised layer. Assuming a totally ionised gas the total pressure of the ionised layer with respect to the internal neutral gas is: \\

\begin{equation}
 P_{T} = P_{i} + \rho_{i} c_{i}^{2} = 2 \rho_{i} c_{i}^{2}
 \end{equation}
 
 where $c_{i}$ is the isothermal sound speed in the ionised interclump gas (assumed to be 11.4 km s$^{-1}$ \citet{Bertoldi}) and $\rho_{i}$ is the density within the IBL.

Values for $\Phi$, $n_{e}$ and $P_{T}$ were calculated using
Eqs.(1)-(3), assuming a boundary layer thickness fraction of
$\eta = 0.2$ \citep{Bertoldi} and an effective electron temperature of $T_{e} = 10^{4}$ K.
As the relatively low angular resolution of the NVSS means that a number of BRCs are either unresolved or marginally resolved at 20 cm the cloud radii were taken from the original BRC survey
paper of \citet{SFO91}).  
Values for $\Phi$, $n_{e}$ and $P_{T}$ are presented in Table \ref{tbl:phi}.\\

In our derivation of the ionising fluxes of these clouds we have implicitly assumed that the bright rims are resolved, this is not always the case (e.g. SFO 15). A comparison of the ratio between predicted and measured ionising fluxes in the resolved (e.g. SFO 38) and non-resolved case does not reveal any consistent effect associated with this beam `dilution'. Our analysis of the cloud SFO 5 reveals an ionising flux of  7.5 10$^{8}$ cm$^{-2}$s$^{-1}$, whereas the (resolved) observations of \citet{Lefloch97} find 4.8 x 10$^{9}$ cm$^{-2}$s$^{-1}$ using the same analysis. The difference in results appears insignificant considering the different frequencies of observation and total area of integrated emission. \citet{lefloch02} find a factor of $\sim$6 between predicted and measured ionising flux in their resolved observations of the Trifid nebula. However, the (also resolved) observations of SFO 5 presented in \citet{Lefloch97} find a small (0.8) difference between their predicted and measured ionising fluxes. While the non-resolution of the bright rims in some cases is certainly undesirable and an important consideration in future observations it does not appear to be a significant factor in the flux comparisons presented here.

\begin{table*}
\begin{center}
\caption{Values for the measured ionising flux, predicted
ionising flux, the
measured electron density and ionised gas pressure for Type 1  radio sources detected in the
survey.}
\label{tbl:phi}
\begin{tabular}{|ccccc|}\hline
 & Measured ionising flux & Predicted ionising flux & Electron Density & Ionised gas
 pressure \\
SFO Object & $\Phi$ (10$^{8}$ cm$^{-2}$\,s$^{-1}$) &$\Phi_{P}$ (10$^{8}$
 cm$^{-2}$\,s$^{-1}$)
 & $n_{\rm e}$ (cm$^{-3}$ )& $P_{i}/k_{B}$ (10$^{5}$ cm$^{-3}$K)\\ \hline

1	 & 18.5 & 43.7   & 872 & 274.3\\ 
	     	  	   	    	       
2	 & 2.7  & 2.1    & 271 & 85.3  \\  
	     	                    	  
4	 & 2.7  & 43.1   & 628 & 197.5 \\  
	     	                    	  
5	 & 7.5  & 61.3   & 364 & 114.4 \\  
	     	                    	  
6	 & 3.1  & 32.1   & 388 & 122.0 \\  
	     	  	            
7	 & 3.6  & 78.7   & 164 & 51.6  \\   
	        
10	 & 9.9  & 10.5   & 516 & 162.3 \\ 
	      	  	        
11	 & 2.3  & 10.2   & 232 & 73.0  \\
	      	  	        
12	 & 5.2  & 44.5   & 350 & 110.1 \\
	      	           	    	    
13	 & 3.2  & 12.4   & 203 & 63.9  \\
	          	   	    	 	   
15	 & 3.0  & 18.8   & 228 & 71.7  \\
	     	  	            	 	       
25	 & 2.2  & 27.1   & 211 & 66.4  \\
	      	  	        
27	 & 5.6  & 1.7    & 427 & 134.3 \\
	          	   	    	 	   
30	 & 22.1 & 4065.9 & 454 & 142.8\\
 	          	   	    	 	   
31	 & 8.7  & 3.5    & 483 & 151.9 \\
	          	   	    	 	   
32	 & 2.2  & 11.4   & 396 & 124.6 \\
	          	   	    	 	   
35	 & 1.2  & 6.9    & 204 & 64.2  \\
	          	   	    	 	   
36	 & 2.4  & 59.6   & 175 & 55.0  \\
	          	   	    	 	   
37	 & 2.0  & 10.2   & 357 & 112.3 \\
 	          	   	    	 	   
38	 & 2.8  & 13.3   & 248 & 78.0  \\
	          	   	    	 	   
40	 & 2.5  & 8.1    & 348 & 109.5 \\
	          	   	    	 	   
41	 & 1.5  & 8.0    & 248 & 78.0  \\

42	 & 1.3  & 7.3    & 250 & 78.6  \\

43	 & 4.8  & 145.8  & 277 & 87.1  \\

44	 & 2.1  & 4.6    & 269 & 84.7  \\\hline

\end{tabular}	
\end{center}
\end{table*}
\subsection{Identification of Ionising Stars}
\label{sec:ionistars}
  We searched the SIMBAD astronomical database in order to identify possible ionising stars of each BRC. All O or B type stars located within each HII region were considered as possible candidate ionising stars. For each star the predicted ionising fluxes incident on the rim of the relevant SFO object were determined using the tables of \citet{Panagia} and the stellar spectral type. If there was disagreement in the literature as to the specification of a star, or it was simply not quoted, it has been assumed that the star in question is an evolved main sequence star (Type V). A list of stars with high predicted ionising fluxes (i.e. nearby OB stars) was drawn up. Stars that contributed less than 50\% of the flux of the most dominant star were discounted. The stars that have been identified as the main ionising stars of each SFO object are listed in Table \ref{tbl:istars1} along with their positions, the HII region with which they are associated and their spectral type.

  \begin{table*}[th!]
 \begin{center}
 \caption{Ionising stars of the SFO objects, their relevant HII regions, positions and ionising fluxes.}
   \begin{tabular}{|ccccccc|}
 
    \hline
      \multicolumn{1}{|c}{SFO object}
    & \multicolumn{1}{c}{HII region}
    & \multicolumn{1}{c}{Ionising Star}
    & \multicolumn{1}{c}{$\alpha$}
    & \multicolumn{1}{c}{$\delta$}
    & \multicolumn{1}{c}{Spectral}
    & \multicolumn{1}{c|}{ Projected Distance}\\
     &{}&{}&(2000)&(2000)&Type & to Bright Rim (Pc) \\
    \hline

1 &(S171/NGC 7822) 	&  BD +66 1675$^{a}$      &  00 02 10.3 	 &     +67 24 32  &   O7V    &  3.7    \\
  	
2 &(S171/NGC 7822) 	&  BD +66 1675	     	  &  00 02 10.3 	 &     +67 24 32  &   O7V    &  16.9   \\
       		   		   		    
3 &(S171/NGC 7822) 	&  BD +66 1675	     	  &  00 02 10.3 	 &     +67 24 32  &   O7V    &  5.3    \\
       		   		   		    
4 &(S185) 	  	&  HD 5394$^{b}$ 	  &  00 56 42.3 	 &     +60 43 00  &   B0IV   &  1.1    \\
       			       
5 &(S190/IC 1805)  	&  BD +60 502$^{c}$  	  &  02 32 42.5 	 &     +61 27 22  &   O5V    &  14.9   \\      
  &			&  BD +60 504$^{c}$       &  02 32 49.4 	 &     +61 22 42  &   O4V    &  16.1   \\  
  &			&  BD +60 507$^{c}$	  &  02 33 20.6 	 &     +61 31 18  &   O5V    &  17.1   \\
       			       
6 &(S190/IC 1805)	&  BD +60 504	     	  &  02 32 49.4 	 &     +61 22 42  &   O4V    &  20.5   \\
       			       
7 &(S190/IC 1805)  	&  BD +60 502	     	  &  02 32 42.5 	 &     +61 27 22  &   O5V    &  14.6   \\
  &			&  BD +60 504	     	  &  02 32 49.4 	 &     +61 22 42  &   O4V    &  16.6   \\
  &			&  BD +60 507	     	  &  02 33 20.6 	 &     +61 31 18  &   O5V    &  11.4   \\
       			       
8 &(S190/IC 1805)  	&  BD +60 502	     	  &  02 32 42.5 	 &     +61 27 22  &   O5V    &  12.4   \\
  &			&  BD +60 504	     	  &  02 32 49.4 	 &     +61 22 42  &   O4V    &  11.3   \\
  &			&  BD +60 507	     	  &  02 33 20.6 	 &     +61 31 18  &   O5V    &  11.1   \\
       	
9 &(S190/IC 1805)  	&  BD +60 502	     	  &  02 32 42.5 	 &     +61 27 22  &   O5V    &  15.0   \\
  &			&  BD +60 504	     	  &  02 32 49.4 	 &     +61 22 42  &   O4V    &  14.4   \\
  &			&  BD +60 507	     	  &  02 33 20.6 	 &     +61 31 18  &   O5V    &  13.1   \\
       			       
10 &(S199/IC 1848) 	&  HD 17505$^{d}$	  &  02 51 08.0 	 &     +60 25 04  &   O6V    &  12.3   \\
       		        	  	
11 &(S199/IC 1848) 	&  HD 17505	     	  &  02 51 08.0 	 &     +60 25 04  &   O6V    &  11.9   \\
       		        	  		    
12 &(S199/IC 1848) 	&  BD +60 586$^{d}$	  &  02 54 10.7 	 &     +60 39 03  &   O8III  &  3.9    \\
       		        	  		    
13 &(S199/IC 1848) 	&  BD +59 578$^{d}$	  &  02 59 23.2 	 &     +60 33 59  &   O7V    &  7.0    \\
       		   		   		    
14 &(S199/IC 1848) 	&  BD +59 578	     	  &  02 59 23.2 	 &     +60 33 59  &   O7V    &  8.6    \\
       			       
15 &(S236/IC 410)  	&  HD 242908$^{e}$	  &  05 22 29.3 	 &     +33 30 50  &   O5V    &  22.1   \\
   &			&  HD 242926$^{e}$	  &  05 22 40.1 	 &     +33 19 10  &   O6V    &  12.0   \\
       		
16 &(S276)	 	&  $\delta$ Ori$^{f}$     &  05 32 04.1 	 &     +00 21 55  &   O9.5I  &  37.3   \\
  &            	    	&  $\theta^{1}$ Ori$^{f}$ &  05 35 15.5 	 &     -05 23 19  &   O7V    &  26.7   \\
  &            	    	&  $\iota$ Ori$^{f}$	  &  05 35 26.8 	 &     -05 54 31  &   O8.5II &  26.9  \\
       			       
17 &(S264/$\lambda$ Ori) &  $\lambda$ Ori$^{f}$   &  05 35 08.2 	 &     +09 56 04  &   O5V    &  16.3   \\
       		   	     	
18 &(S264/$\lambda$ Ori) &  $\lambda$ Ori         &  05 35 08.2 	 &     +09 56 04  &   O5V    &  17.0   \\
       		   	     		
19 &(S277/IC 434) 	&  HD 37468$^{g}$  	  &  05 38 44.8 	 &     -02 36 01  &   O9.5V  &  13.3   \\
       		   	     		
20 &(S277/IC 434) 	&  HD 37468  	     	  &  05 38 44.8 	 &     -02 36 01  &   O9.5V  &  8.3    \\
       		   	     		
21 &(S277/IC 434) 	&  HD 37468  	     	  &  05 38 44.8 	 &     -02 36 01  &   O9.5V  &  7.1    \\
       			       
22 &(S281) 		&  $\theta^{1}$ Ori$^{h}$ &  05 35 15.5 	 &     -05 23 19  &   O7V    &  6.5    \\
   &			&  $\iota$ Ori$^{h}$ 	  &  05 35 26.8 	 &     -05 54 31  &   O8.5II &  8.1   \\
	\hline  			

\end{tabular}
    \label{tbl:istars1}
    \end{center}
    \end{table*}
 \setcounter{table}{2}

 \begin{table*}[th!]
  \begin{center}
 \caption{\bf{(cont.)} \sf Ionising stars of the SFO objects, their relevant HII regions, positions and ionising fluxes.}
  \begin{tabular}{|ccccccc|}
 
    \hline
      \multicolumn{1}{|c}{SFO object}
    & \multicolumn{1}{c}{HII region}
    & \multicolumn{1}{c}{Ionising Star}
    & \multicolumn{1}{c}{$\alpha$}
    & \multicolumn{1}{c}{$\delta$}
    & \multicolumn{1}{c}{Spectral}
    & \multicolumn{1}{c|}{Projected Distance}\\
     &{}&{}&(2000)&(2000)&Type &to Bright Rim (Pc)\\
    \hline
23 &(S249) 		&  BD +22 1303$^{i}$       &  06 22 58.2 	 &     +22 51 46  &   O9V    &  8.3  \\
	  	   		   		     
24 &(S275/NGC 2244) 	&  HD 46223$^{j}$	   &  06 32 09.3 	 &     +04 49 25  &   O5V    &  19.3 \\
	  	   		   		     
25 &(S273/NGC 2264) 	&  HD 47839$^{k}$	   &  06 40 58.6 	 &     +09 53 45  &   O7V    &  4.7  \\
	  	   		   		     
26 &(S296/CMaOBI) 	&  HD 54662$^{l}$	   &  07 09 20.2 	 &     -10 20 48  &   O7III  &  29.7 \\
	  			
27 &(S296/CMaOBI) 	&  HD 53456$^{l}$   	   &  07 04 38.3  	 &     -11 31 27  &   BOV    &  3.9  \\
 	  			
28 &(S296/CMaOBI) 	&  HD 53367$^{l}$          &  07 04 25.5 	 &     -10 27 16  &   B0IV   &  2.1  \\
	  	   		   		     
29 &(S296/CMaOBI) 	&  HD 53975	     	   &  07 06 36.0 	 &     -12 23 38  &   O7.5V  &  9.5  \\
       		   		   	
30 &(S49) 		&  BD -13 4927$^{e}$	   &  18 18 40.1         &     -13 45 18  &   O5V    &  1.0  \\
	  	   		   	           
31 &(S117) 		&  HD 199579$^{m}$	   &  20 56 34.7      	 &     +44 55 29  &   O6V    &  20.5 \\
	  	   		   	           
32 &(S131/IC 1396) 	&  HD 206267$^{n}$	   &  21 38 57.6      	 &     +57 29 21  &   O6V    &  11.3 \\
	  	   		   		     
33 &(S131/IC 1396) 	&  HD 206267	    	   &  21 38 57.6      	 &     +57 29 21  &   O6V    &  10.5 \\
	  	   		    	   		   
34 &(S131/IC 1396) 	&  HD 206267	    	   &  21 38 57.6      	 &     +57 29 21  &   O6V    &  12.1 \\
	  	   		   
35 &(S131/IC 1396) 	&  HD 206267	    	   &  21 38 57.6      	 &     +57 29 21  &   O6V    &  14.5 \\
	  	   		   
36 &(S131/IC 1396) 	&  HD 206267	    	   &  21 38 57.6      	 &     +57 29 21  &   O6V    &  4.9  \\
	  	   		   
37 &(S131/IC 1396) 	&  HD 206267	    	   &  21 38 57.6      	 &     +57 29 21  &   O6V    &  11.9 \\
	  	   		   
38 &(S131/IC 1396) 	&  HD 206267	    	   &  21 38 57.6      	 &     +57 29 21  &   O6V    &  10.4 \\
	  	   		   
39 &(S131/IC 1396) 	&  HD 206267	    	   &  21 38 57.6      	 &     +57 29 21  &   O6V    &  12.6 \\
       		   		   
40 &(S131/IC 1396) 	&  HD 206267	    	   &  21 38 57.6      	 &     +57 29 21  &   O6V    &  13.4 \\
       		   		   
41 &(S131/IC 1396) 	&  HD 206267	    	   &  21 38 57.6      	 &     +57 29 21  &   O6V    &  13.5 \\
       		   		   
42 &(S131/IC 1396) 	&  HD 206267	    	   &  21 38 57.6      	 &     +57 29 21  &   O6V    &  14.1 \\
       		   		   
43 &(S142/NGC 7380) 	&  HD 215835	    	   &  22 46 54.2      	 &     +58 05 04  &   O5V    &  5.4  \\
       			       
44 &(S145) 	 	&  HD 213023$^{o}$	   &  22 26 52.4 	 &     +63 43 05  &   O9V    &  8.5  \\
    &	  		&  BD 62 2078$^{p}$        &  22 25 33.6 	 &     +63 25 03  &   O7.5V  &  13.8 \\       
       \hline			       
\end{tabular}\\
\scriptsize 
Stars found to be at similar distances to, formally declared to be associated with or described as `probable members' of relevant HII regions in the following references:
$^{a}$\citet{CF}; $^{b}$\citet{SDBB}; $^{c}$\citet{I}; $^{d}$\citet{HVV}; $^{e}$\citet{H}; $^{f}$\citet{RO}; $^{g}$\citet{K}; $^{h}$\citet{OD}; $^{i}$\citet{HSG}; $^{j}$\citet{OI}; $^{k}$\citet{G}; $^{l}$\citet{C}; $^{m}$\citet{BS}; $^{n}$\citet{M}; $^{o}$\citet{Ch};
$^{p}$\citet{S}

\end{center} 
    \end{table*}

\subsection{Flux Comparison}
A comparison of the 20 cm flux measured from the NVSS images with the predicted Lyman continuum flux allows us to determine the likelihood that the optical bright rims are due to ionisation by the nearby stars identified in Section \ref{sec:ionistars}. From the tables of \citet{Panagia} we have identified the fluxes of Lyman continuum photons associated with each star and thus calculated the subsequent flux we expect to impinge on the optical bright rims. We have assumed that there are no losses due to absorption by intervening material between the star and cloud, also, the star-cloud distance used in our calculations is that seen in projection in the plane of the sky and together these assumptions lead to the predicted ionising flux being a strict upper limit. It should be noted that a misclassification of an ionising star may lead to large differences in the ionising flux that we predict, a misclassification of half a spectral class may lead to an increase or decrease of a factor of two in the predicted Lyman photon flux. There are sometimes disagreements in the literature on the spectral classification of a particular star, two examples of this are the stars HD 5394 and BD +60 502 that we believe to be ionising the surfaces of SFO 4 and SFO 5 respectively. The former has been identified as BOIV by \citet{Morgan55} but as B3IV by \citet{Racine68} and the latter as 05 by \citet{Conti71} and B8 by \citet{Boulon58}. In the cases of disagreement we have taken the type specified as the most reliable by the SIMBAD database.\\

  The predicted ionising fluxes are presented along with the measured ionising fluxes in Table \ref{tbl:phi}. It can be seen in the majority of cases that $\Phi_{P}$ $>$ $\Phi$ as we expect, however, the difference between $\Phi_{P}$ and $\Phi$ in some cases is much larger than we might expect. The disparity between $\Phi_{P}$ and $\Phi$ in these cases is most likely due to the  assumption that there is negligible absorption between the ionising star and the bright rim. \citet{lefloch02} find an attenuation of a factor of $\sim$6 over an assumed distance of $\sim$1 pc. The assumed distances in our sample range from 1 to 37 pc and we thus expect attenuation to have a significant effect upon the ratio of predicted to measured ionising fluxes. It is worth noting that the projected star-cloud distance is an underestimate. The combination of these two facts illustrates that theoretical predictions of $\Phi$ based upon the spectral type and projected distance of the ionising star often overestimate the true ionising photon flux illuminating these clouds. Another effect which may affect our results is that the 20 cm emission from the bright rims was measured using an interferometer which acts as a high pass filter and may thus filter out flux from large scale structures.   In two cases (SFO 27 and SFO 31) the  flux that has been predicted is significantly less than the flux that is observed. In the case of SFO 27 the 20 cm flux is confused with an unassociated source and so the measured 20 cm flux and hence the calculated value of $\Phi$ are highly uncertain. In the case of SFO 31 the direction of the suspected ionising star is not supported by the morphology of the cloud and no other possible ionising stars have been identified in the region.

\subsection{Upper Limits to the Ionising Flux for Non-Detections}
\label{sec:upplim}
There are a total of 12 clouds in our survey where no 20 cm emission was detected to a level of 3$\sigma$. We have checked the upper limits of any possible 20 cm emission from these clouds for consistency with the flux predicted from our candidate ionising stars. The predicted ionising flux was determined as in Section \ref{sec:rims} using the stars and spectral types as laid out in Table \ref{tbl:istars1}. The 3$\sigma$ upper limits, subsequent maximum possible observed  (i.e 3$\sigma$)  fluxes and predicted ionising fluxes are presented in Table \ref{tbl:upplim}. It can be seen that $\Phi_{P}$ $>$ $\Phi_{max}$ in all but one case and that $\Phi_{P}$ $>$ $\Phi_{max}$ by a factor of $\sim$10 in a large number ($\sim$40\%) of cases. These predictions are obviously inconsistent with the NVSS observations. In these cases the reasons for the discrepancies between theoretical prediction and measurement may mean that the predicted ionising star may be misclassified or misidentified,  that extinction between the suspected ionising star and bright rim is significant,  or that the projected distances between star and cloud are incorrect.

\begin{table*}
\begin{center}
\caption{Values for the 3$\sigma$ levels, maximum possible measured ionising flux and predicted
ionising flux, for radio sources not detected in the survey.}
\label{tbl:upplim}
\begin{tabular}{|cccc|}\hline
 & 3$\sigma$ flux upper limit & Max. observed ionising flux & Predicted ionising flux \\
SFO Object & (mJy) & $\Phi_{max}$ (10$^{8}$ cm$^{-2}$\,s$^{-1}$) & $\Phi_{P}$ (10$^{8}$ cm$^{-2}$\,s$^{-1}$) \\ \hline

3	 & 2.04  & 2.02  & 21.15 \\ 
	     	   				    
8	 & 1.52  & 1.78  & 118.31\\  
	     	   			       
9	 & 2.10  & 2.02  & 78.22 \\	 
	     	   				     
19	 & 0.75  & 0.51  & 0.57  \\   
	     	   
20	 & 1.40  & 1.47  & 1.47  \\ 
	     	   		 
22	 & 1.78  & 1.93  & 22.83 \\
	     	   		 
23	 & 1.91  & 2.07  & 2.52  \\
	     	   				 
24	 & 2.69  & 2.17  & 11.52 \\
	     	   					
26	 & 1.37  & 1.22  & 1.06  \\
	     	   					
33	 & 1.55  & 1.13  & 13.29 \\
	     	   					 
34	 & 1.40  & 1.48  & 9.98  \\
	     	   					    
39	 & 1.72  & 1.85  & 9.18  \\\hline

	\end{tabular}	
\end{center}
\end{table*}

\subsection{Candidate Ultra-Compact HII Regions}
\label{sec:HII}

  We have identified one 20 cm source in this survey as a Type 3 source, SFO 14. The 20 cm NVSS emission from SFO 14 follows a potential ionised cloud rim, although there is a localised peak coincident with the IRAS position which is clearly identifiable in the MSX image. 
  
SFO 14 has the characteristics of a compact or ultra-compact HII region: it is infrared-luminous (L$_{IR}$ $\sim$10$^3$ - 10$^4$ L$_{\odot}$) and its IRAS colours match the criteria suggested by \citet{wc89} for ultra-compact HII regions (namely that log F$_{60}$/log F$_{12}$ $\ge$ 1.3 and log F$_{25}$/log F$_{12}$ $\ge$ 0.57). To further investigate the possibility of this object being an embedded compact or ultra-compact HII region within the BRC we have used the Far InfraRed (FIR) and radio luminosity of the source to estimate the spectral class of any YSO that may be present. The FIR luminosity of an embedded YSO is almost entirely due to the luminosity of the YSO itself, reprocessed by the circumstellar dust in the region and re-emitted in the FIR. 
  
  In addition to HIRES IRAS fluxes taken from images obtained from the NASA/IPAC Infrared Science Archive (http://irsa.ipac.caltech.edu) we have used the 450 $\mu$m and 850 $\mu$m fluxes presented in \citet{tmw} as well as the 2 mm flux presented in \citet{S} to estimate the FIR luminosities of SFO 14. The luminosity was estimated by integrating under a grey-body fit using the IRAS 60 $\mu$m flux together with the 450 $\mu$m, 850 $\mu$m and 2 mm flux measurements, the distance (1.9 kPc) was taken from \citet{SFO91}. Further details about the SCUBA observations and grey-body modelling may be found in \citet{tmw}.
   The assumption has been made that all of the luminosity arises from a single embedded YSO. This assumption may be rather crude but enables a first-order estimate. \citet{wc89} showed that, for a realistic initial mass function, the spectral type of the most massive member in a cluster is only 1.5 - 2 spectral classes lower than that derived for the single embedded star case. 
  
  In a compact or ultra-compact HII region, the 20 cm flux is predominantly due to thermal free-free emission. If this emission is assumed to be optically thin and we assume that the region is in photoionisation equilibrium then we can relate the integrated radio flux of the region to the total number of ionising photons from the embedded YSO via Eq.(7) of \citet{Carpenter}.\\

\begin{equation}  
  N_{i}=7.7 \times 10^{43} ~S_{\nu} ~D^2 ~\nu^{0.1} 
\end{equation}  
  
  where N$_{i}$ is the total number of photons per second ionising the cloud material, S$_{\nu}$ is the integrated radio flux density in mJy, D is the distance to the source in kpc and $\nu$ is the frequency of the observation in GHz. Note that the 5 GHz term of Eq.(7) of \citet{Carpenter} has been removed and the equation coefficient adjusted accordingly. Using the results of this equation we can use the tables of \citet{Panagia} to find the spectral type of any central YSO. The results of the above analyses are presented in Table \ref{tbl:lum}. There is reasonably good agreement between the spectral type predicted through the different methods suggesting that there is a B0.5 - B1.5 YSO embedded within the molecular material of the clouds. 
  
A literature search was carried out in order to identify the nature of the candidate ultra-compact HII region. SFO 14 is a well studied region within the W3/W4/W5 molecular cloud complex and is associated with the infrared stellar cluster AFGL 4029 \citep{Deharveng}. \citet{Snell} report the presence of a strong molecular outflow and \citet{Carpenter00} find that the region contains an embedded cluster containing 240$\pm$10 stars. The region has been found to contain a luminous red YSO invisible in the optical, associated with a cluster of massive red stars; an optical and IR reflection nebula; a high velocity ionised stellar wind; an optical jet and a bright H$_{2}$ emission knot \citep{Deharveng}. The spectral type of the most massive star in this cluster inferred from the NVSS 20 cm flux is consistent with that derived by \citet{Deharveng} from infrared observations.\\
  
The BRC identified as a potential high mass star-forming site has been confirmed as containing an embedded cluster. The spectral class of the star within this cluster has been determined as being early B type and thus relatively young with a main sequence lifetime of $\sim$10$^{6}$ years. A molecular outflow is present also \citep{Snell} and so star formation is still occurring in this region. The region of emission is larger than the NVSS 45$^{\prime\prime}$ beam (corresponding to 0.4 pc at the assumed distance) at the 3$\sigma$ level and so is not an ultra-compact HII region, the main sequence lifetime for the most massive predicted YSO is $\sim$10$^{6}$ years and so we can assume an age of the embedded cluster that is younger than this. 
 The radii of the 3$\sigma$ contours of the embedded cluster is 1.1 pc. Given a propagation speed of 0.903 pc Myr$^{-1}$ \citep{Lefloch94} we find that the radio region would have taken $\sim$1.2x10$^{6}$ years (similar to the main sequence lifetime of a BO.5 star) to have formed as a consequence of the embedded clusters' influence. Using conservative estimates the cluster is sufficiently old that it may have been influenced by the propagation of photoionisation shocks into the cloud. The distance of the core from the optical bright rim is 0.56 pc, assuming a shock velocity of 1 km s$^{-1}$ \citep{searching,White99} we find a shock crossing time of $\sim$5x10$^{5}$ years. This is not conclusive evidence that the star exciting the bright rim of this cloud has affected the star-forming processes within the cloud, however, neither can the possible influence of this star be ruled out.\\
 
   Accurate ages are needed both for the star embedded within the cluster and for the star(s) that are ionising the region, which will allow a more certain link between the development of the embedded cluster and the ionising star(s) to be drawn. Theoretical work is underway to analyse and re-create the large scale properties of the clusters, this will lead to conclusive support for or against the RDI scenario. Deeper observations at mid-infrared wavelengths are needed to identify any embedded clusters which are still unknown and molecular observations will help to provide better constraints on the internal pressures of a larger sample of BRCs.

\begin{table*}
\begin{center}
\caption{Infrared and radio-derived spectral types for the Type 3 radio sources detected in the survey.}
\label{tbl:lum}
\begin{tabular}{|ccccccc|}\hline
SFO Object & IRAS PSC ID & IR Luminosity & Spectral Type & Flux Density & Ionising Photon Flux & Spectral Type \\
 &  & (L$_{IR}$/L$_{\odot}$) & (IR) & S$_{\nu}$ (mJy) & Log(N$_{i}$)(s$^{-1}$) & (Radio) \\ \hline

14	 & 02575+6017 & 8321  & B0.5 & 427.7  & 47.1 & B1.5 \\\hline        
	     	                     	       
\end{tabular}	
\end{center}
\end{table*}
  
\section{Discussion}
  25 clouds from this survey were classified as Type 1 radio sources, i.e. with 20 cm emission clearly associated with their photoionised cloud rims. The pressures from their IBLs were estimated in Section \ref{sec:rims} and, together with data on the molecular interior of the clouds these allow us to determine the pressure balance between the IBL and the internal molecular material of the clouds. In the RDI scenario a photoionisation driven shock propagates into the molecular cloud and if the internal pressure of the cloud is greater than or equal to the pressure in the developing IBL, the shock stalls at the surface of the cloud \citep{Lefloch94}. Evaporation of the cloud ensures that eventually (provided that the source of ionisation is maintained) the surface area of the cloud will decrease, leading to a relative increase in the IBL pressure and hence allowing the propagation of a D-critical ionisation front. \\
  
  If the clouds that we observe are currently overpressured with respect to their IBL then we do not expect there to be (or have been) photoionisation shocks propagating into the cloud and it is therefore unlikely that any existing star formation in these clouds could have been caused via the RDI scenario.  
If the clouds are underpressured with respect to the IBL then we may expect photoionisation shocks to be propagating into the cloud and the possibility exists that RDI is the cause of star formation in the region. Thus the pressure balance between the IBL and molecular material of the clouds acts as a simple diagnostic of the clouds whose star formation may have been triggered by their UV illumination.\\

  The data that we require on the molecular interior of these clouds may be derived through molecular line observations. There are a small number of such observations that have previously been published and these form the basis for our comparison of internal molecular pressures with those determined for IBLs in this work. 

\subsection{Molecular Pressures}
\label{sec:pressures}

  The pressure of the molecular gas within the clouds in our sample is a result both of turbulent velocities within the cloud interiors and the thermal pressure. As the clouds are composed mostly of cold gas there is a negligible thermal contribution to the internal pressure or any observed linewidths relating to the molecular material of these clouds.
Given molecular line observations we may derive the internal molecular pressures of our cloud sample from the turbulent velocity dispersion ($\sigma^{2}$) and the molecular gas density $\rho_{m}$: We use the relation between molecular pressure P$_{m}$, the square of the turbulent velocity dispersion $\sigma^{2}$ and the density of the molecular gas $\rho_{m}$; i.e. P$_{m}$ $\simeq$ $\sigma^{2}$$\rho_{m}$.  The turbulent velocity dispersion may be written in terms of the observed linewidth $\Delta\nu$ as $\sigma^{2}$ = $<$$\Delta\nu$$>$$^{2}$/(8$ln$2).  
   
Observations of line emission from the northern SFO objects were collected via a SIMBAD search specific to the SFO catalogue.
  We have used the linewidth taken from the C$^{18}$O($J=1\rightarrow0$) observations of \citet{Jansen94} and their value for H$_2$ density to determine an internal molecular pressure for SFO 4, the values for the internal molecular pressure and density of SFO 5 are derived from the results of the C$^{18}$O($J=1\rightarrow0$) observations of \citet{Lefloch97}. Values for the internal pressure and density of SFO 11 have been taken from \citet{searching}. 
 
 \citet{Devries} conducted a millimetre and submillimetre molecular line survey of BRCs selected from the SFO catalogue. In their survey they observed C$^{18}$O, HCO$^{+}$ and N$_{2}$H$^{+}$ (in addition to other species) emission from the clouds. The core densities of the clouds SFO 16, 18 and 25 were determined from the N$_{2}$H$^{+}$ observations of \citet{Devries} as  it is likely to be optically thin in our cloud sample and hence a good tracer of high column density. We have assumed spherical geometry and used the radii from \citet{SFO91}. SFO 13 was not detected in N$_{2}$H$^{+}$ by \citet{Devries} and its density was thus determined from their HCO$^{+}$ observations. Internal molecular pressures for these four clouds were then determined from the linewidths of the C$^{18}$O($J=1\rightarrow0$) observations presented in the same work.
 C$^{18}$O($J=1\rightarrow0$) linewidths and core densities from \citet{Cernicharo1992} and \citet{Duvert1990} were used to calculate internal pressures for SFO 20 and 37 respectively in preference to the observations of \citet{Devries} due to their superior angular resolution.

 \begin{table*}[!hp]
\begin{center}
\caption{Ionised boundary layer pressures and internal pressures and densities of clouds observed in molecular line wavelengths. A dash indicates non-detection in the NVSS.}
\label{tbl:Pint}
\begin{tabular}{|ccccc|}\hline
SFO Object & IBL pressure & &Internal Pressure & Internal Molecular Density \\
 & (10$^{5}$ cm$^{-3}$ K/k$_{B}$) & & (10$^{5}$ cm$^{-3}$ K/k$_{B}$) &  $(10^3 cm^{-3})$\\ \hline

4$^{a}$	 & 197.5 &$>>$& 8.2   & 50.0   \\
	   	  		      
5$^{b}$	 & 114.4 &    & 32.3  & 11.0  \\	    

11$^{c}$ & 82.0  &    & 25.0  & 2.5    \\

13$^{d}$ & 63.9  &$>>$& 16.3  & 2.0    \\

16$^{d}$ &  -	 &    & 0.3   & 2.0    \\

18$^{d}$ &  -	 &    & 7.8   & 16.6   \\

20$^{e}$ &  -	 &    & 6.9   & 20.0   \\
		       
25$^{d}$ & 66.4  &    & 12.5  & 7.4    \\

37$^{f}$ & 112.3 &$>>$& 1.5   & 3.0    \\\hline	    
	     	                     	       
\end{tabular}\\
$^{a}$\citet{Jansen94};$^{b}$\citet{Lefloch97}; $^{c}$\citet{searching}; $^{d}$\citet{Devries}; $^{e}$\citet{Cernicharo1992}; $^{f}$\citet{Duvert1990}	
\end{center}
\end{table*}

The IBL pressures calculated from the NVSS data range from 51.6 to 274.3 10$^{5}$ cm$^{-3}$ K/k$_{B}$ with a mean of 107.7 10$^{5}$ cm$^{-3}$ K/k$_{B}$ while the molecular line data that has been collected show a range of internal molecular pressures of 0.3 - 32.3 10$^{5}$ cm$^{-3}$ K/k$_{B}$ with a mean of 12.3 10$^{5}$ cm$^{-3}$ K/k$_{B}$ (Table \ref{tbl:Pint}).

Factors contributing to the uncertainties in our values for internal molecular pressure include the accuracy involved in measuring linewidths and the uncertainty for radii for these clouds taken from \citet{SFO91} (including the assumption of spherical symmetry made in determining volume densities from column densities). We estimate that uncertainties in the internal molecular pressures presented total no more than a factor of 5. 

\begin{center}
\begin{figure}[!hp]
\begin{center}
\includegraphics*[scale=0.40]{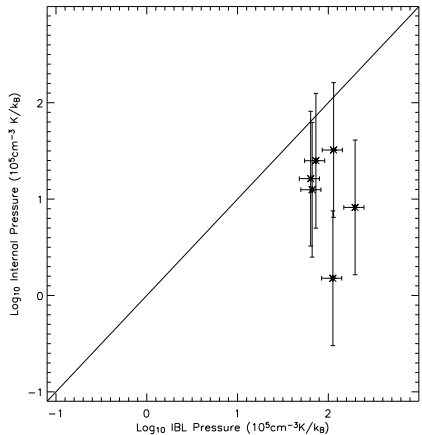}
\caption{Graph of internal molecular pressure versus IBL pressure. Error bars in IBL pressure represent an approximate 75\% uncertainty in the assumption that T$_{e}$ = 10$^{4}$. The solid line indicates pressure balance between the internal and external pressures.}
\label{fig:Puncert}
\end{center}
\end{figure}
\end{center}

The masses derived from the column densities given in \citet{Devries} concur well within these uncertainties with values for cloud masses derived from submillimetre observations presented in \citet{tmw}. Of the six clouds for which we have analysed the pressure balance three (SFO 4, SFO 13 and SFO 37) exhibit large enough differences in the pressures determined due to their internal turbulent motions and due to the IBL that it is highly likely that they are in a state of photoionisation induced pressure imbalance (see Fig. \ref{fig:Puncert}), even accounting for the large errors inherent within this study. These clouds are therefore excellent candidates for testing the theories of RDI and comparison to BRCs which are known to be in pressure balance (or imbalance)with their IBLs.\\
 In addition to the clouds that we have identified here as being in pressure imbalance with their IBL's, \citet{Cernicharo1992} present observational evidence that SFO 20 (ORI-I-2) is undergoing RDI. They find a bipolar outflow emanating from the source and while their evidence is not definitive they conclude that the UV field present due to the illuminating star $\sigma$-Ori has strongly modified the evolution of the primitive globule and strongly increased its star formation potential.\\

The mean internal pressure of the clouds for which we have molecular line data is 12.3 10$^{5}$ cm$^{-3}$ K/k$_{B}$ (consistent with the results of \citet{tuw} who find an internal molecular pressure of 15 10$^{5}$ cm$^{-3}$ K/k$_{B}$), the remainder of our clouds for which we have 20 cm data but no molecular data have been found to have ionised gas pressures higher than this without exception. All of the clouds within our sample are therefore candidates for the possible propagation of photoionisation induced shocks into their interiors, although as this determination is based upon a global mean of internal pressure no definitive statements can be made. However, with pressures of only 2 10$^7$ cm$^{-3}$ K/k$_{B}$ the BRCs SFO 1 and SFO 10 are highly likely to be underpressured with respect to their IBLs. Together with SFO 7 and SFO 36, which are the two clouds most likely to be overpressured with respect to their IBLs with IBL pressures only a factor of $\sim$4 times the global mean, these clouds represent good candidates for future molecular observations to accurately determine their internal pressures.
   
   \section{Summary and Conclusions}
   From archival studies of radio, optical and infrared observations we have analysed 44 BRCs for signs of triggered star formation. We detected 20 cm emission to a level of 3$\sigma$ or greater in 32 of the sample and using a comparison of various wavelength images have sorted the clouds into the four distinct types identified by \citet{tuw}. i.e. Type 1 BRCs, Type 2 broken-rimmed clouds, Type 3 embedded radio sources and Type 4 unassociated radio sources.
   We have compared the ionising fluxes from the candidate ionising star(s) and that determined from the 20 cm emission of the rim itself for our Type 1 sources. The number of Type 1 sources identified within our sample is 25 and the conditions prevailing within their IBLs have been analysed, including the incident ionising fluxes, electron density and ionised gas pressures. This analysis has made a significant contribution to (almost doubled) the number of BRCs that have known IBL conditions. Through comparison of IBL pressures to interior molecular gas pressures we have identified three clouds which are highly likely to be in a state of pressure imbalance and thus are ideal candidates for future observations and theoretical studies in RDI. Observations of higher resolution and greater sensitivity are required in order to identify possible further embedded sources within our sample as well as further molecular observations to determine internal pressures for a greater sample.\\
  A comparison of the results that we have derived for SFO 5 with those of \citet{Lefloch97} reveals a difference in derived IBL pressure of a factor of $\lesssim$3, this is believed to be due to the differences in observational data (frequency, integration area, etc.), we do not feel that these differences unduly affect our conclusions as the other errors inherent in our derivation are as large or larger than that arising from this beam dilution.

We draw the following conclusions:\\
   
\begin{enumerate}
\item   A comparison of individual pressures determined for IBLs from the NVSS data and internal molecular pressure of the BRCs leads us to believe that the majority of our sample is in approximate pressure equilibrium. It is therefore likely that photoionisation induced shocks are propagating into the interior of the clouds. Of the clouds for which we have detected 20 cm radio emission and for which molecular line data exist the three clouds SFO 4, SFO 13 and SFO 37 have the highest likelihood of being underpressured with respect to their IBLs. \\
\item In many cases the predictions of ionising flux from our candidate ionising stars are inconsistent with the measurements of 20 cm flux that we have measured. Discrepancies may be due to the assumption that there is negligible absorption being invalid, the misclassification of the spectral type of the ionising star and the possible existence of as-yet undiscovered additional OB stars within the region or an inaccuracy in the distance measurements to the objects.\\
\item The Type 3 radio source in our survey has been identified with a known young embedded cluster containing massive stars. With the 20 cm NVSS data we have confirmed that this cluster contains early B type stars. Further observations and modelling are required to investigate whether this cluster may have been induced to form by the presence of nearby, external massive stars.   
\end{enumerate}   

$Acknowledgments$ The authors would like to thank an anonymous referee for their suggestions which considerably improved this paper.
 LKM and JSU are both supported by a PPARC
doctoral studentship. This research would not have been possible without the
SIMBAD astronomical database service operated at CCDS, Strasbourg, France and
the NASA Astrophysics Data System Bibliographic Services. The Digitized Sky
Survey was produced at the Space Telescope Science Institute under U.S.
Government grant NAG W-2166. The images of these surveys are based on
photographic data obtained using the Oschin Schmidt Telescope on Palomar
Mountain and the UK Schmidt Telescope. The plates were processed into the
present compressed digital form with the permission of these institutions.
This research made use of data products from the Midcourse Space 
Experiment.  Processing of the data was funded by the Ballistic 
Missile Defense Organization with additional support from NASA 
Office of Space Science. This research has also made use of the 
NASA/ IPAC Infrared Science Archive, which is operated by the 
Jet Propulsion Laboratory, California Institute of Technology, 
under contract with the National Aeronautics and Space 
Administration.

\begin{center}
\begin{figure*}[p]
SFO 1
\includegraphics*[scale=0.50]{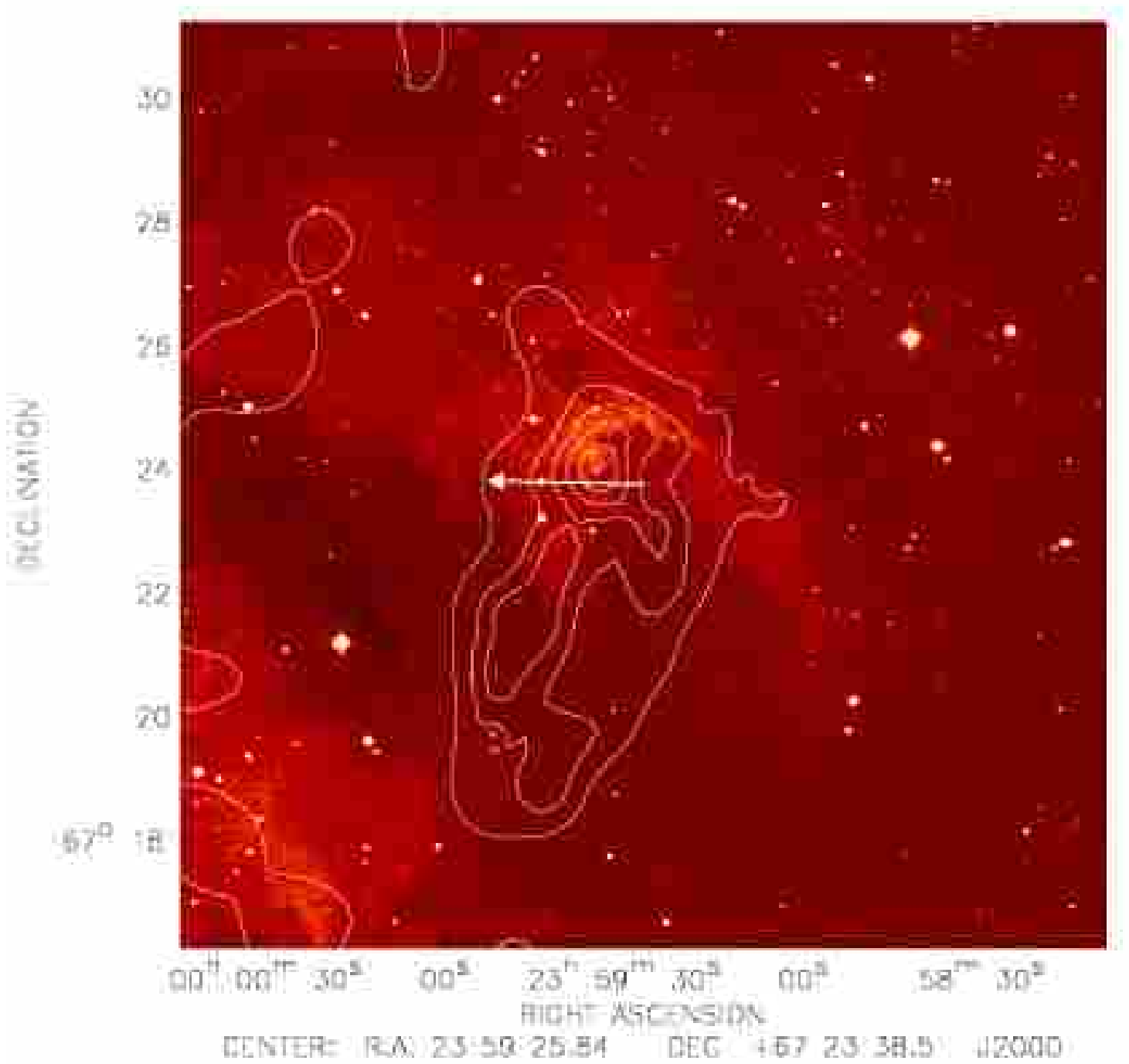}
\includegraphics*[scale=0.50]{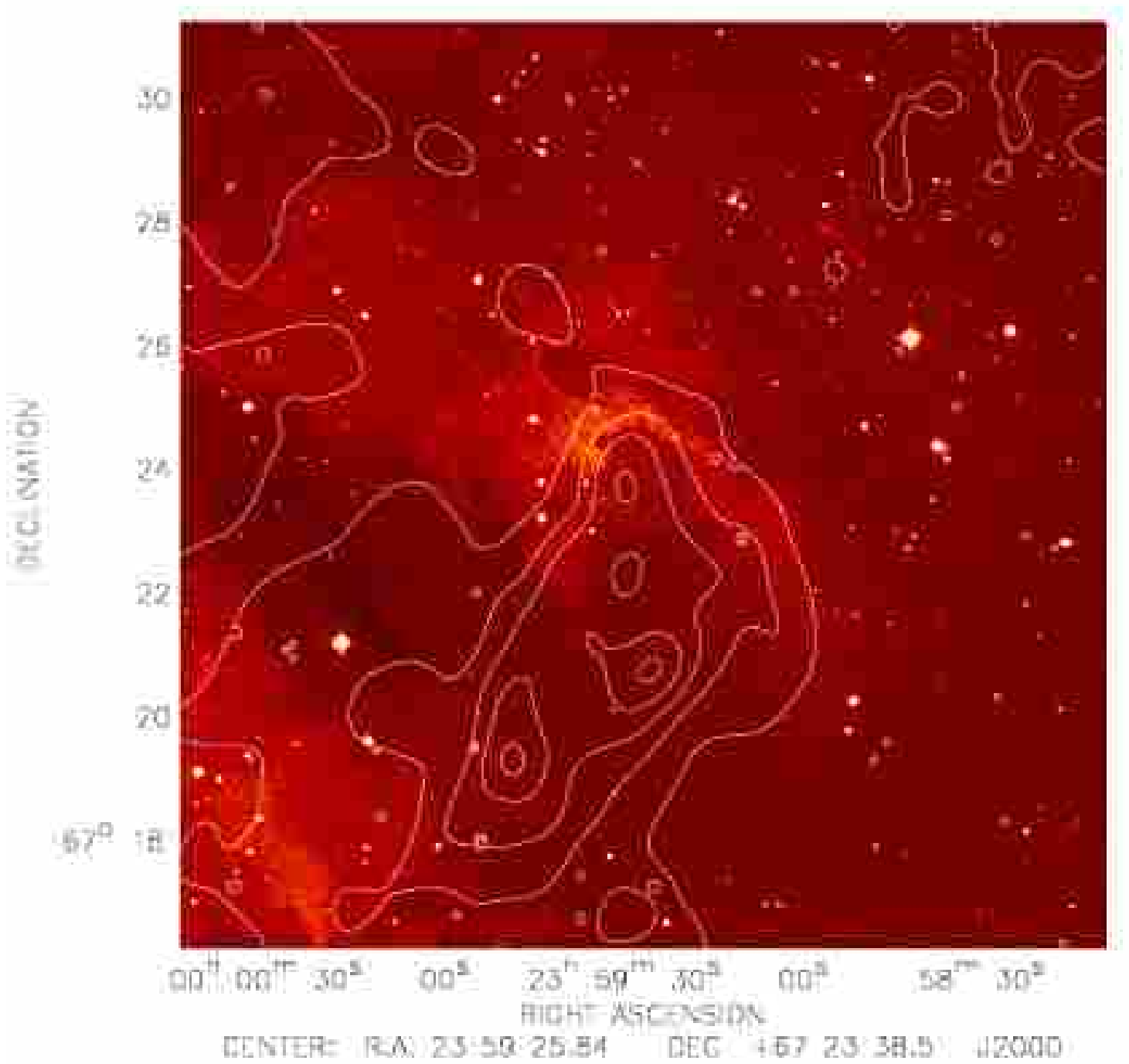}\\
SFO 4
\includegraphics*[scale=0.50]{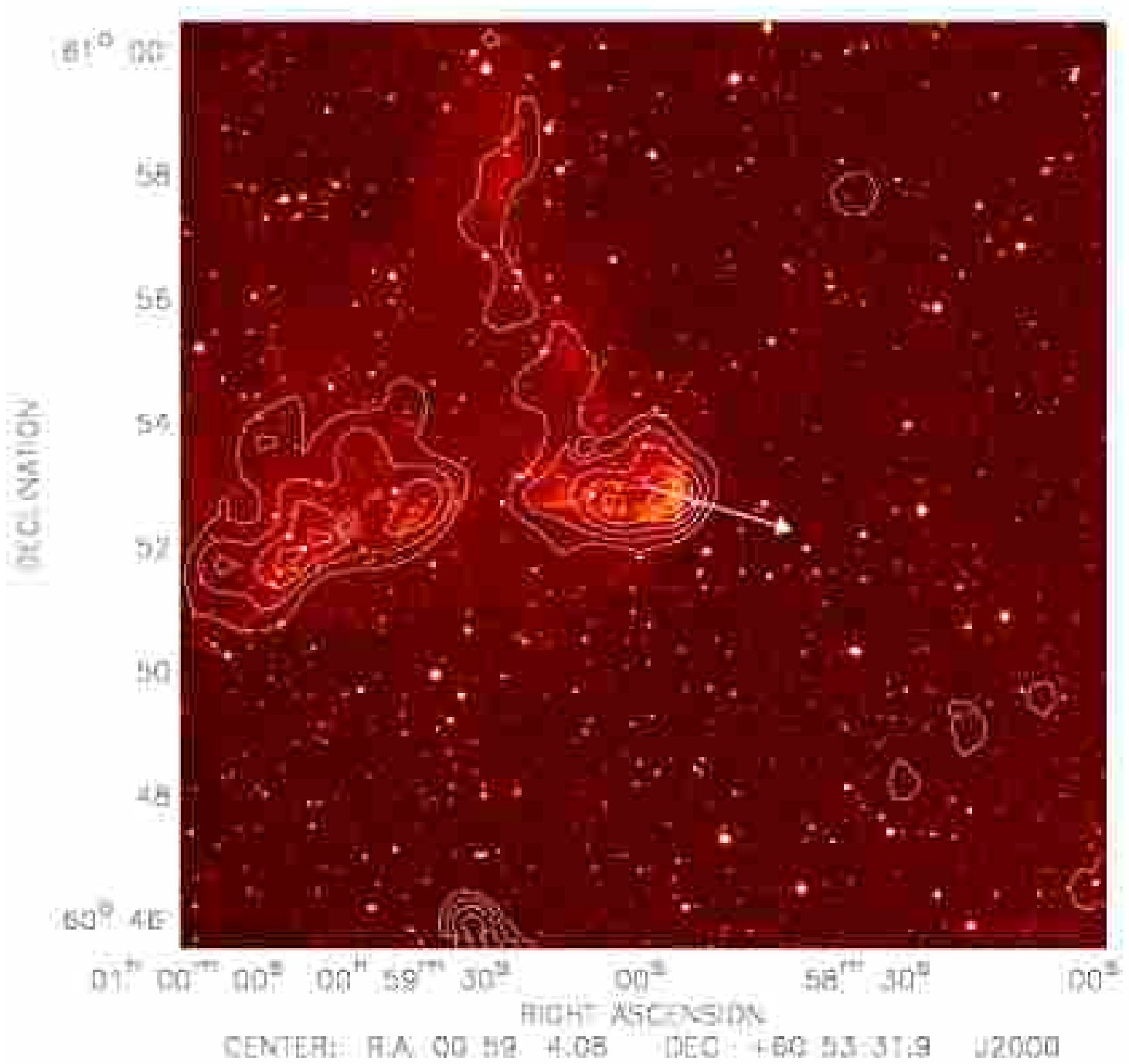}
\includegraphics*[scale=0.50]{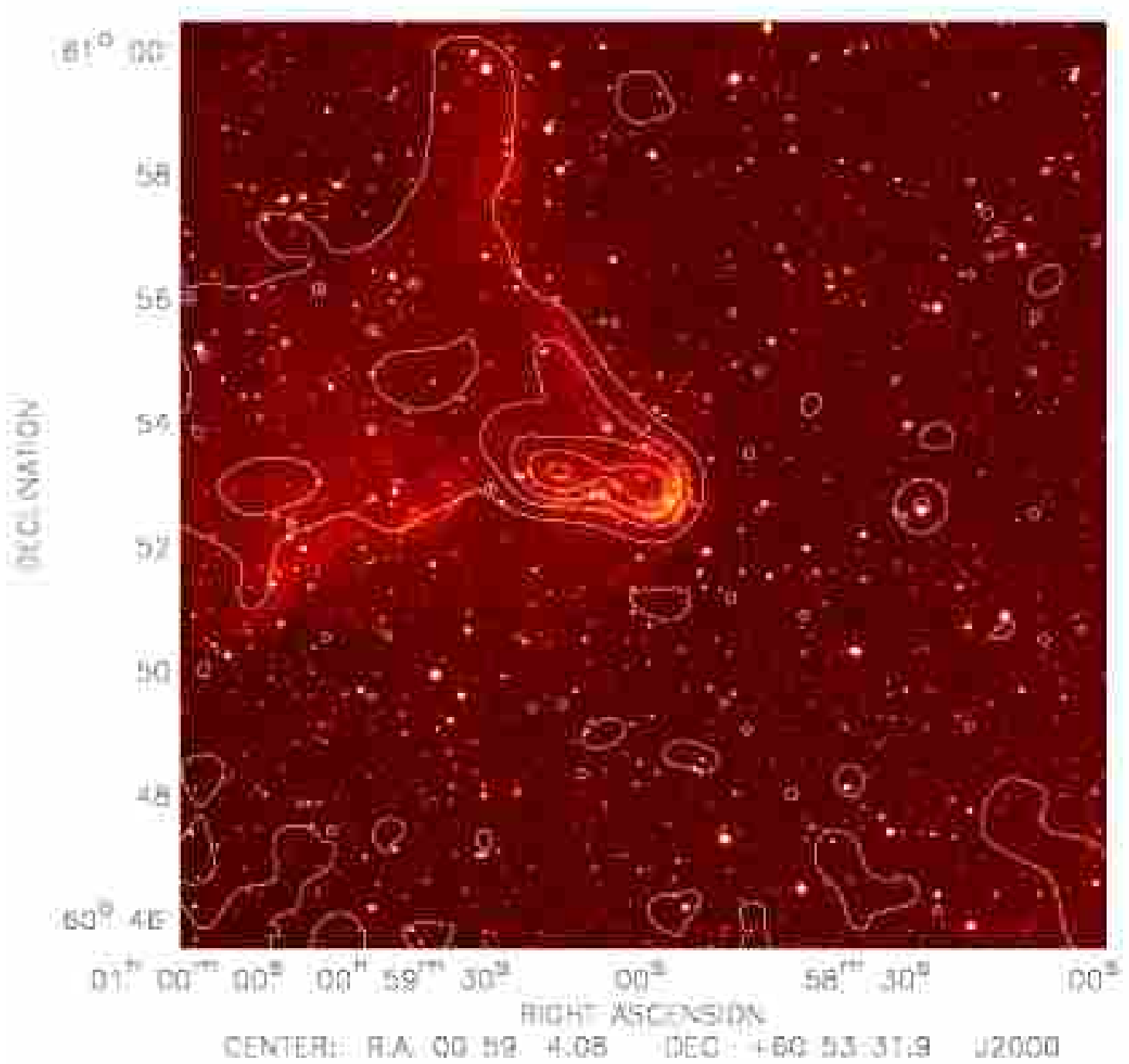}\\
SFO 5
\includegraphics*[scale=0.50]{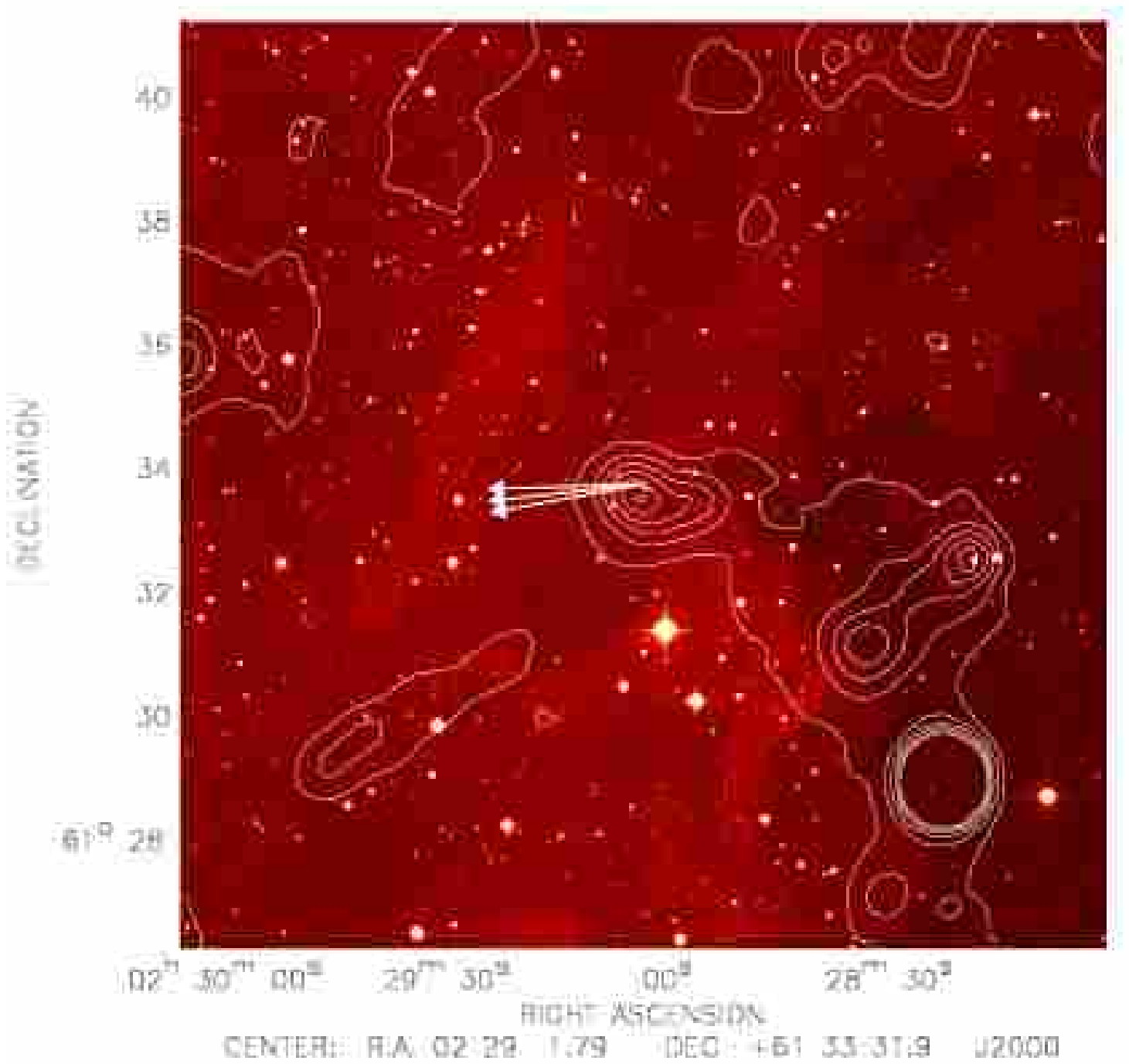}
\includegraphics*[scale=0.50]{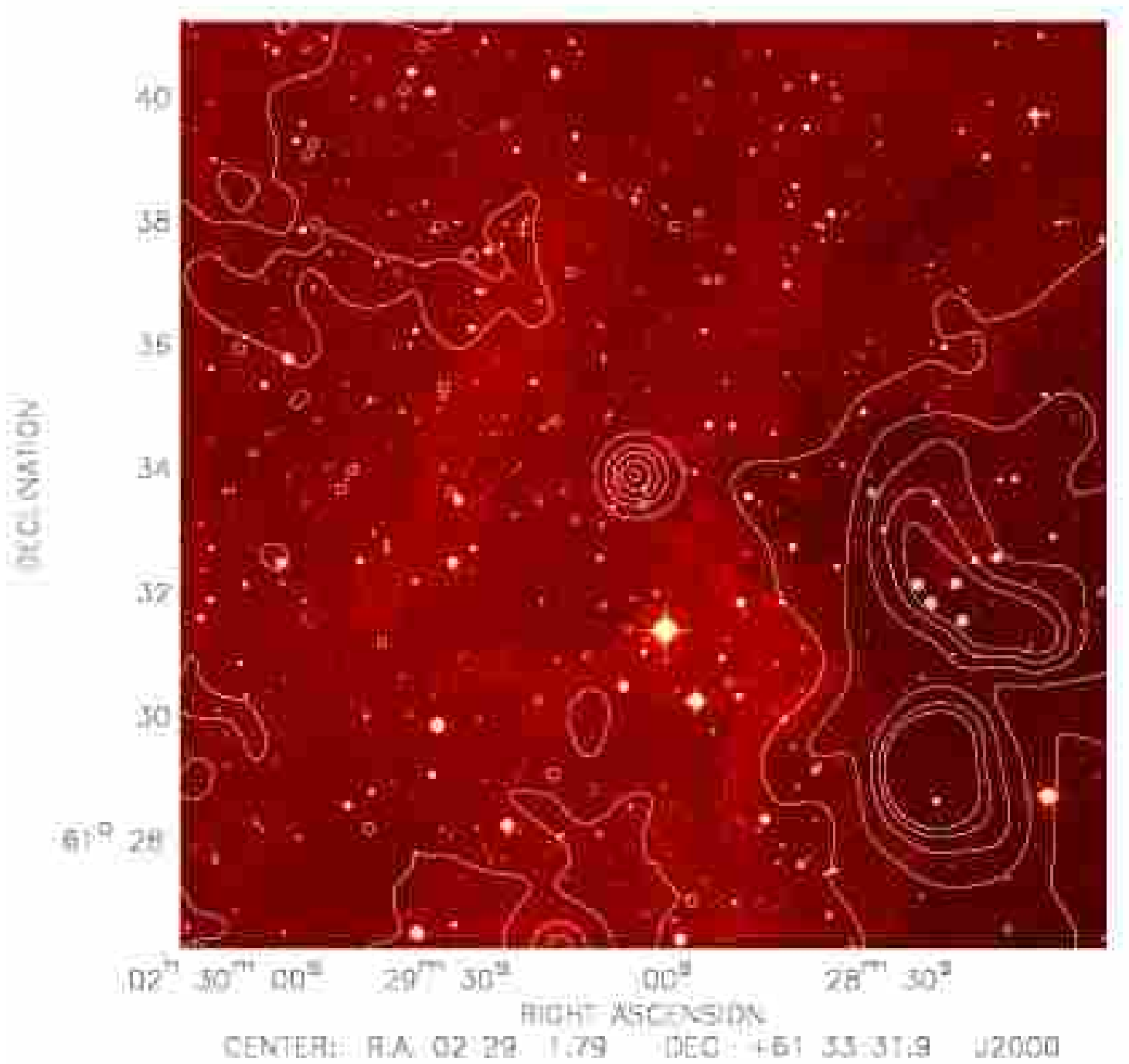}\\
\caption{Images from the survey of clouds with radio detections. Each cloud is represented
by a pair of images from the Digitised Sky Survey. The left images has NVSS 20 cm contours overlaid, while the right image has MSX 8.3 $\mu$m
emission contours overlaid. The MSX images used have been smoothed to the same resolution as the NVSS images, 45$^{\prime\prime}$. The arrows show the direction of the primary ionising sources with the base of the arrow placed at the coordinates
of the IRAS source catalogued in \citet{SFO91}.NVSS contours start at 3$\sigma$ and increase in increments of 20\% of the peak value. 
MSX contours start at 3$\sigma$ unless otherwise specified and increase in increments of 20\% of the peak value. }
\label{fig:images}
\end{figure*}
\end{center}
\setcounter{figure}{1}
\begin{center}
\begin{figure*}
SFO 6
\includegraphics*[scale=0.50]{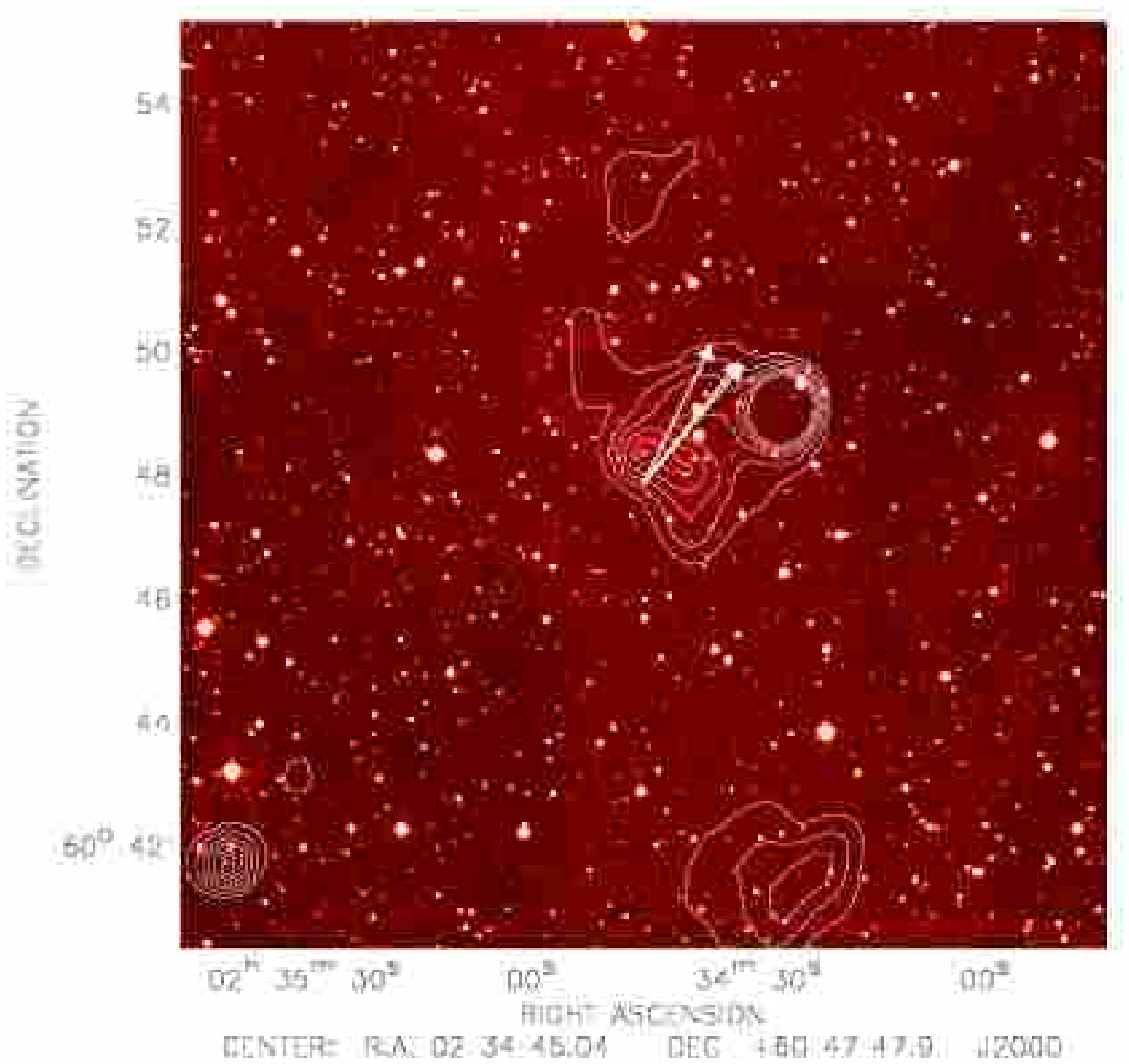}
\includegraphics*[scale=0.50]{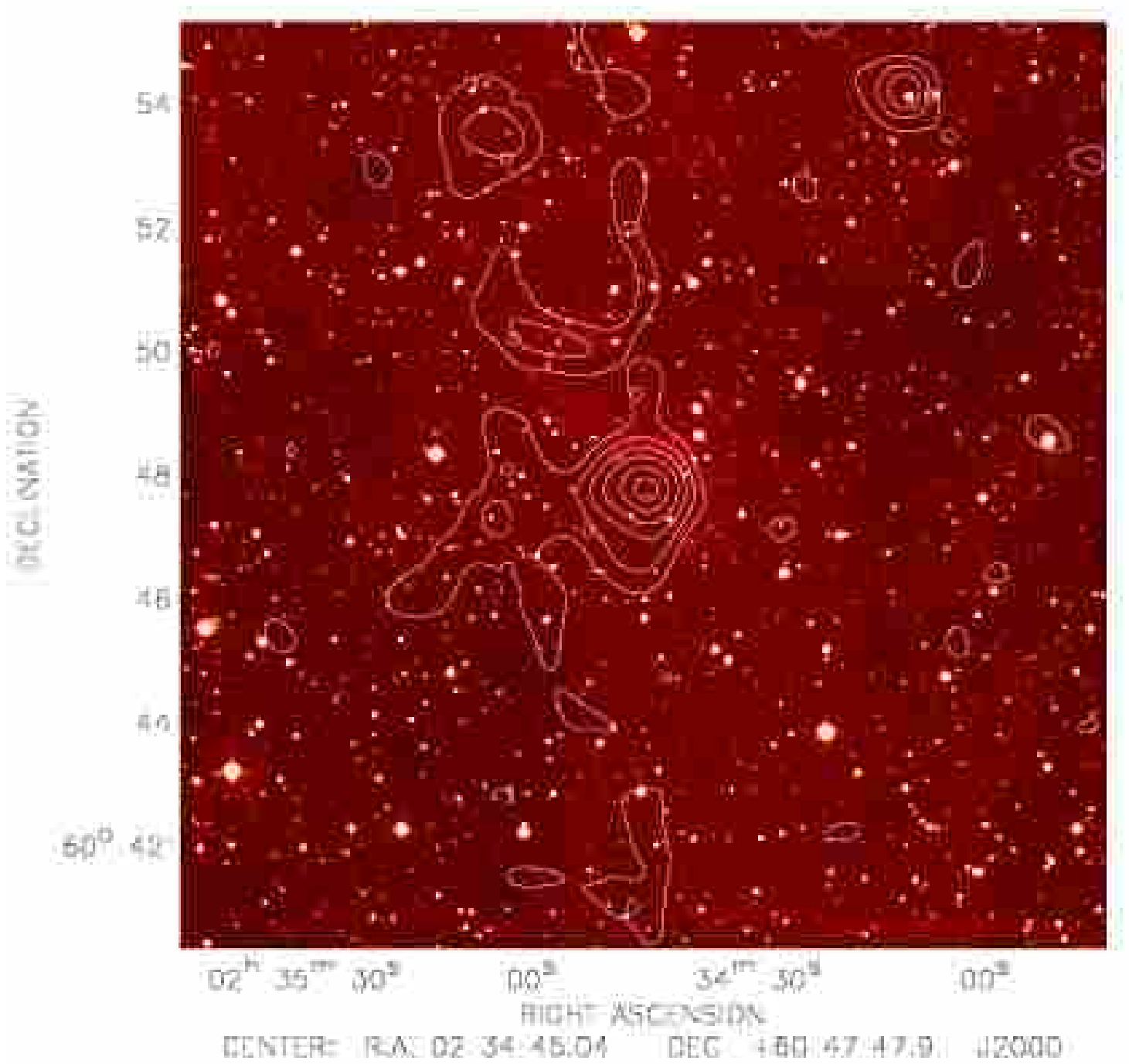}\\
SFO 7 %MSX contours start at 6$\sigma$ for clarity
\includegraphics*[scale=0.50]{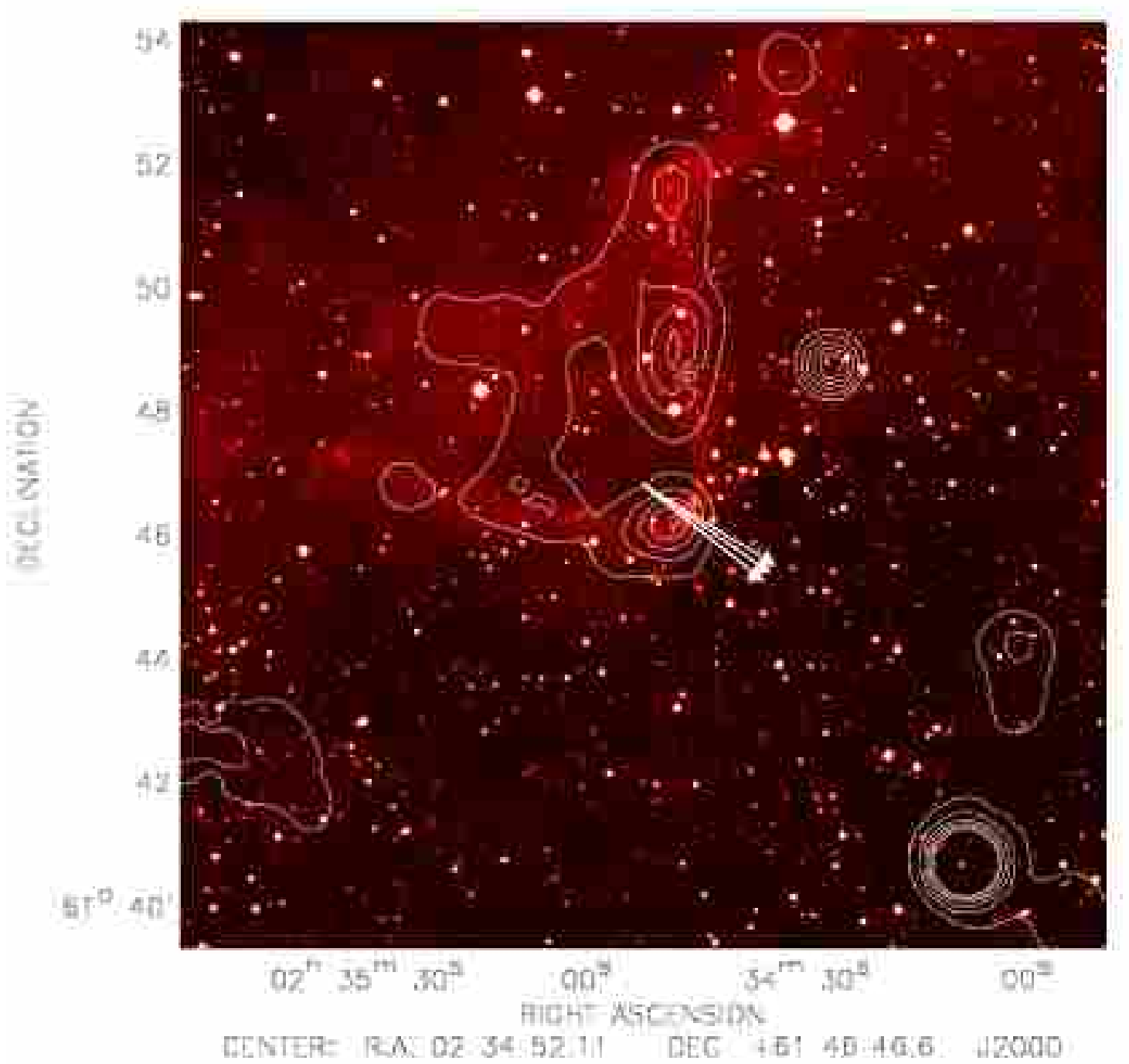}
\includegraphics*[scale=0.50]{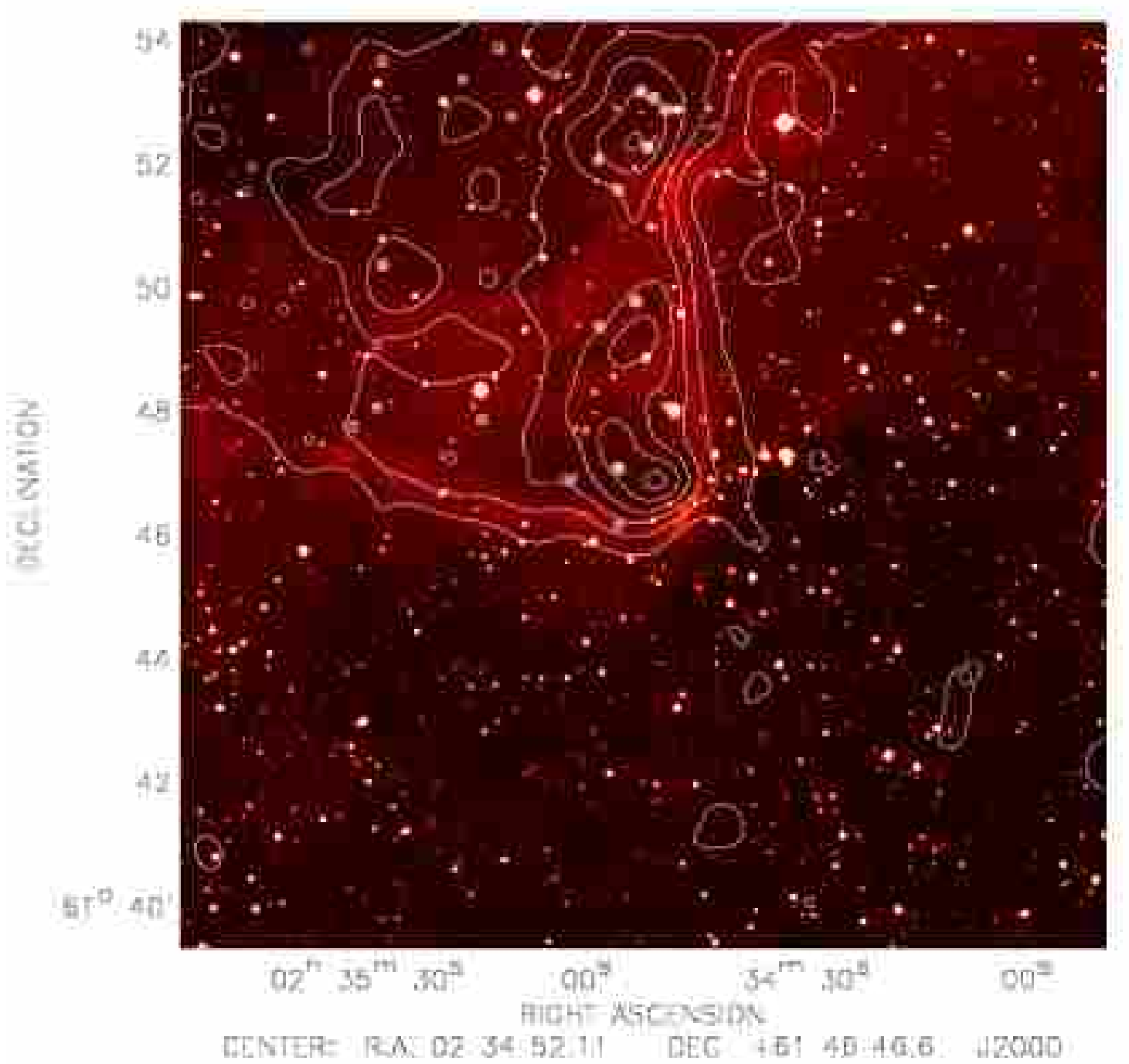}\\
SFO 10 %MSX contours start at 9$\sigma$ for clarity
\includegraphics*[scale=0.50]{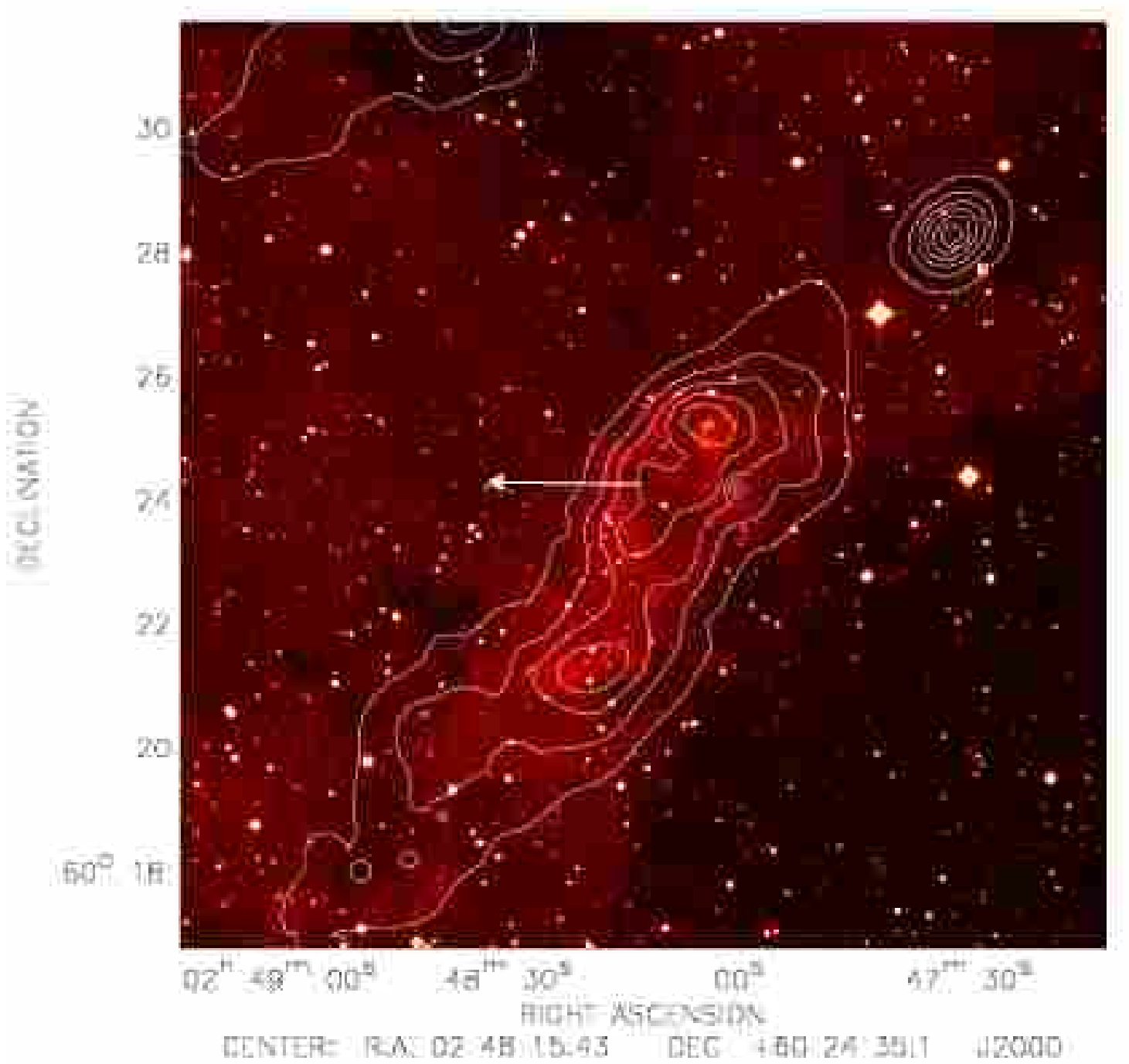}
\includegraphics*[scale=0.50]{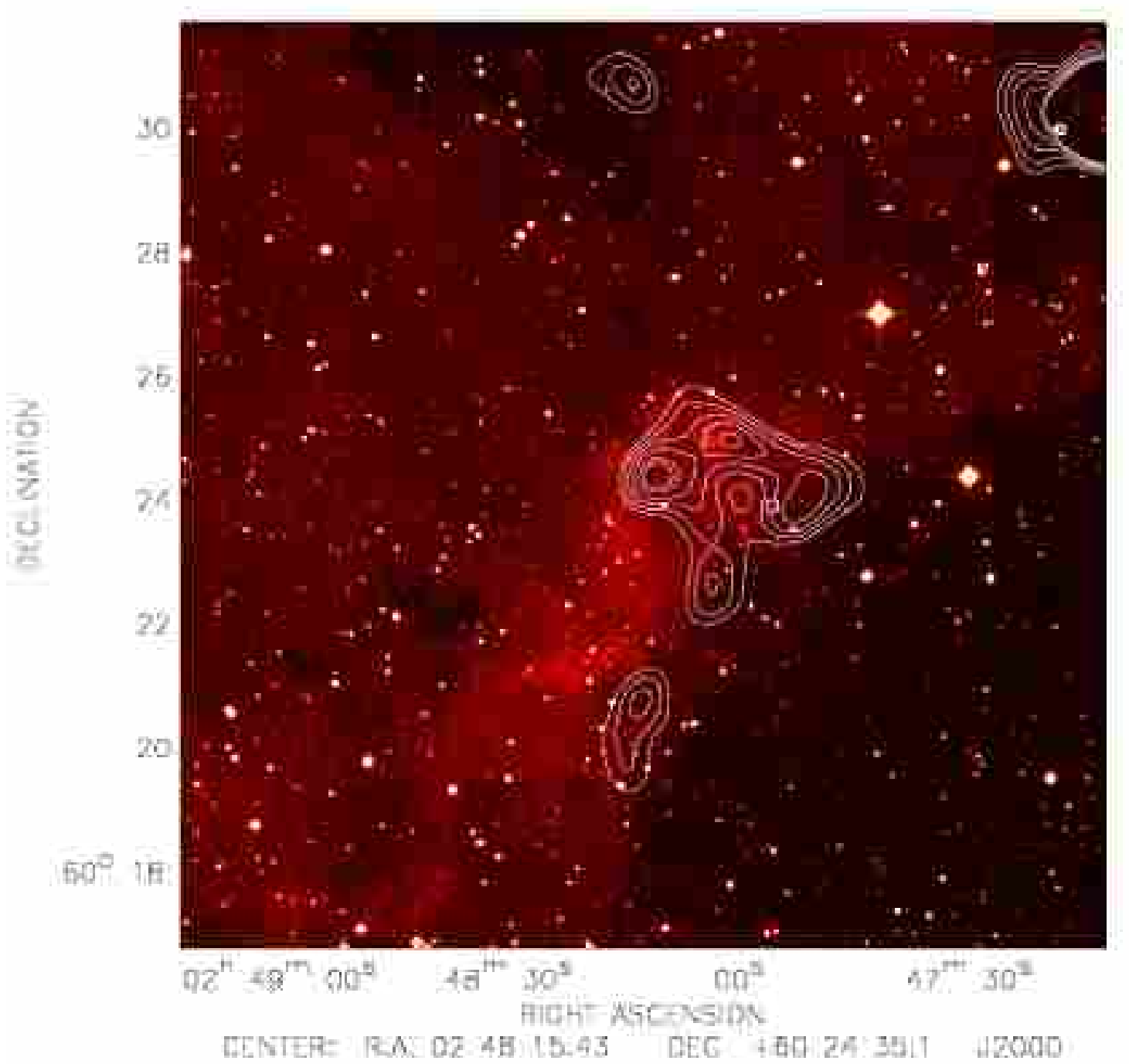}\\
\caption{\bf{(cont.)} \sf SFO 6 MSX contours start at 6$\sigma$, SFO 7 MSX contours start at 6$\sigma$ and SFO 10 MSX contours start at 9$\sigma$ for clarity.}
\end{figure*}
\end{center}
\setcounter{figure}{1}
\begin{center}
\begin{figure*}
SFO 11 %MSX contours start at 9$\sigma$ for clarity
\includegraphics*[scale=0.50]{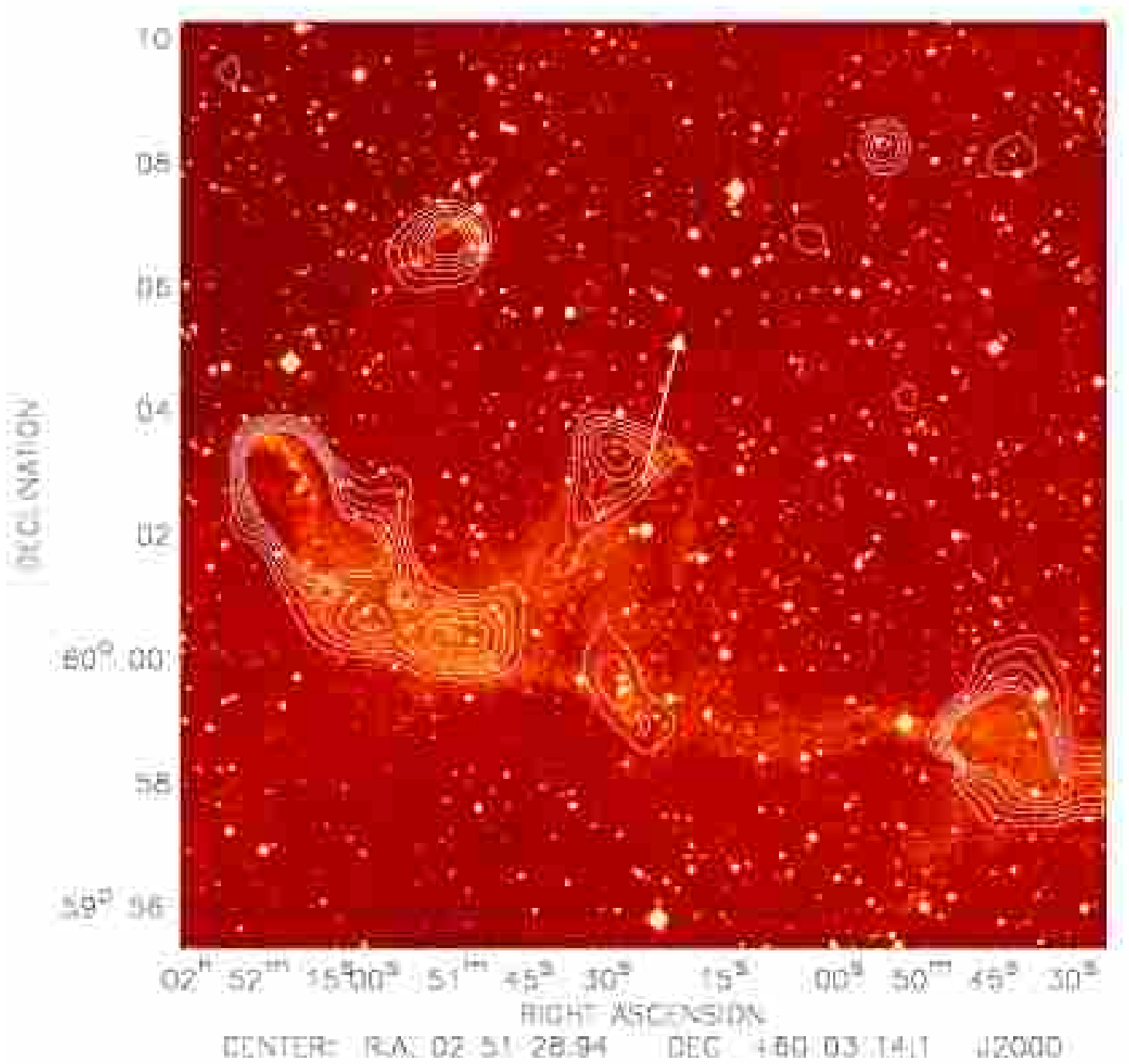}
\includegraphics*[scale=0.50]{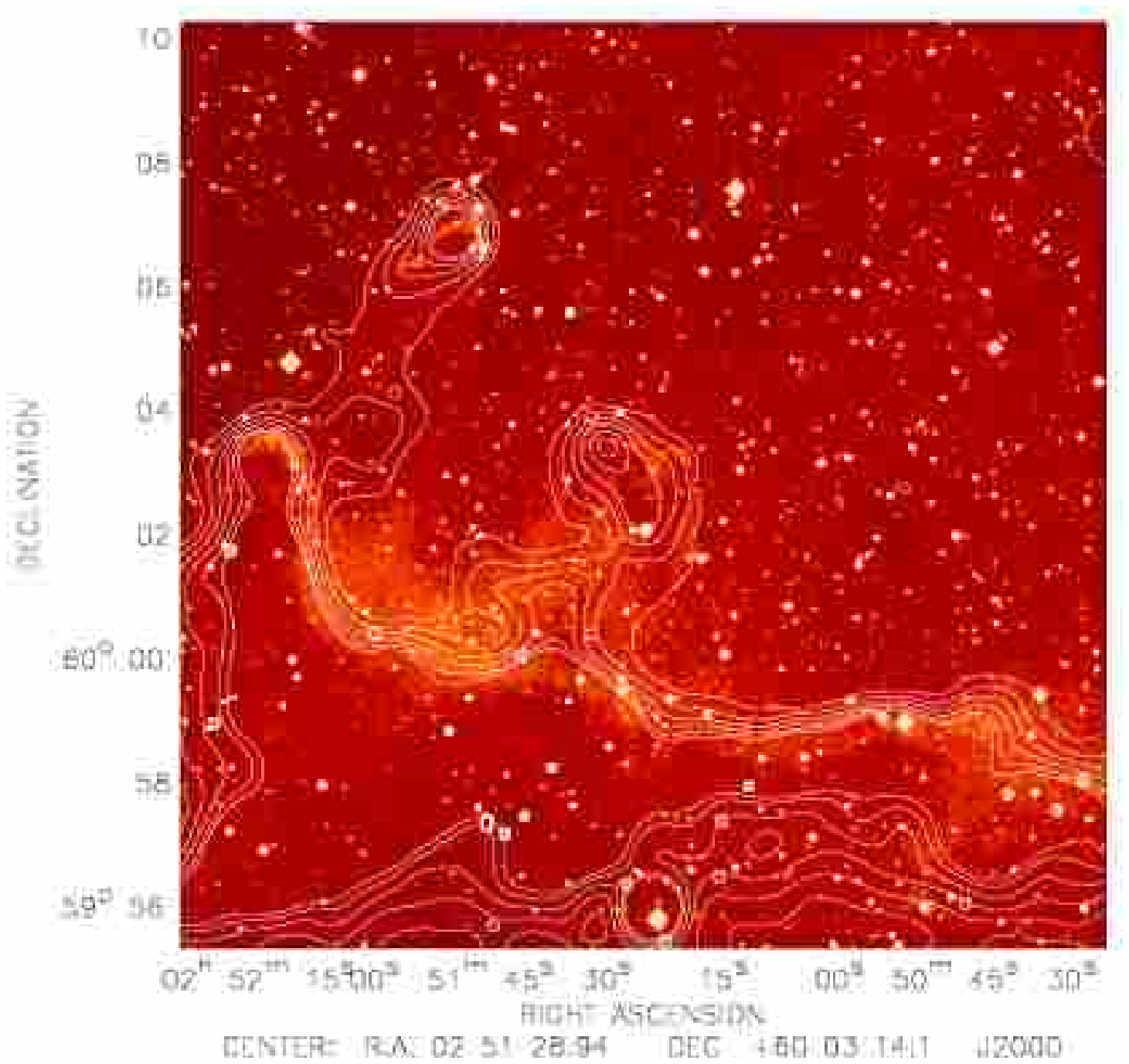}\\
SFO 12
\includegraphics*[scale=0.50]{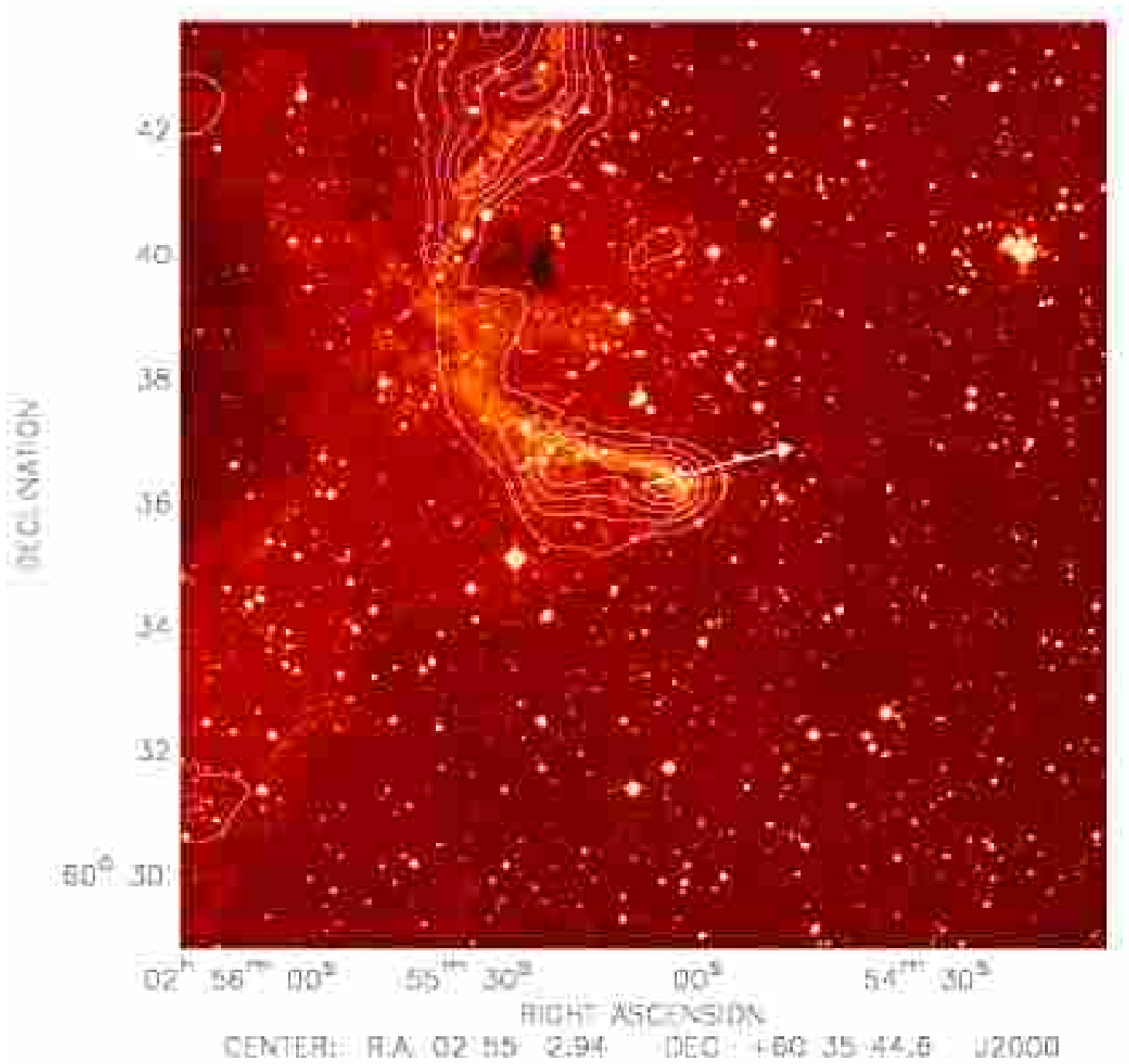}
\includegraphics*[scale=0.50]{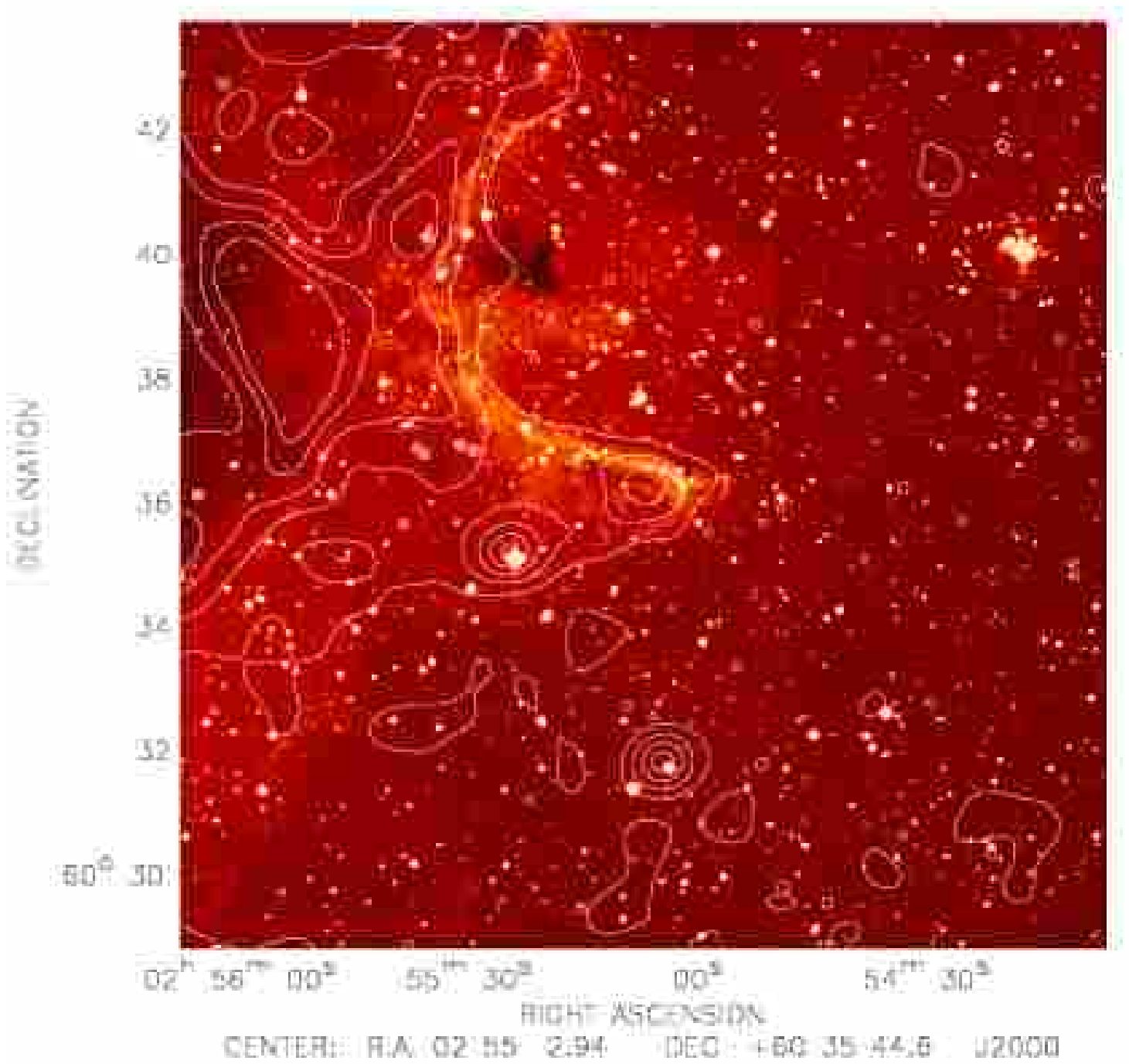}\\
SFO 13
\includegraphics*[scale=0.50]{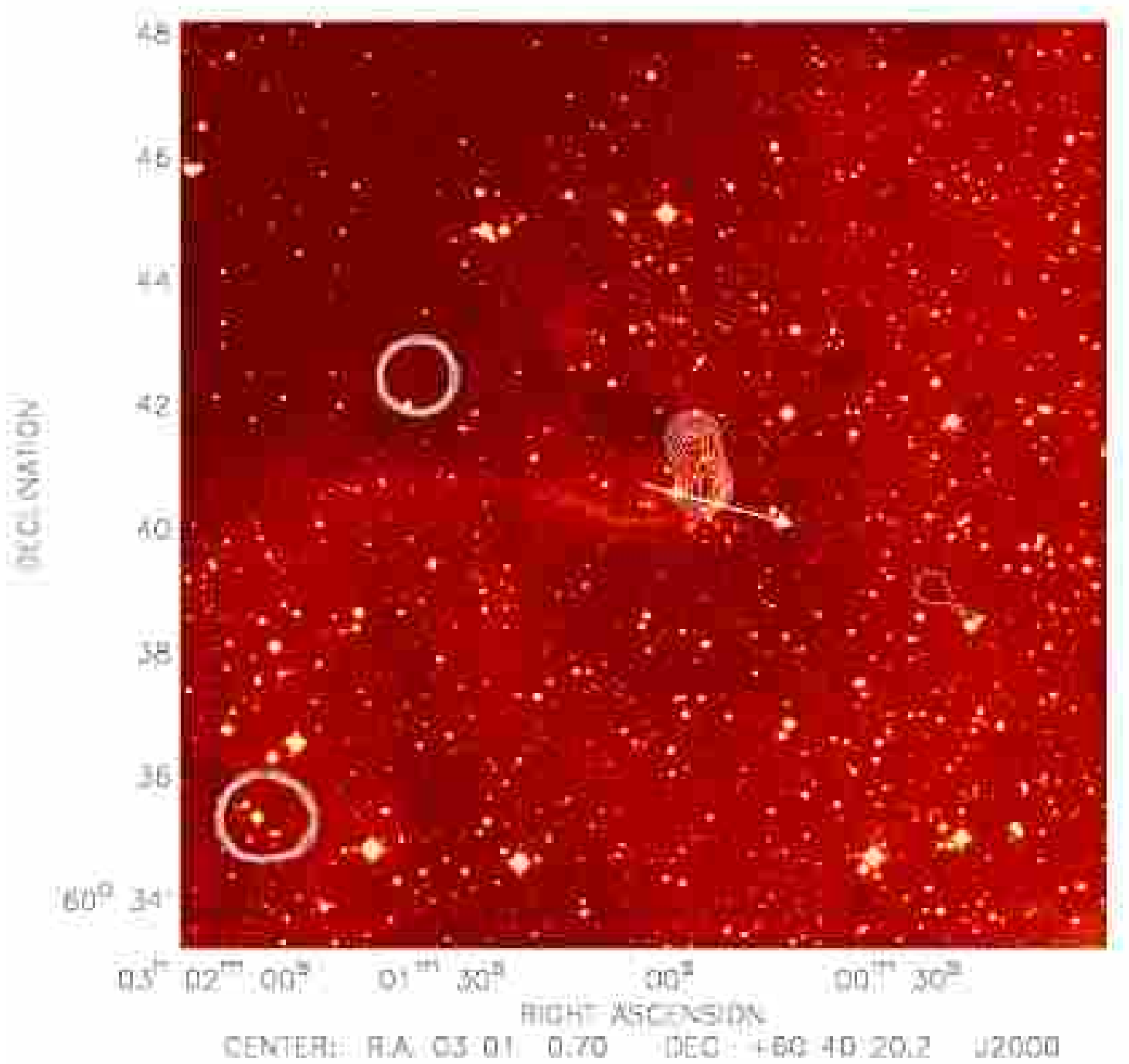}
\includegraphics*[scale=0.50]{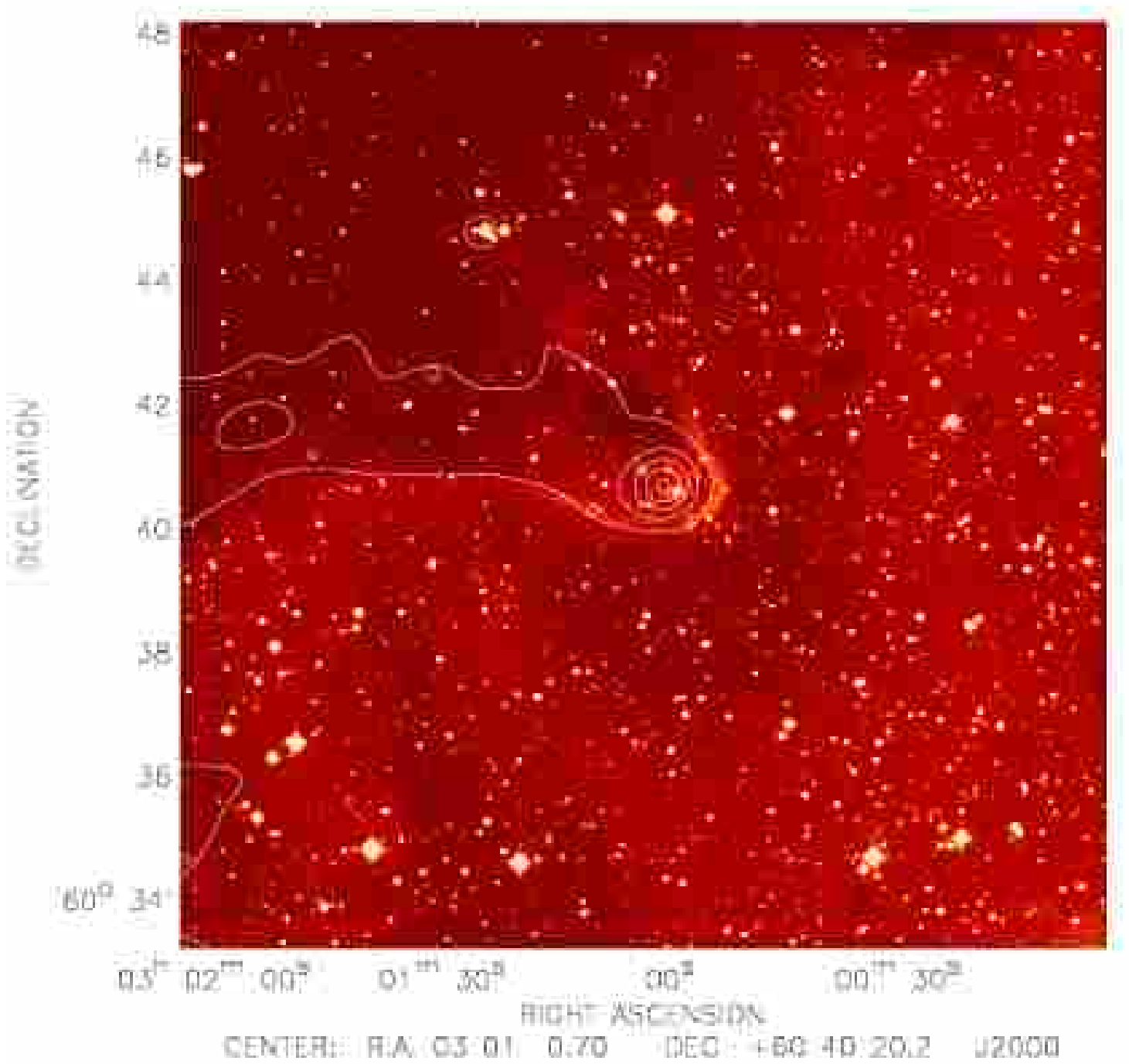}\\
\caption{\bf{(cont.)} \sf SFO 11 MSX contours start at 9$\sigma$ for clarity.}
\end{figure*}
\end{center}
\setcounter{figure}{1}
\begin{center}
\begin{figure*}
SFO 14
\includegraphics*[scale=0.50]{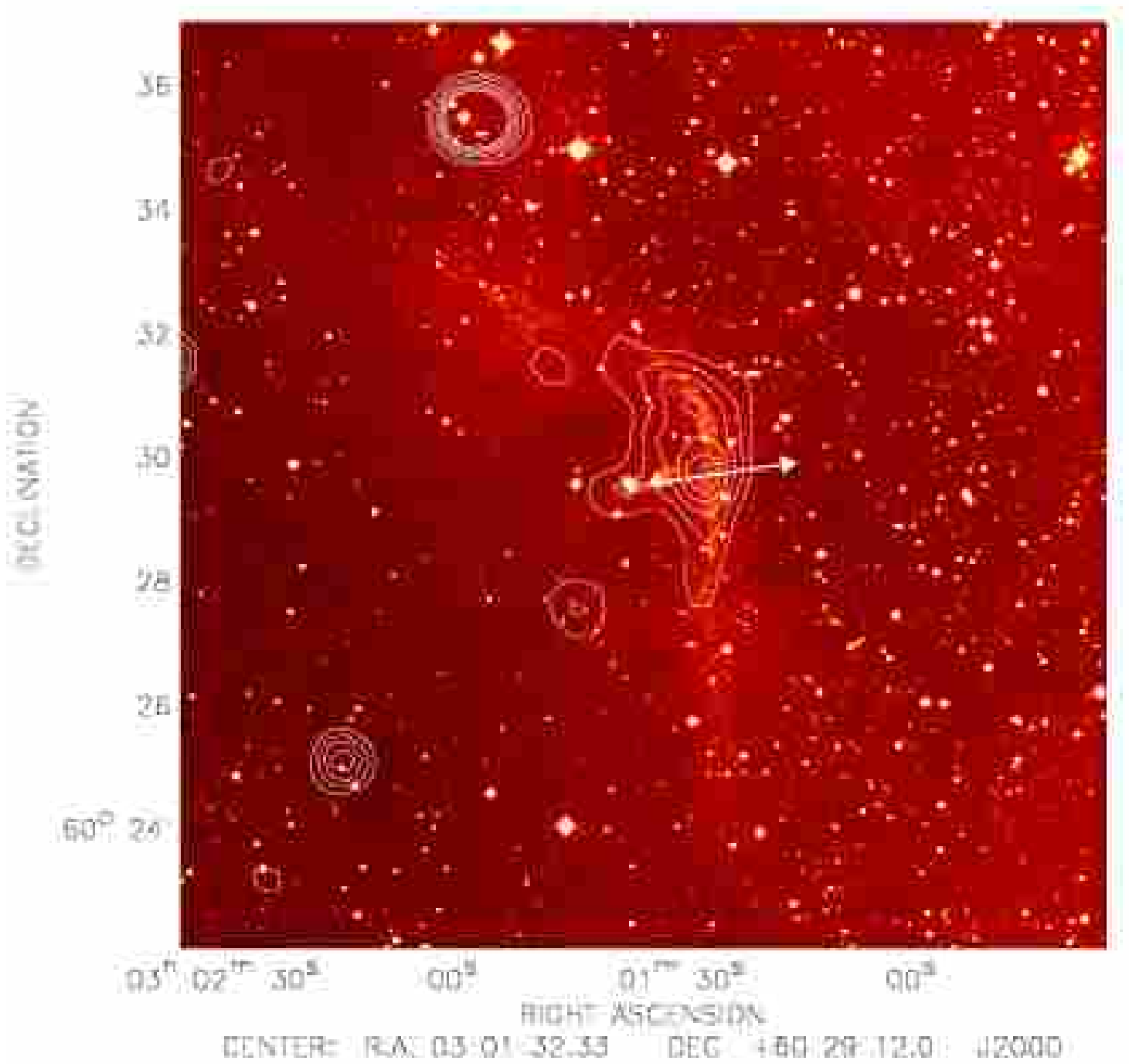}
\includegraphics*[scale=0.50]{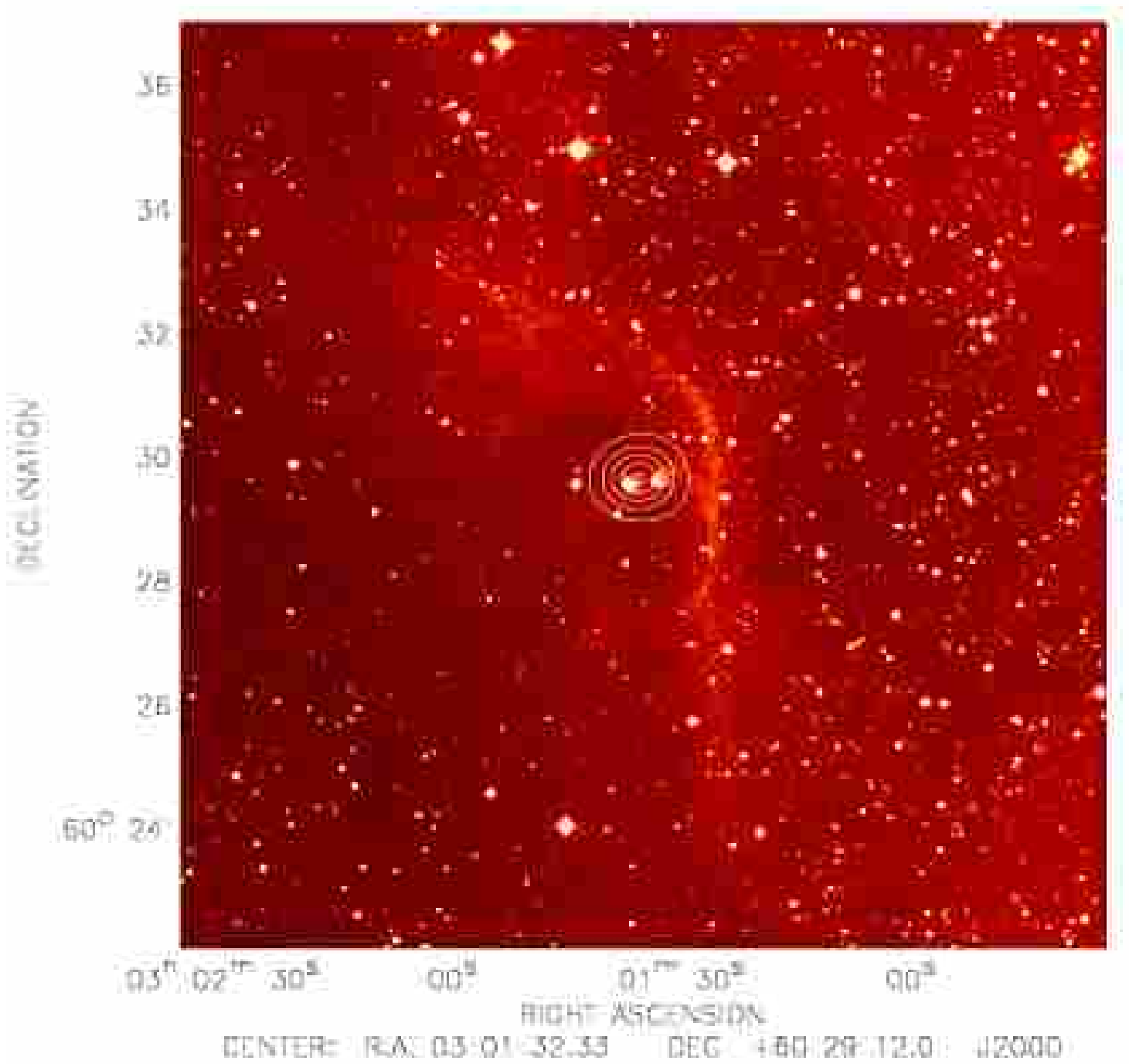}\\
SFO 15 
\includegraphics*[scale=0.50]{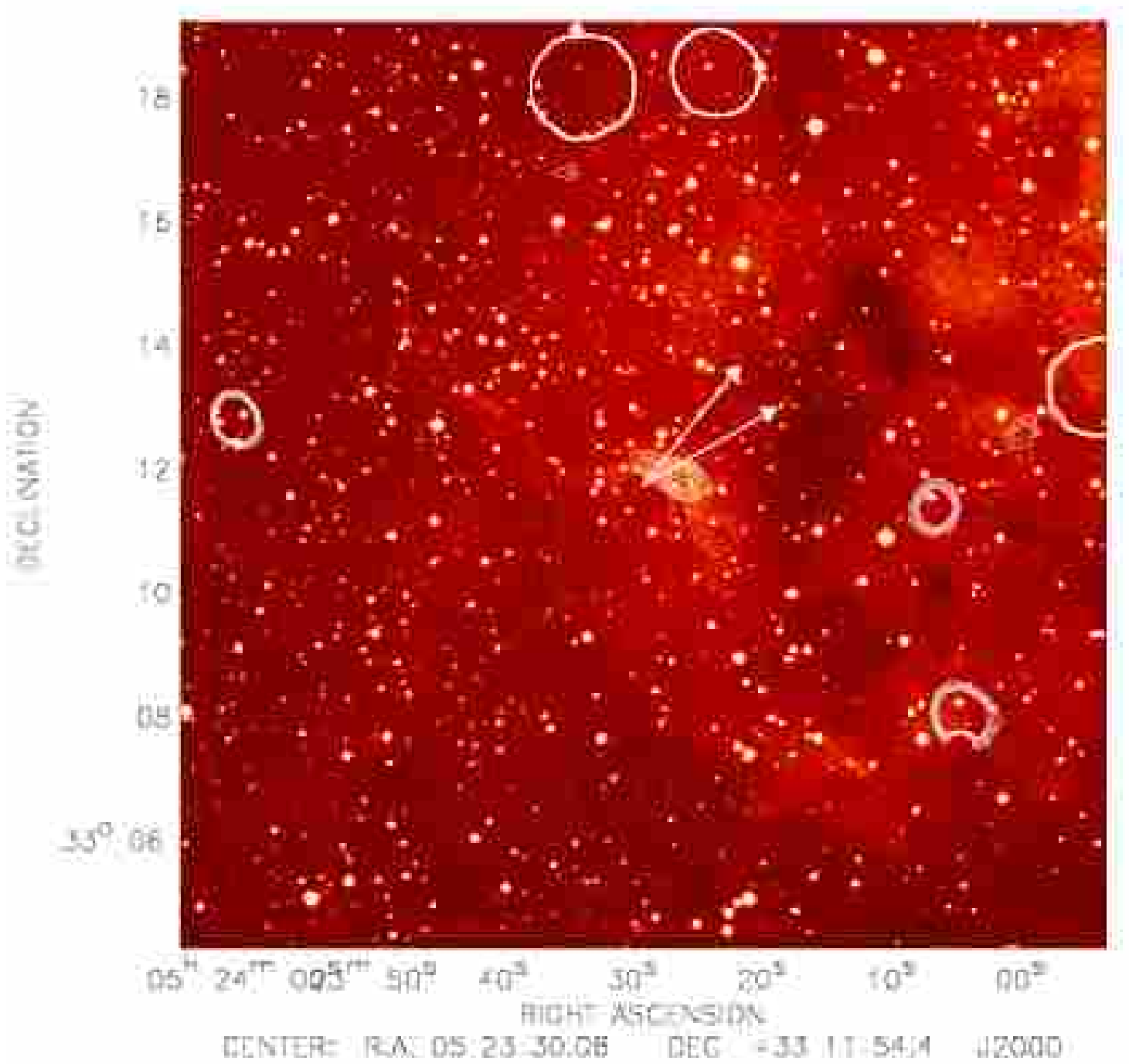}
\includegraphics*[scale=0.50]{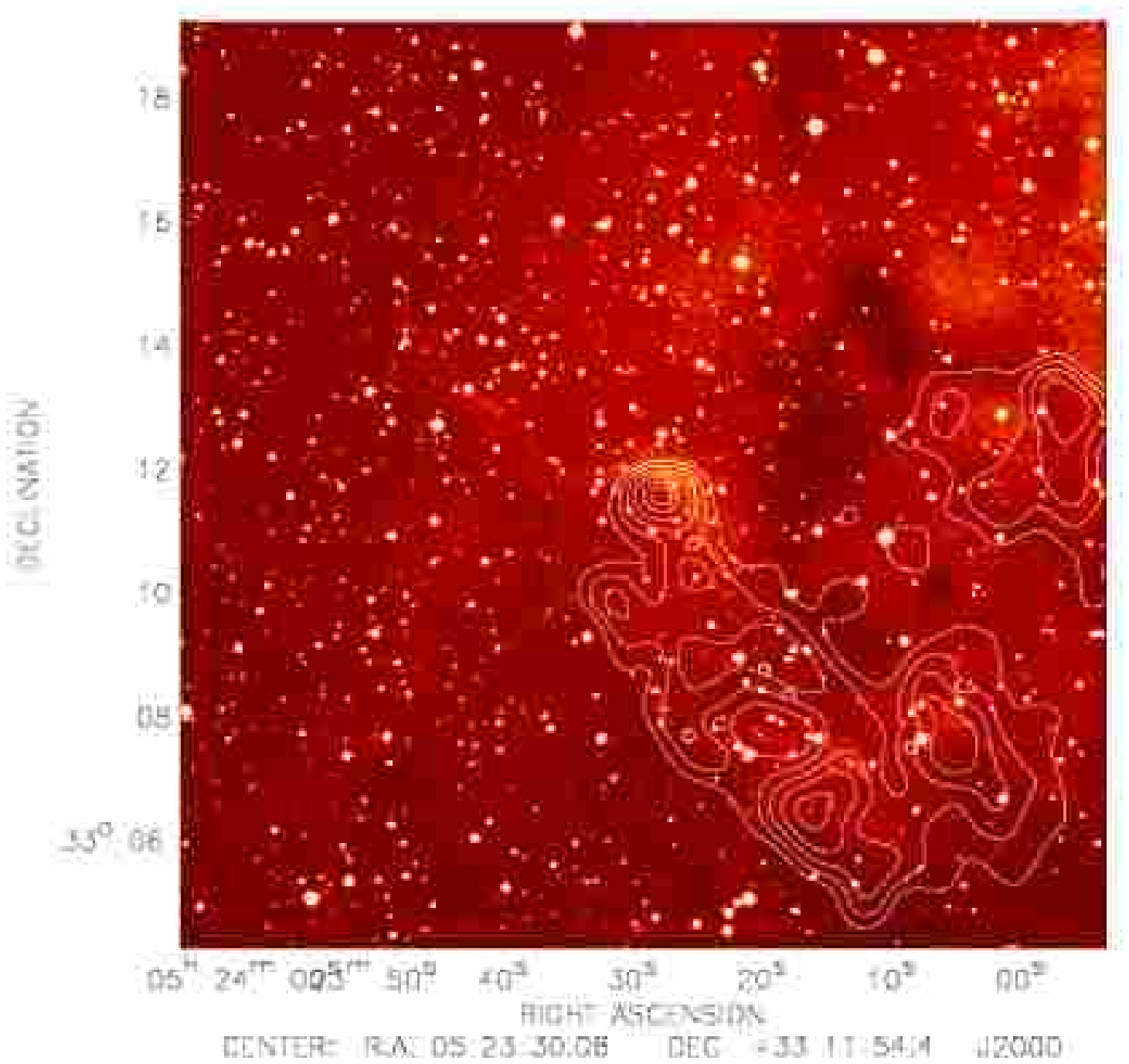}\\
SFO 25
\includegraphics*[scale=0.50]{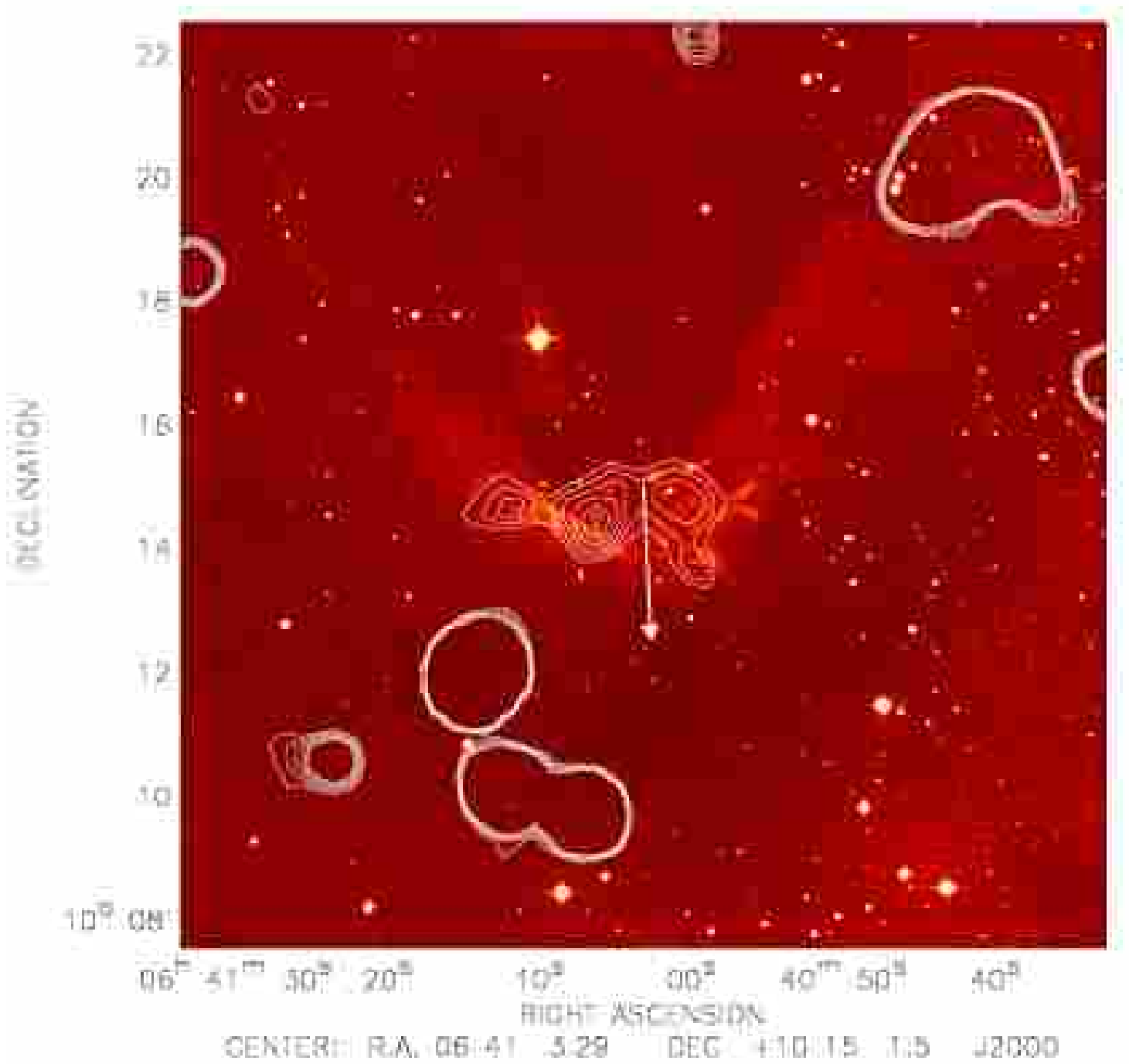}
\includegraphics*[scale=0.50]{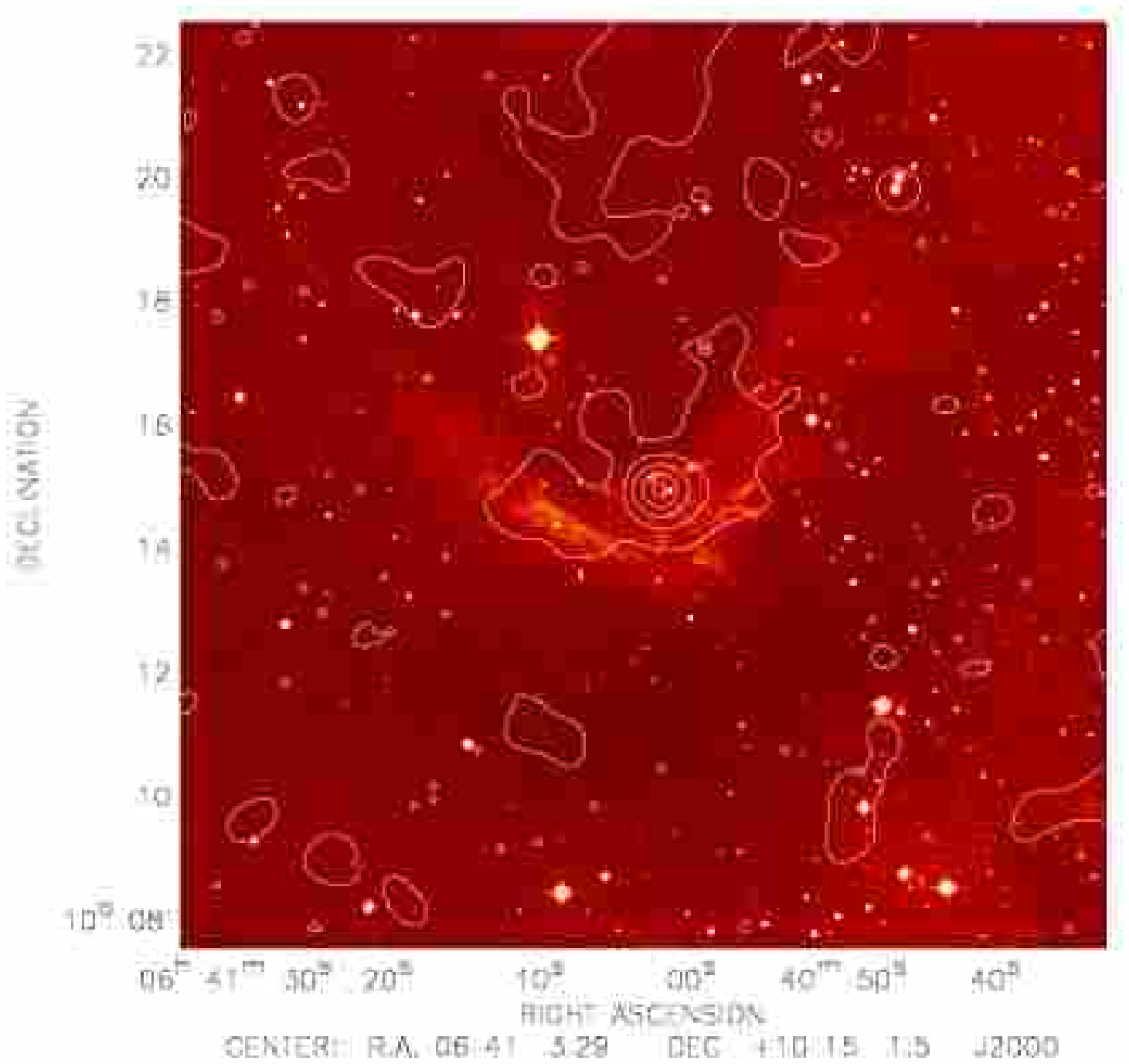}\\
\caption{\bf{(cont.)} \sf SFO 15 MSX contours start at 12$\sigma$ for clarity.}
\end{figure*}
\end{center}
\setcounter{figure}{1}
\begin{center}
\begin{figure*}
SFO 27
\includegraphics*[scale=0.50]{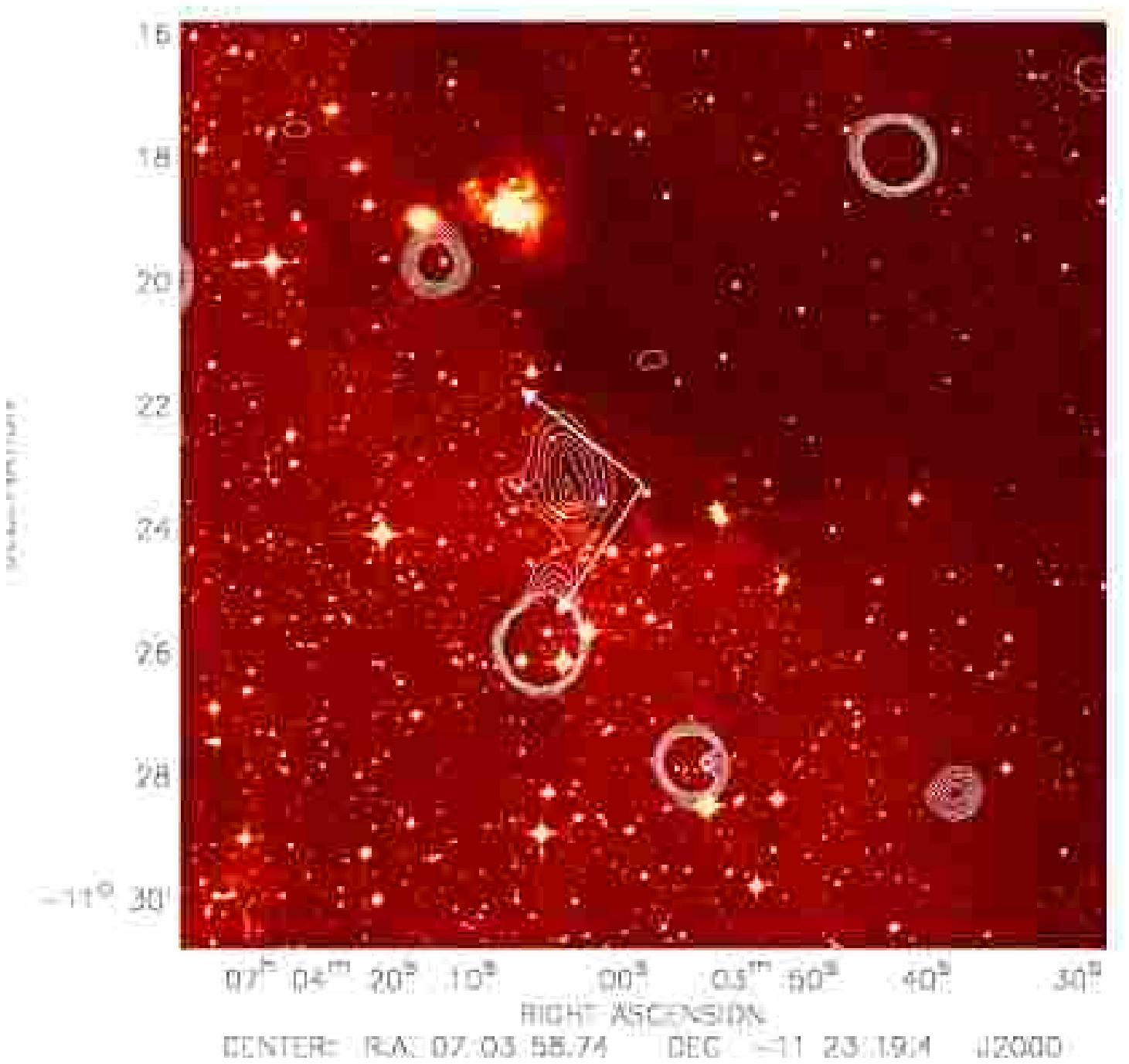}
\includegraphics*[scale=0.50]{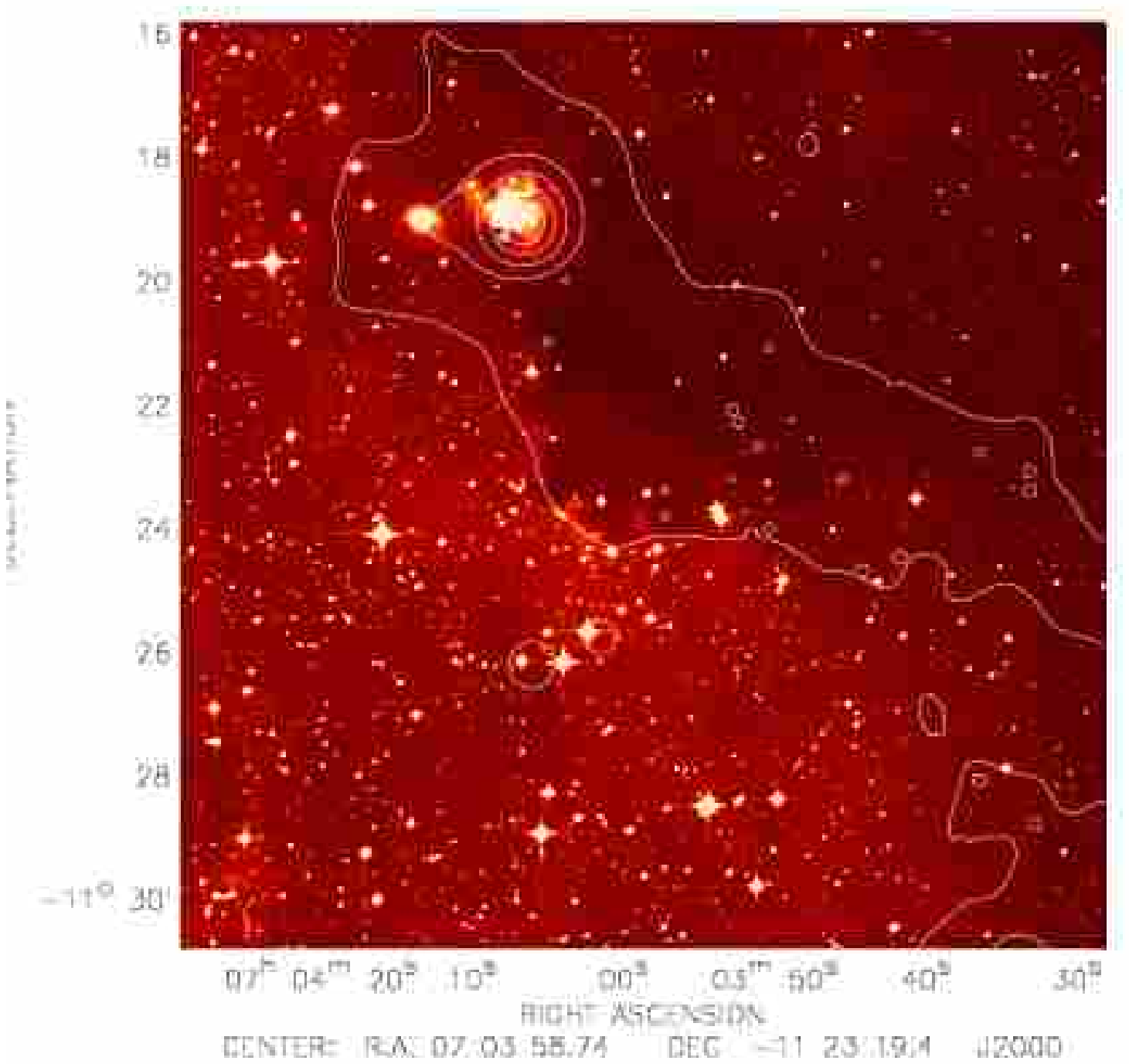}\\
SFO 30
\includegraphics*[scale=0.50]{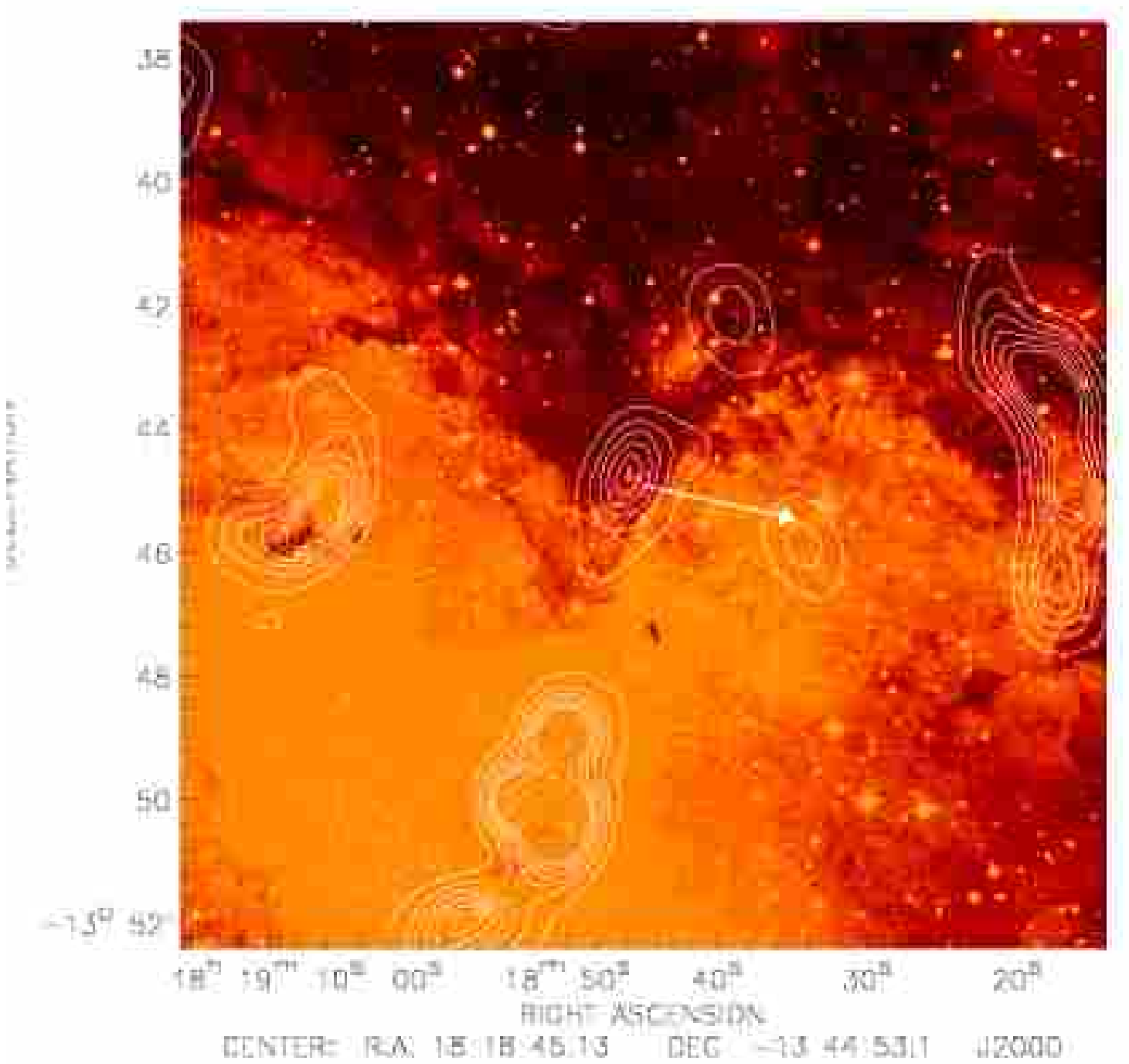}
\includegraphics*[scale=0.50]{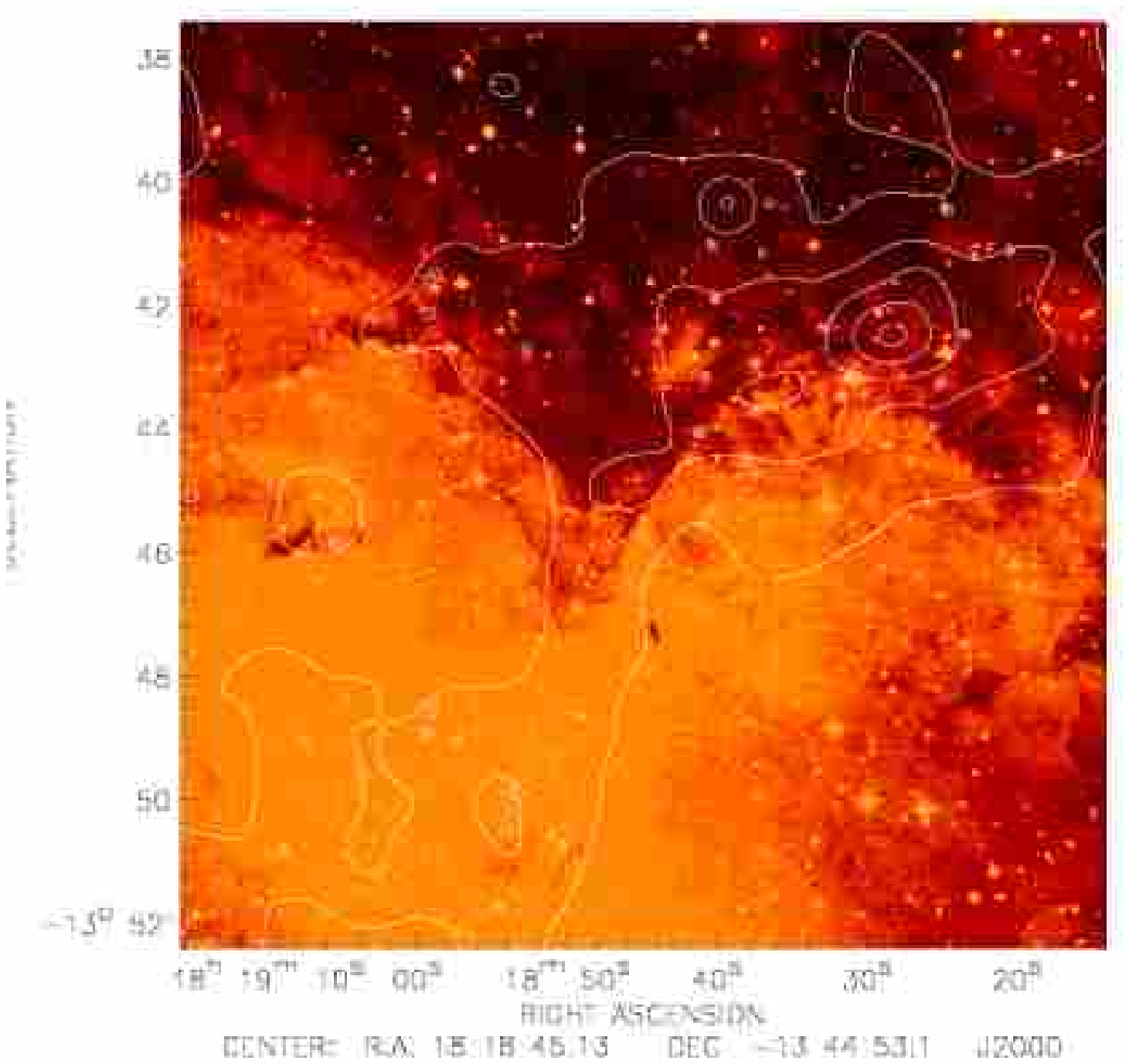}\\
SFO 31 
\includegraphics*[scale=0.50]{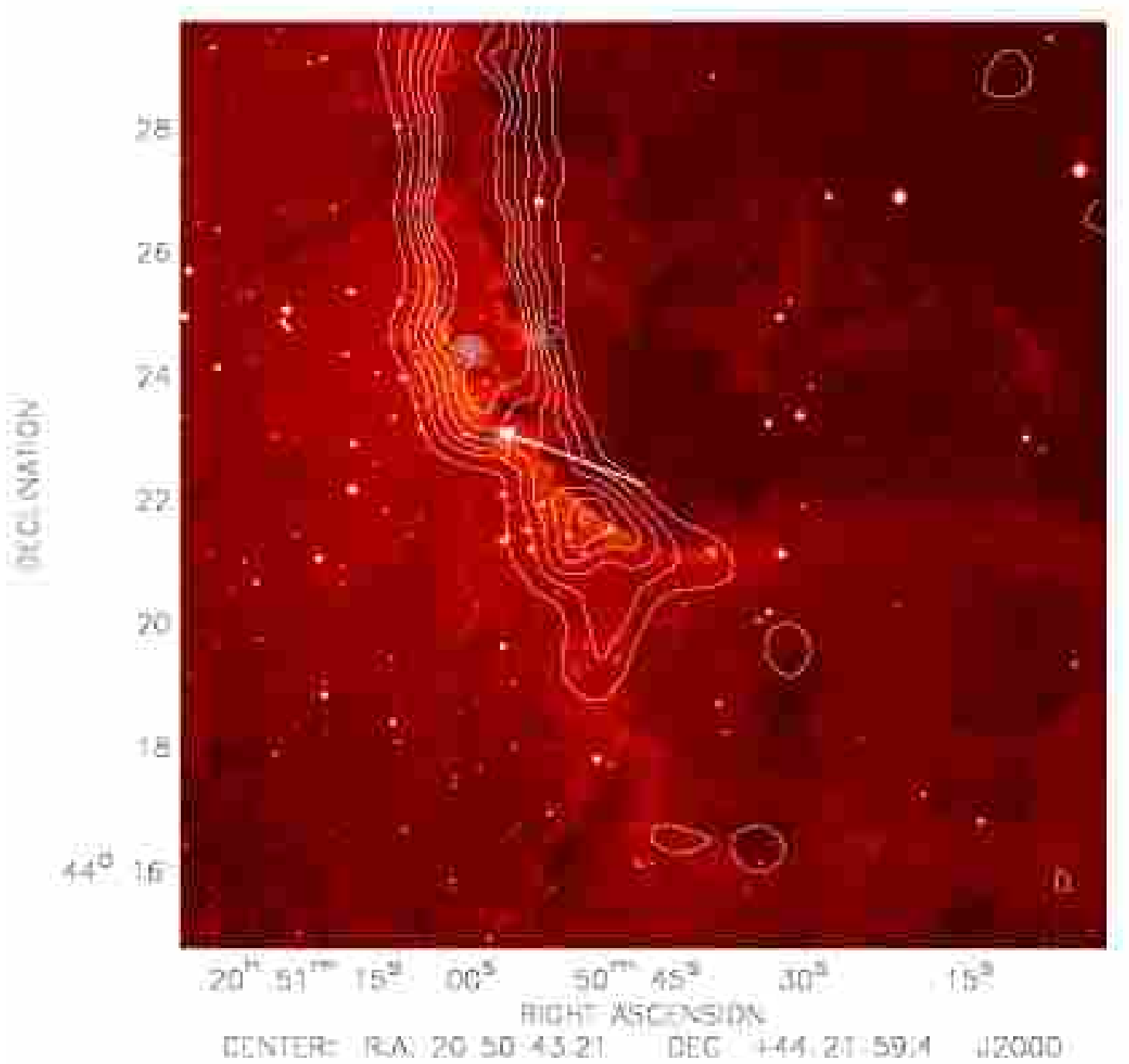}
\includegraphics*[scale=0.50]{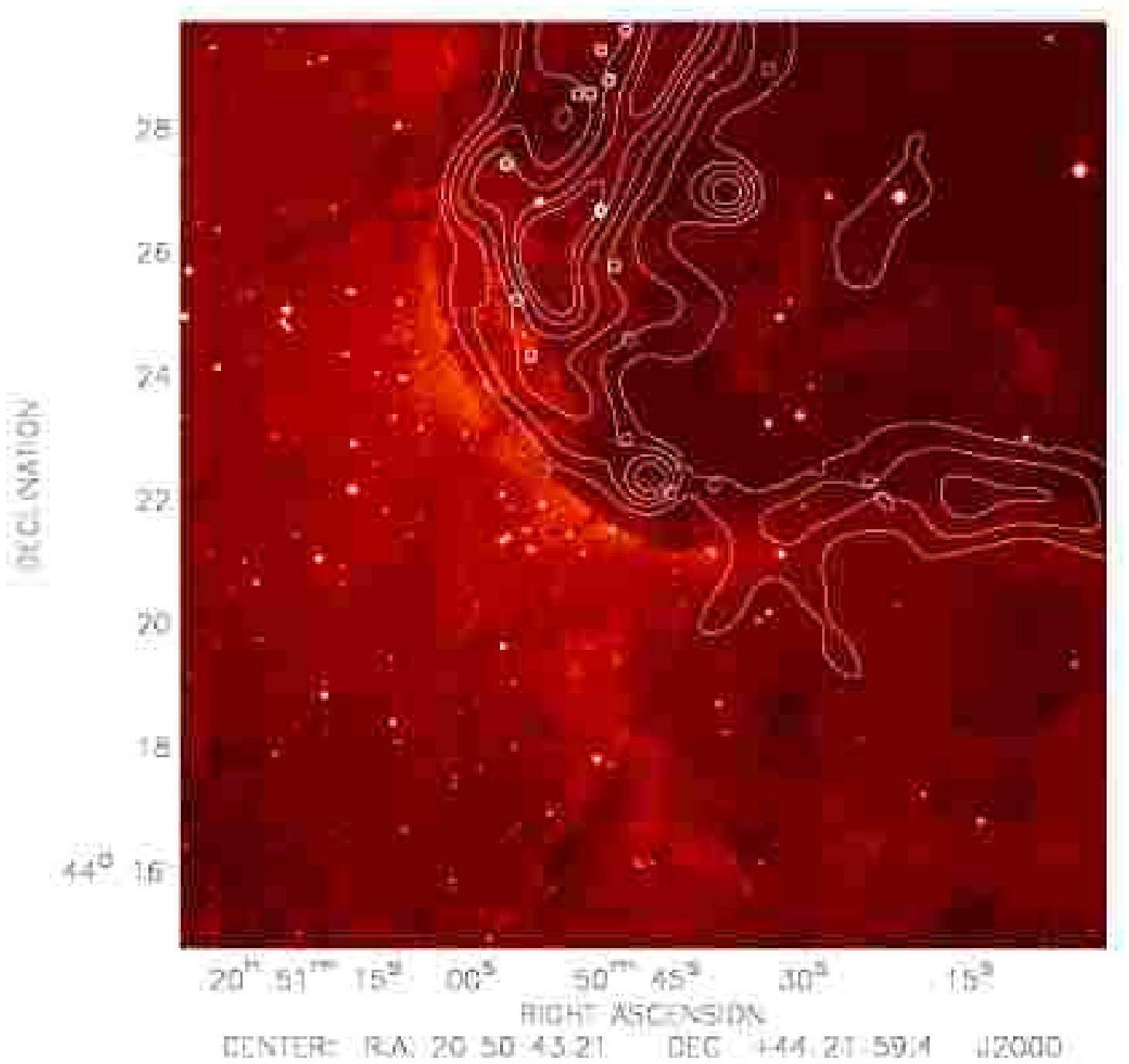}\\
\caption{\bf{(cont.)} \sf SFO 27 MSX contours start at 18$\sigma$ and SFO 31 MSX contours start at 21$\sigma$ for clarity.}
\end{figure*}
\end{center}
\setcounter{figure}{1}
\begin{center}
\begin{figure*}
SFO 32 
\includegraphics*[scale=0.50]{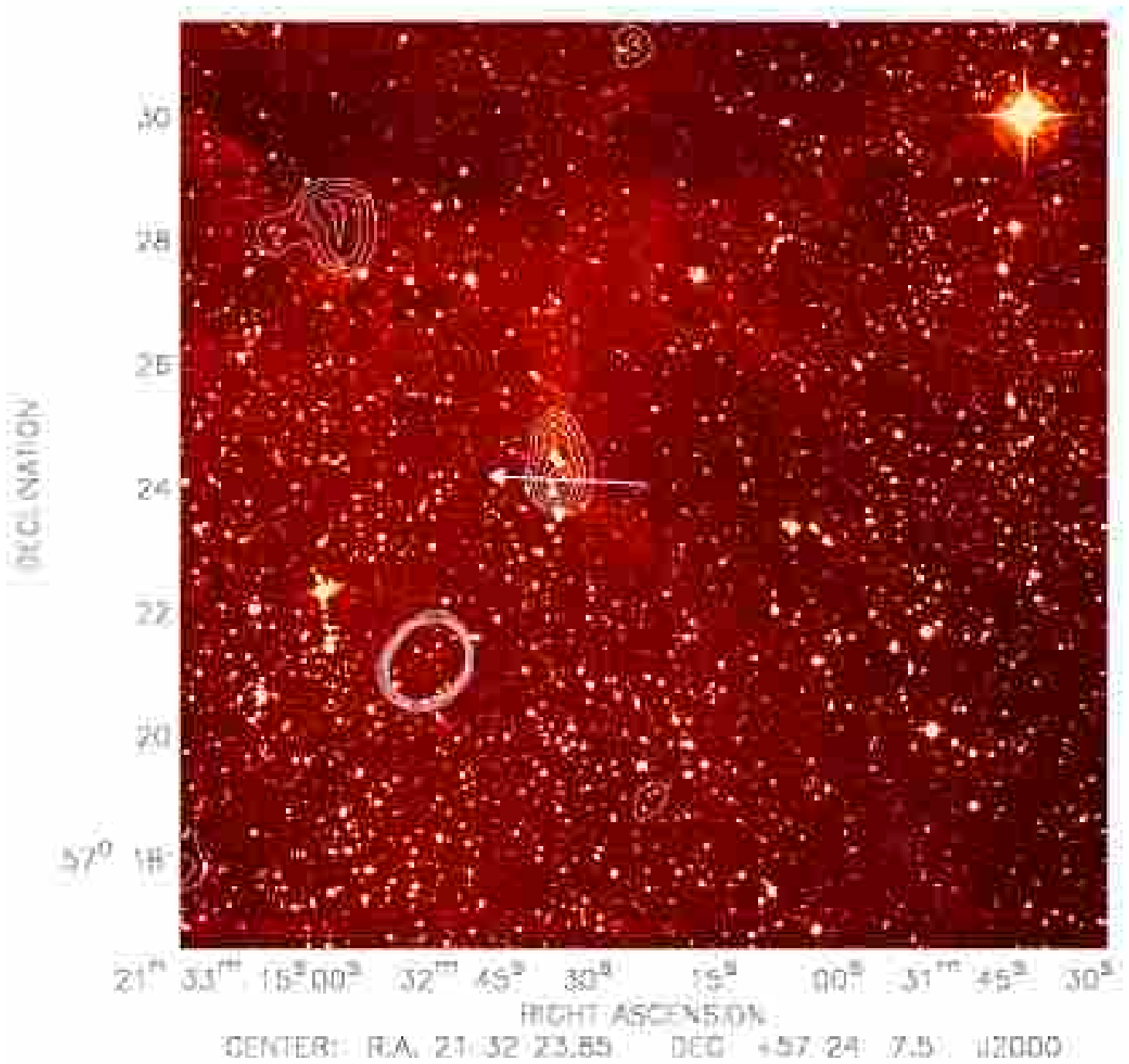}
\includegraphics*[scale=0.50]{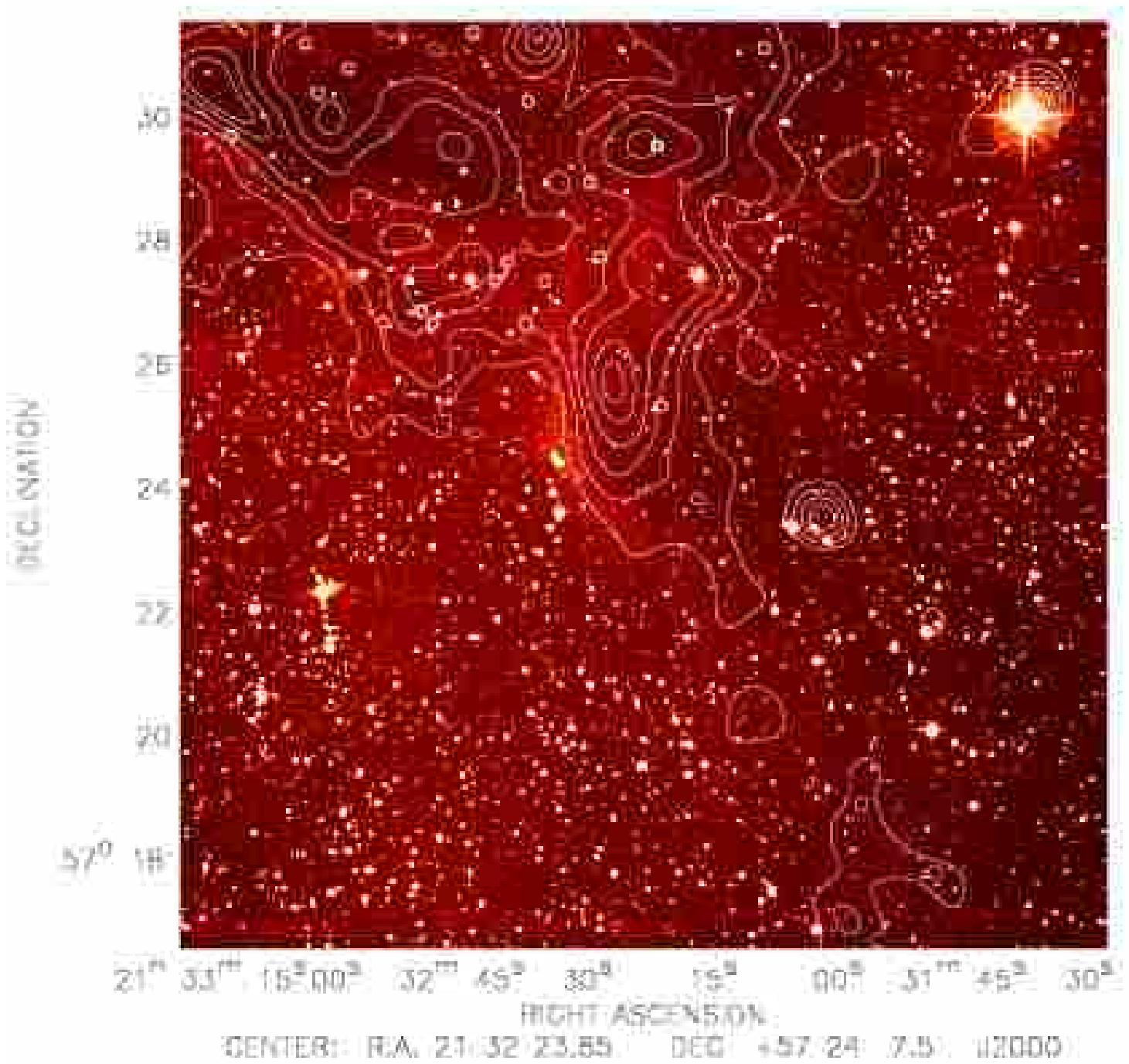}\\
SFO 35 
\includegraphics*[scale=0.50]{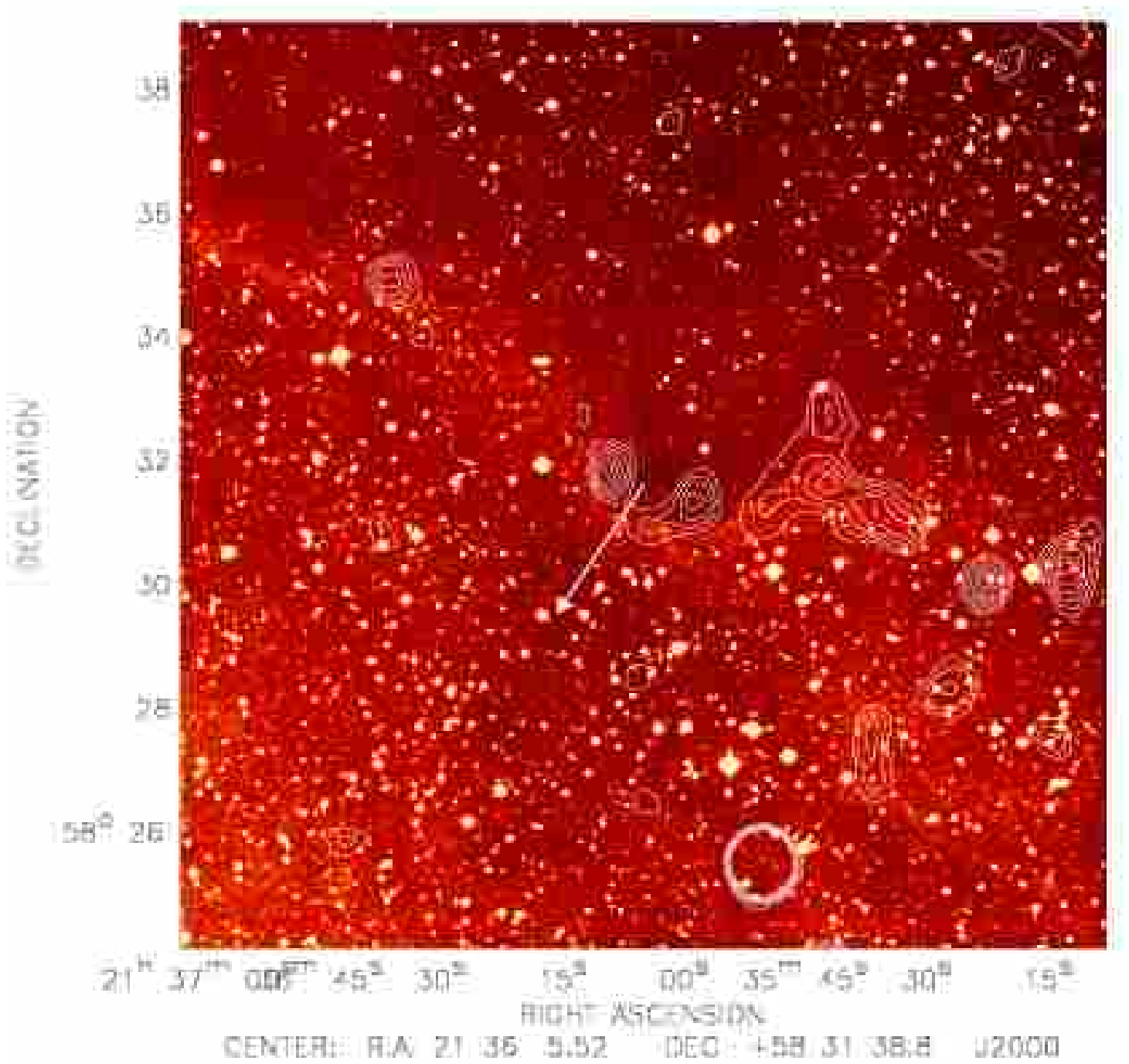}
\includegraphics*[scale=0.50]{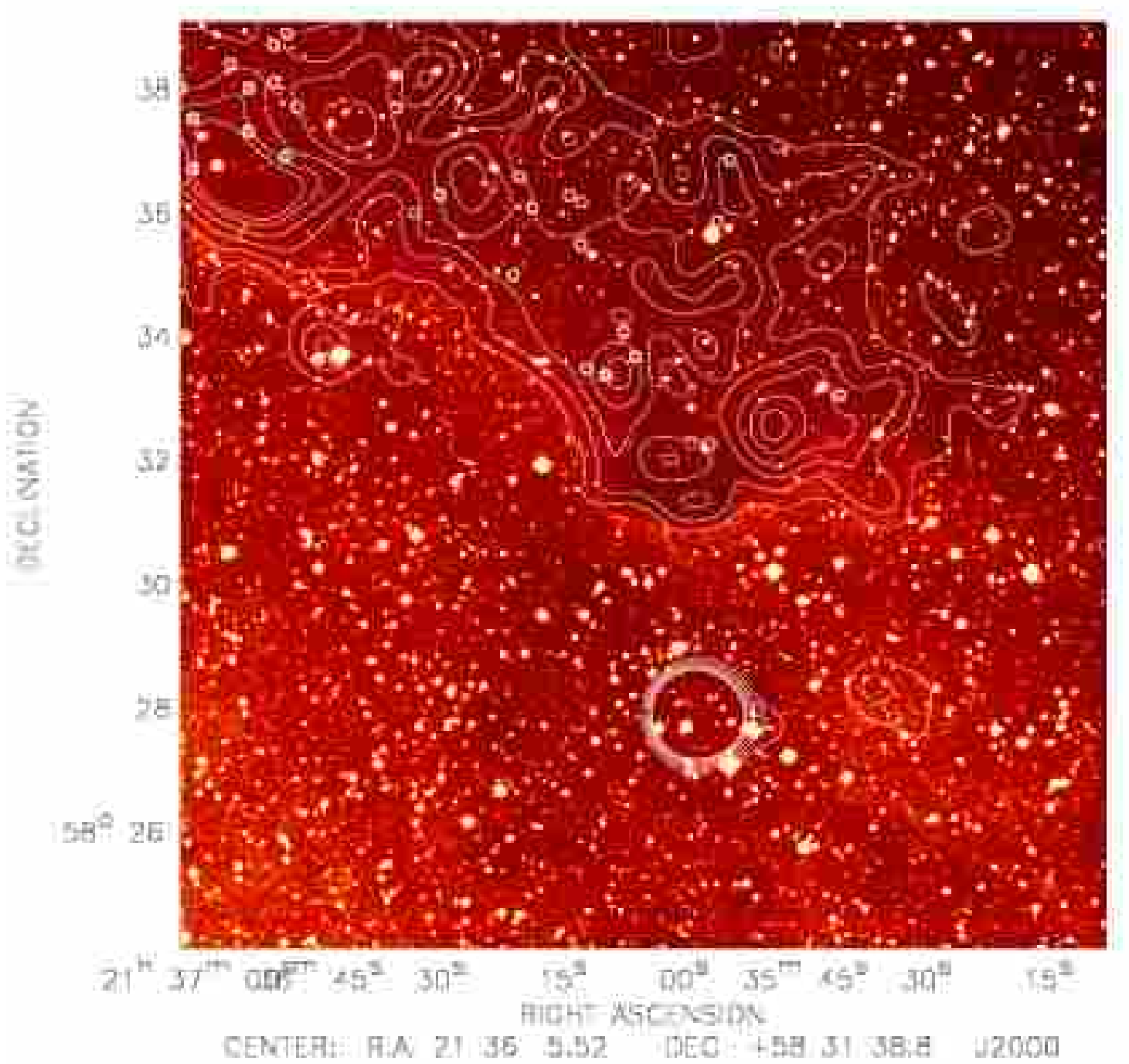}\\
SFO 36
\includegraphics*[scale=0.50]{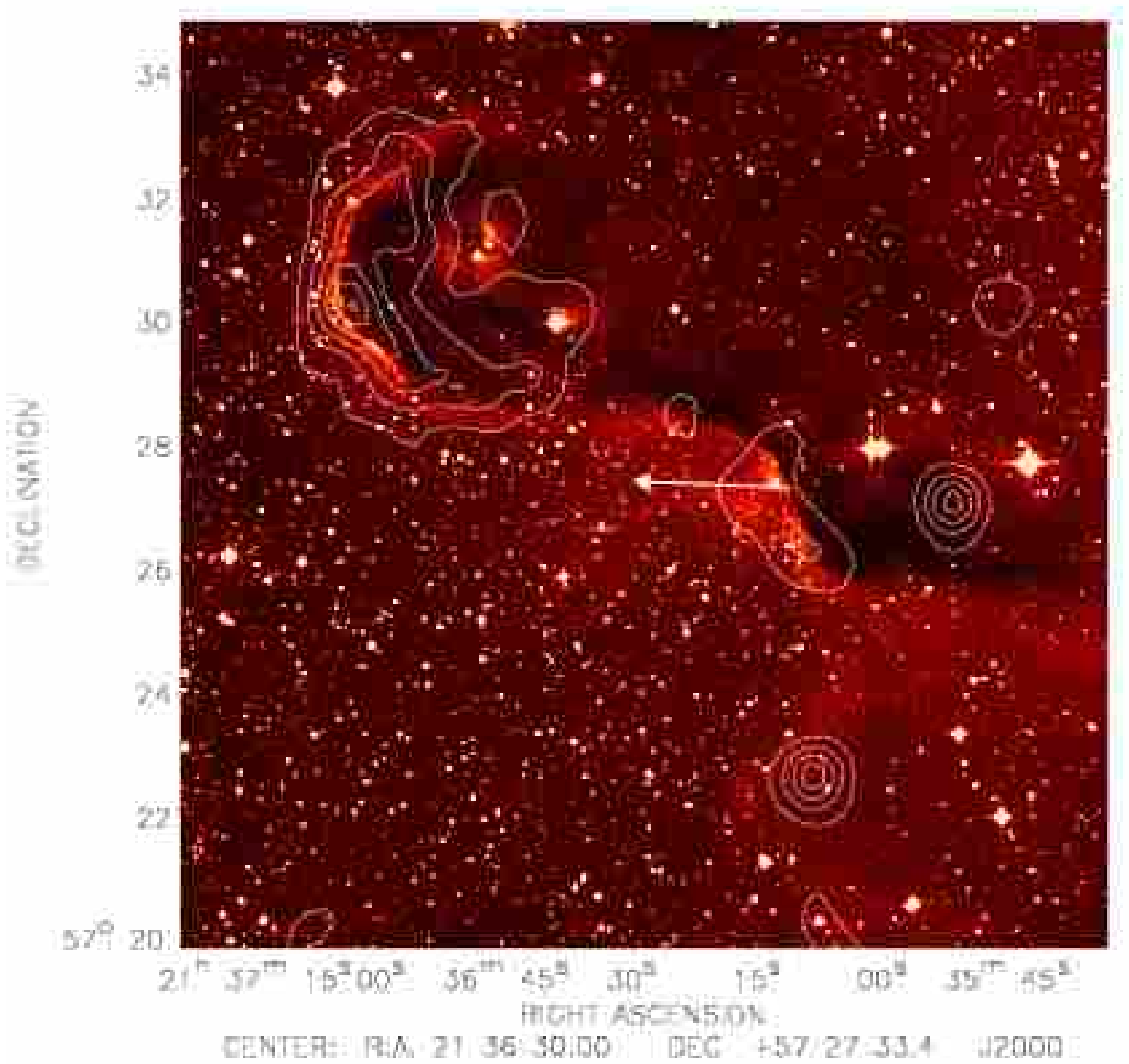}
\includegraphics*[scale=0.50]{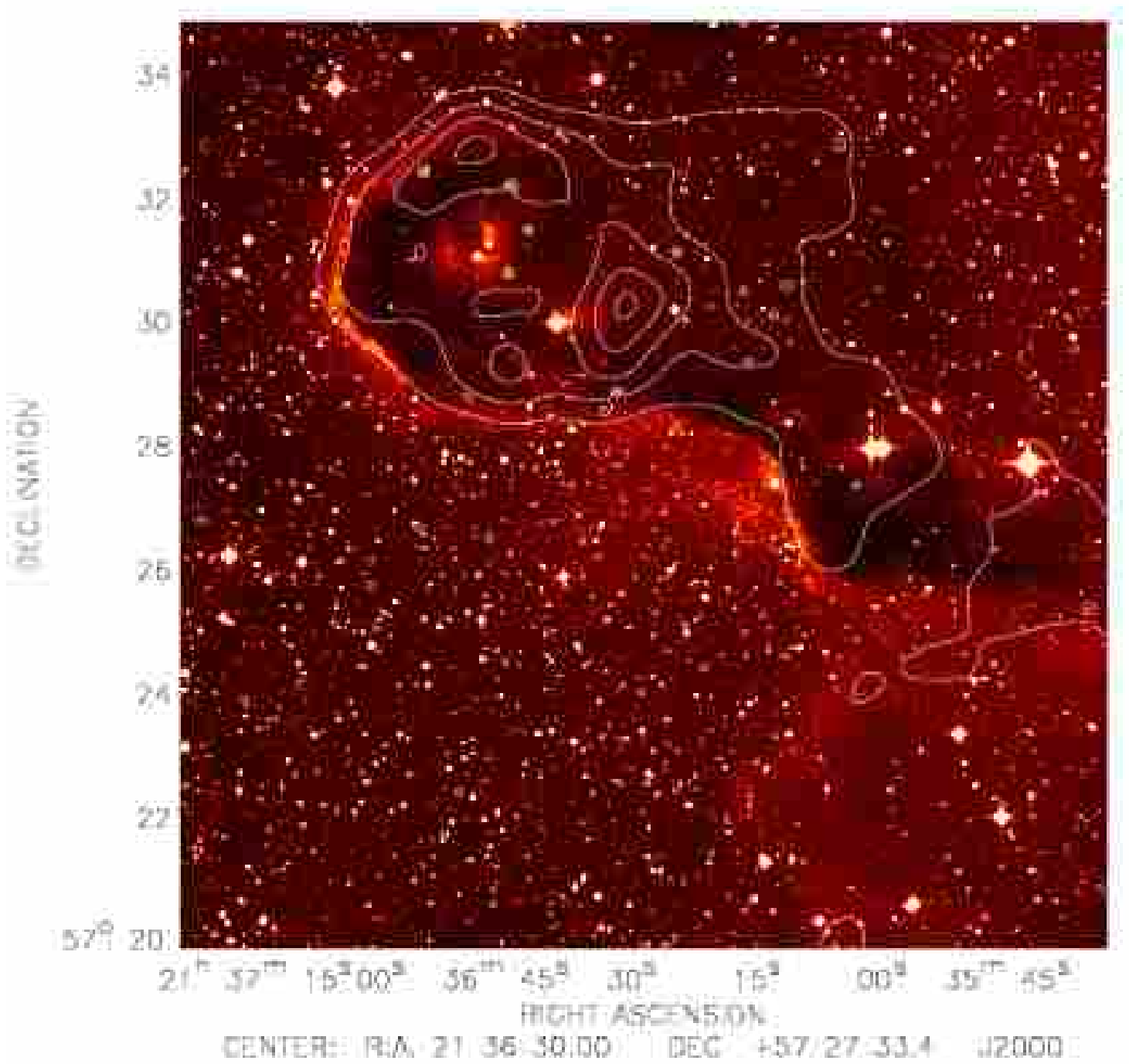}\\
\caption{\bf{(cont.)} \sf  SFO 32 MSX contours start at 9$\sigma$ and SFO 35 MSX contours start at 9$\sigma$ for clarity }
\end{figure*}
\end{center}
\setcounter{figure}{1}
\begin{center}
\begin{figure*}
SFO 37
\includegraphics*[scale=0.50]{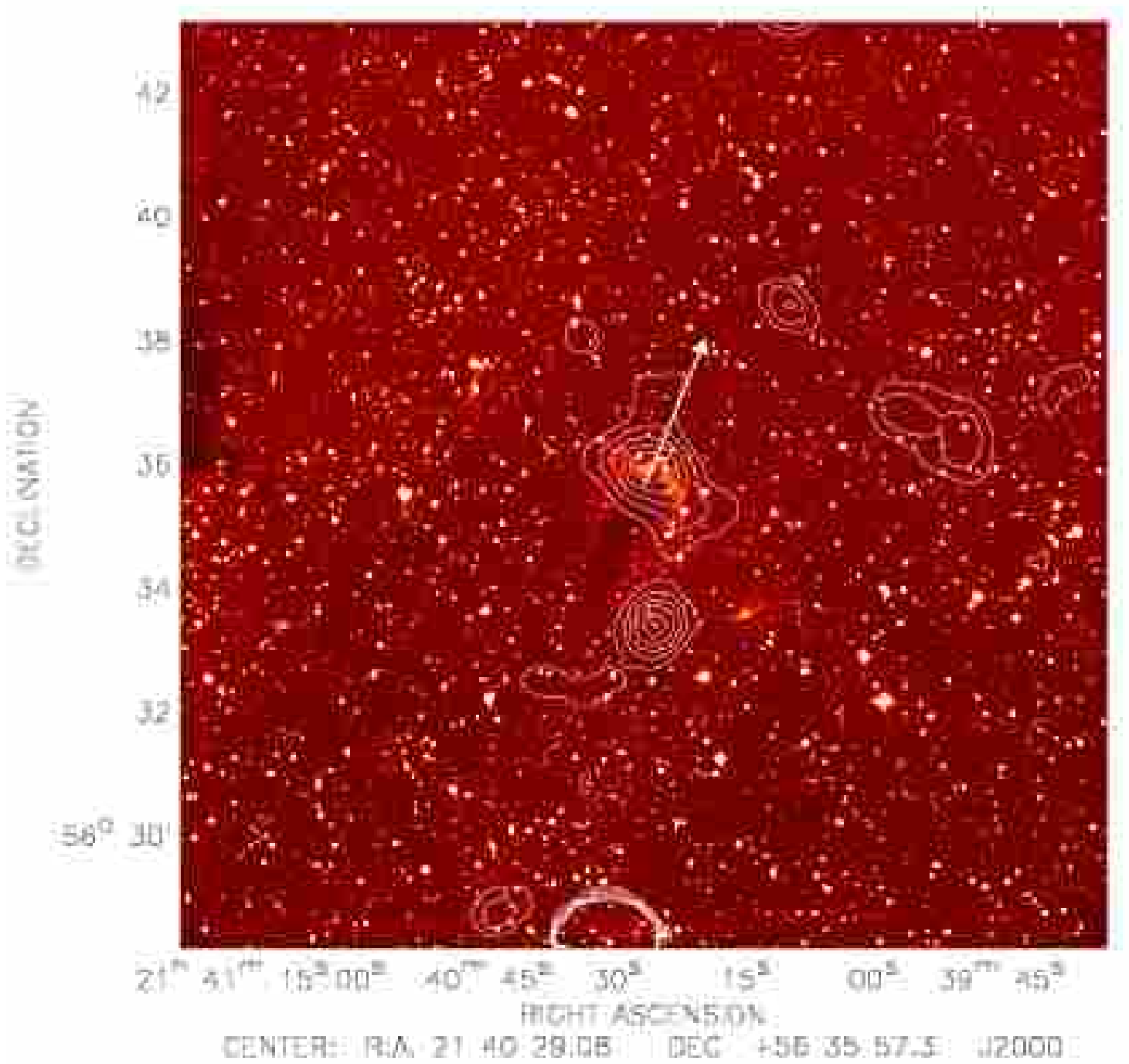}
\includegraphics*[scale=0.50]{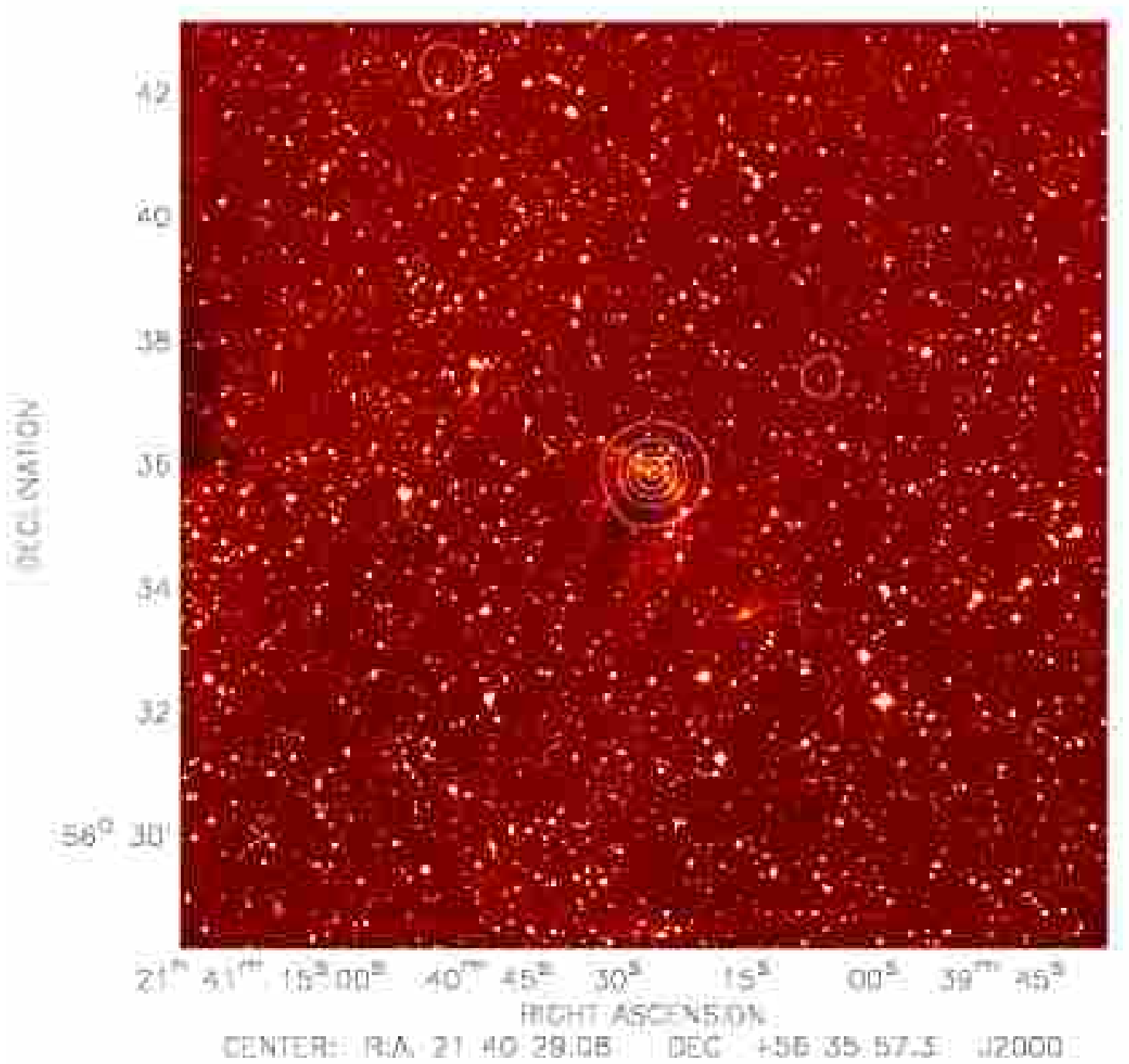}\\
SFO 38
\includegraphics*[scale=0.50]{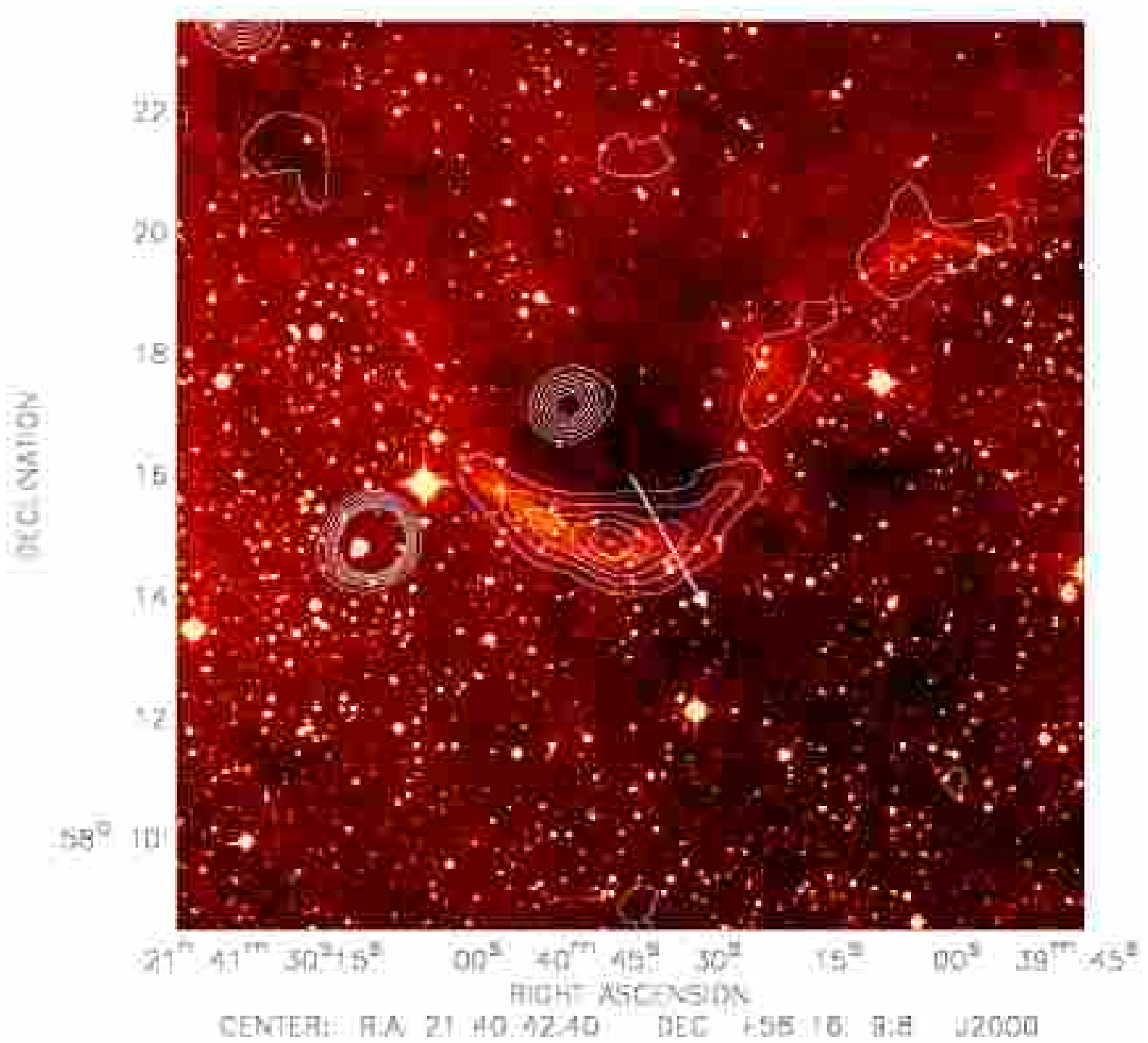}
\includegraphics*[scale=0.50]{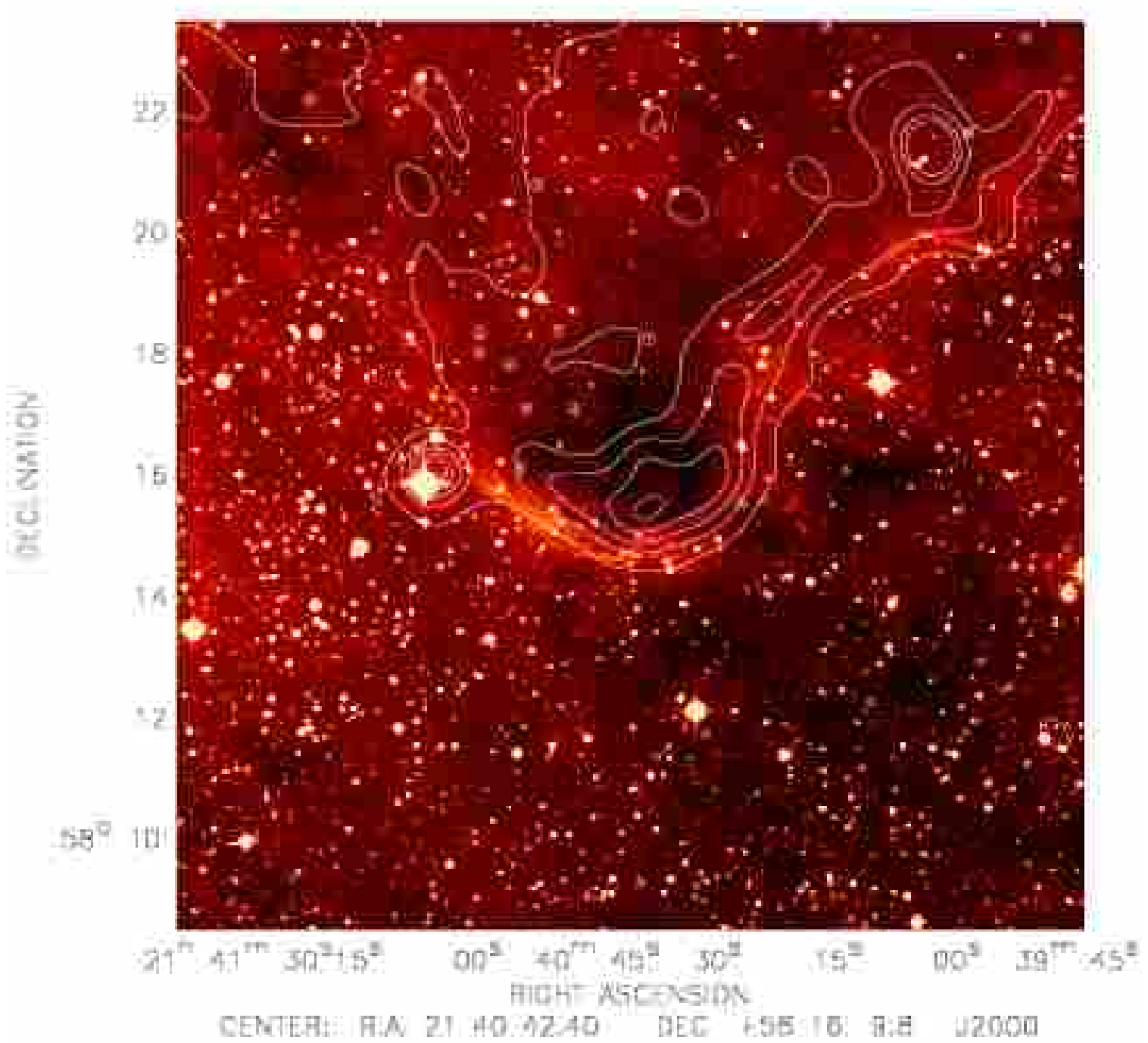}\\
SFO 40
\includegraphics*[scale=0.50]{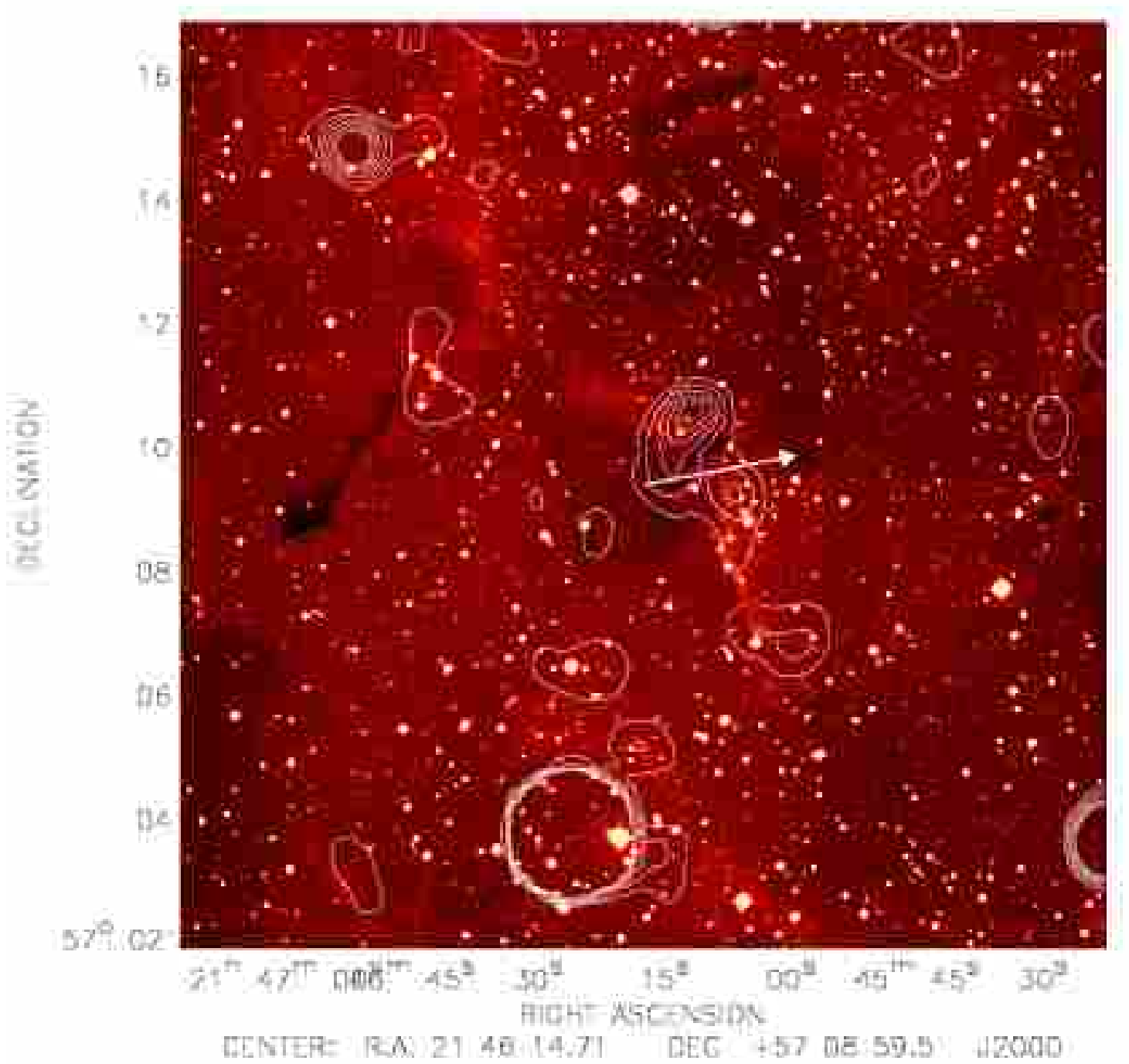}
\includegraphics*[scale=0.50]{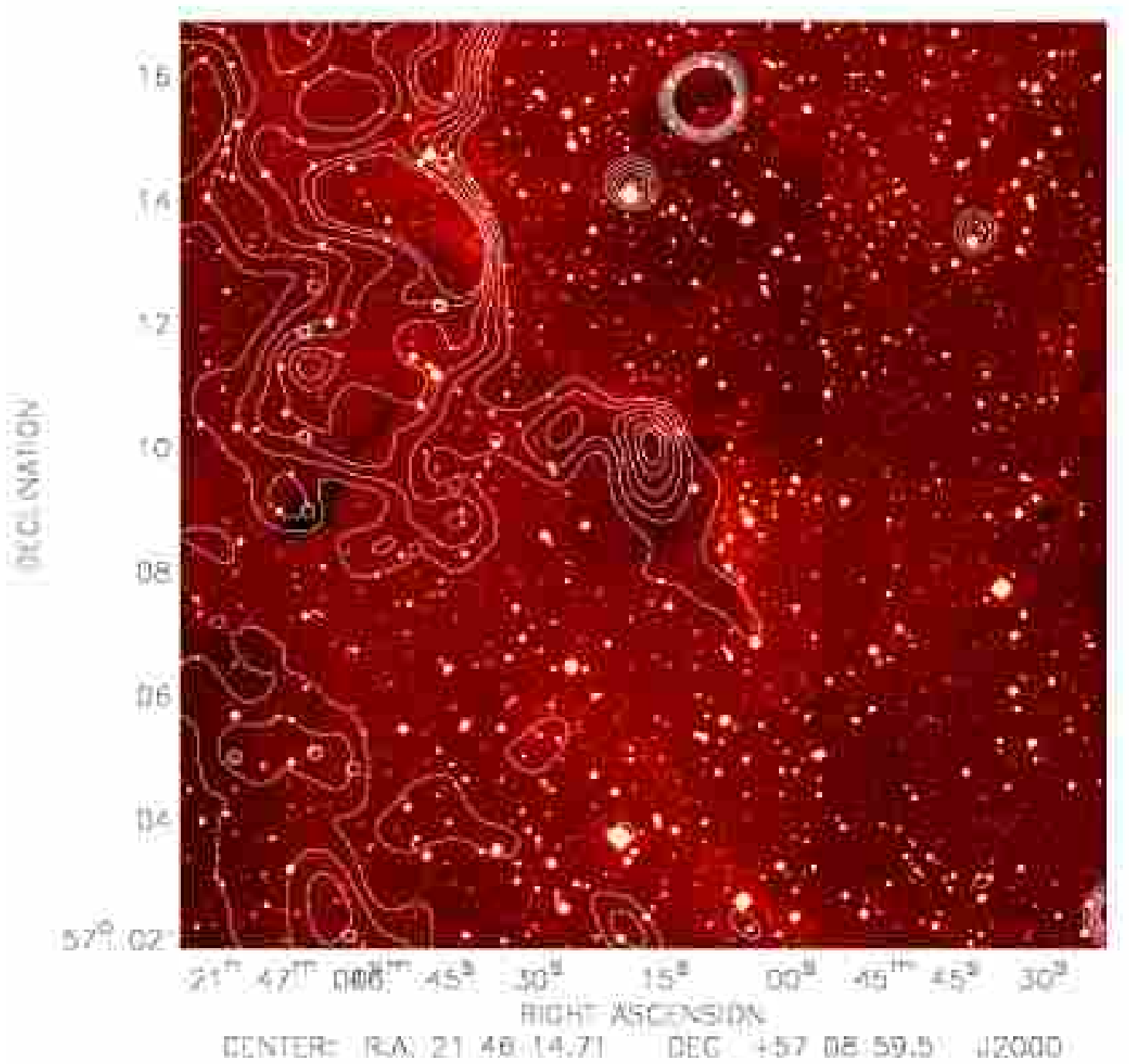}\\
\caption{\bf{(cont.)} \sf SFO 40 MSX contours start at 12$\sigma$ for clarity}
\end{figure*}
\end{center}
\setcounter{figure}{1}
\begin{center}
\begin{figure*}
SFO 41
\includegraphics*[scale=0.50]{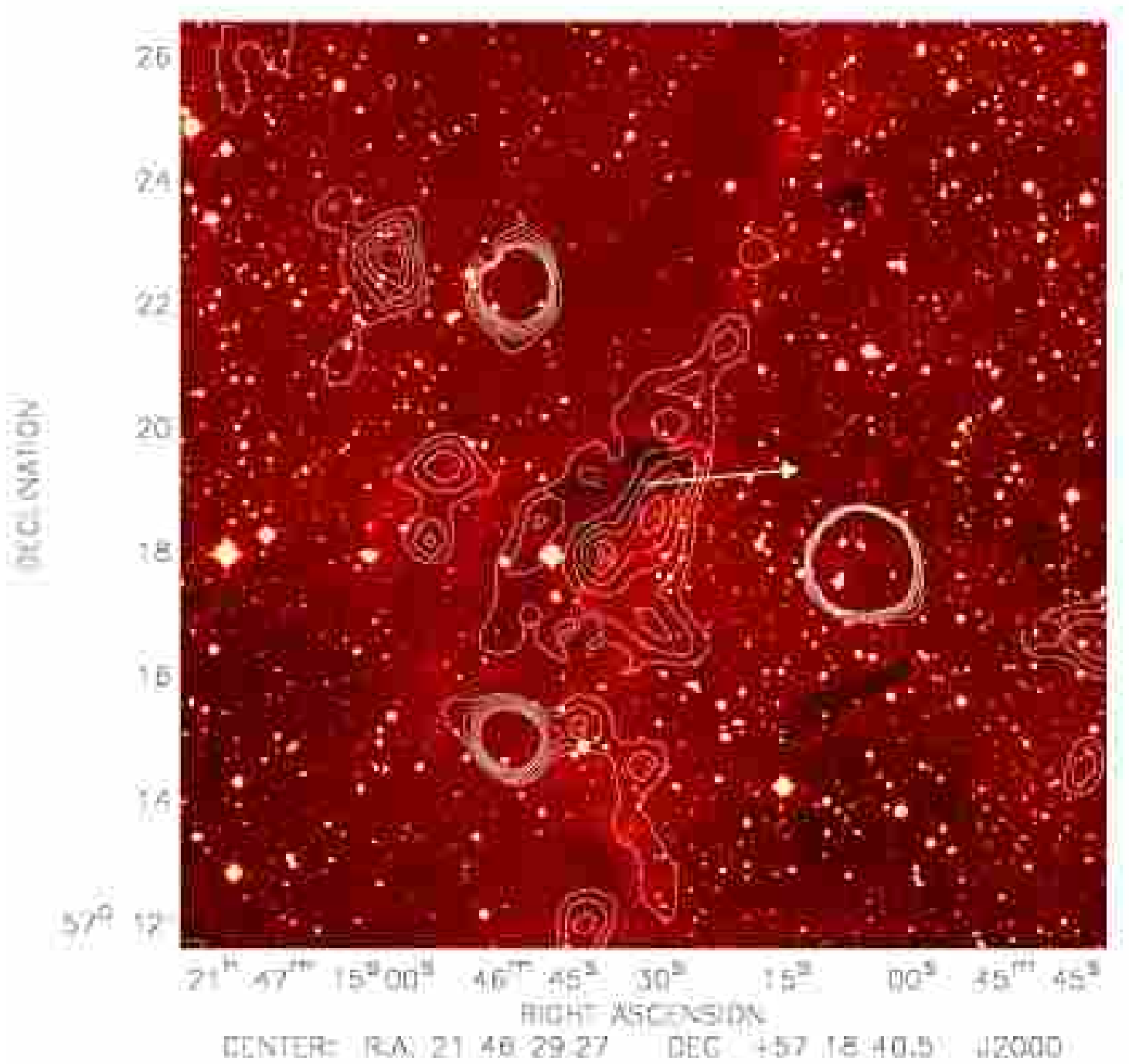}
\includegraphics*[scale=0.50]{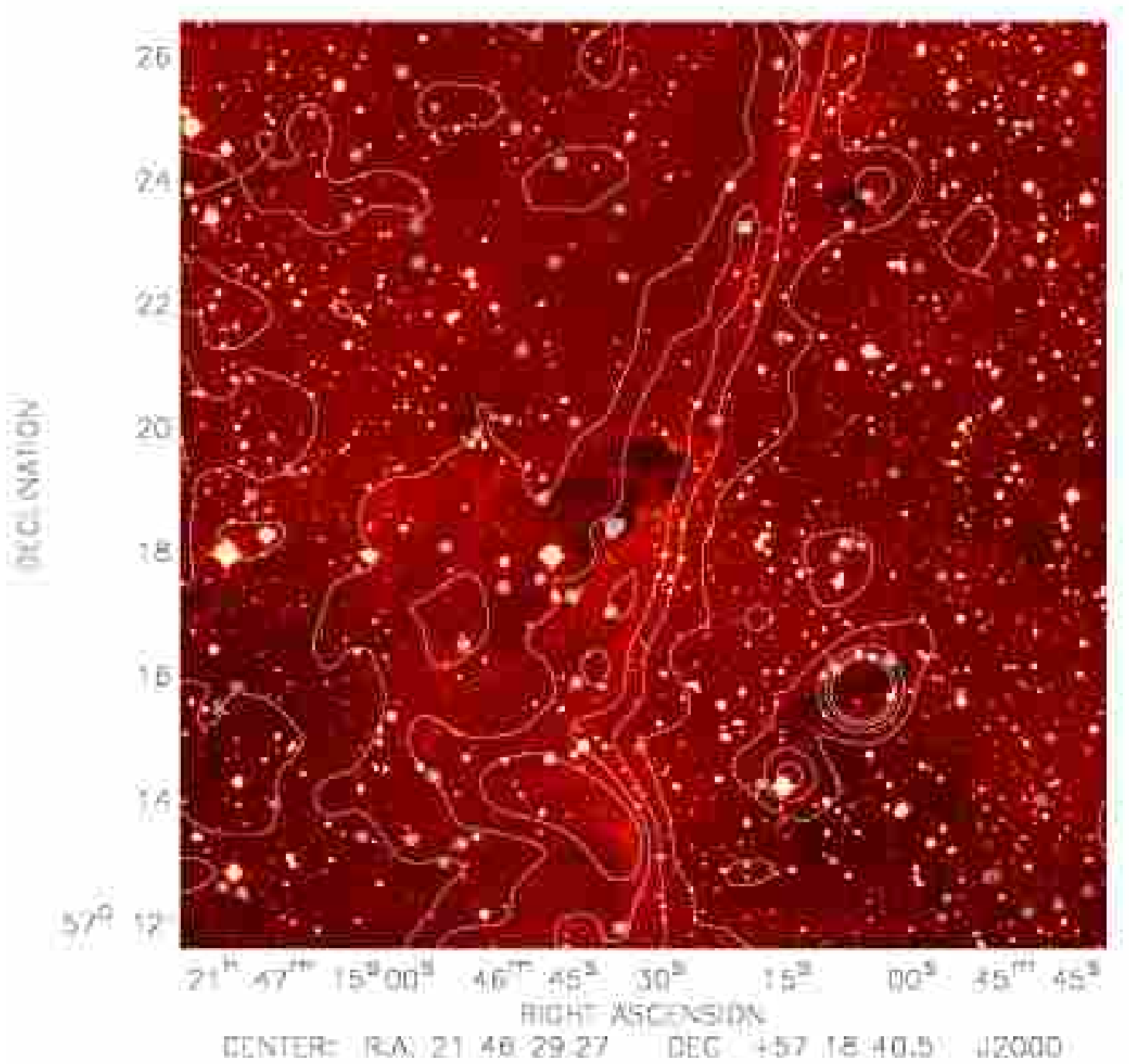}\\
SFO 42 
\includegraphics*[scale=0.50]{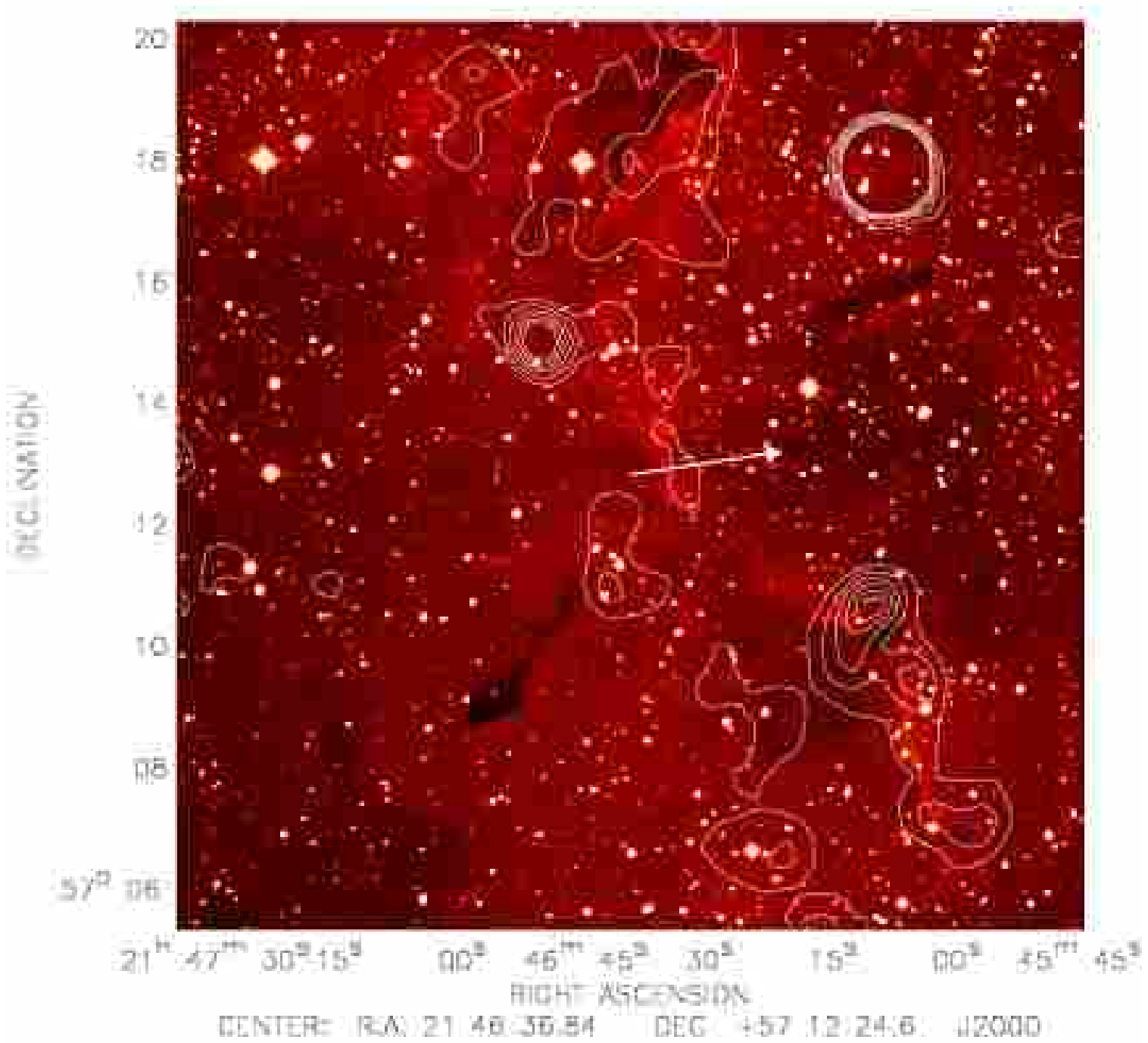}
\includegraphics*[scale=0.50]{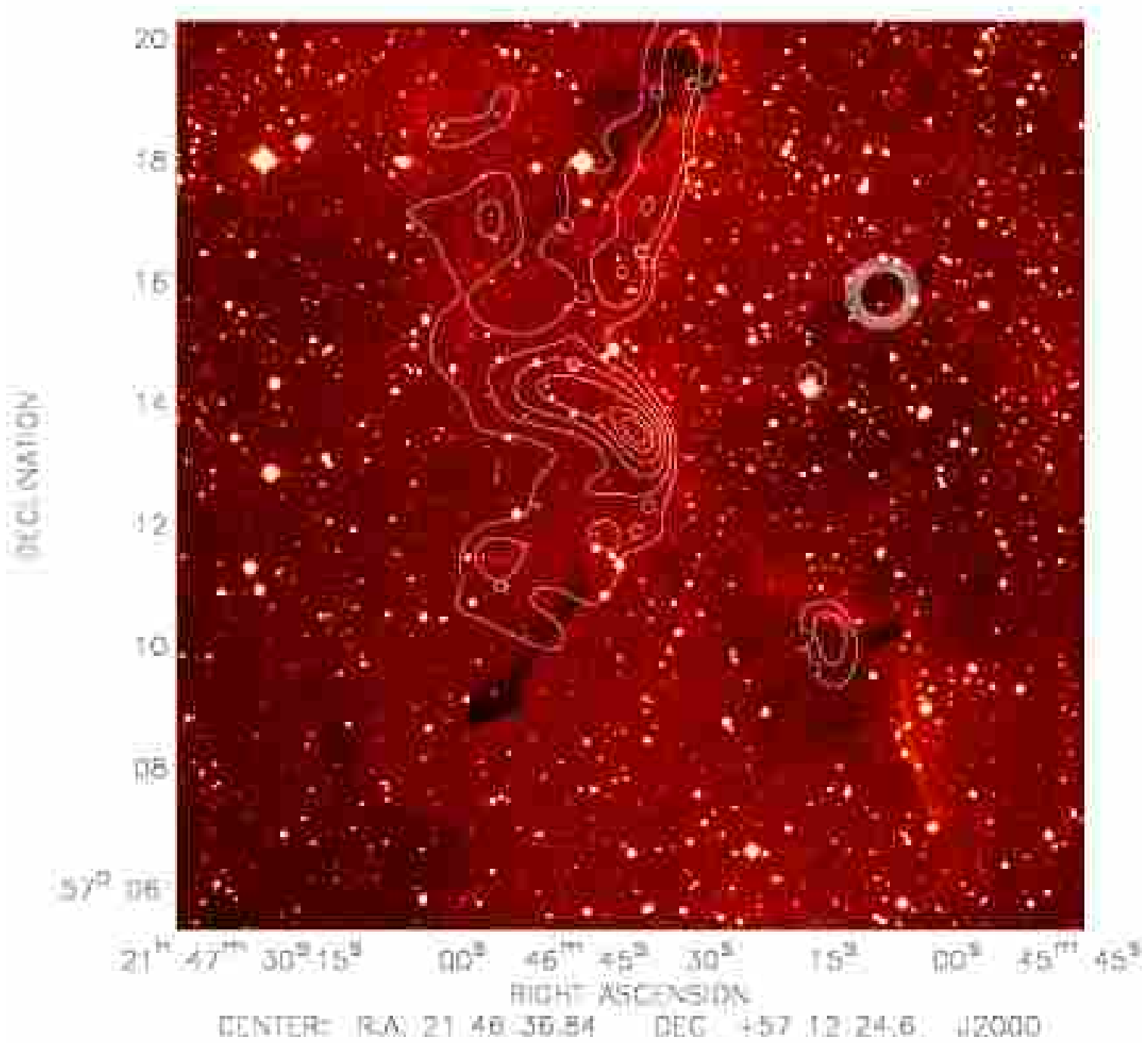}\\
SFO 43
\includegraphics*[scale=0.50]{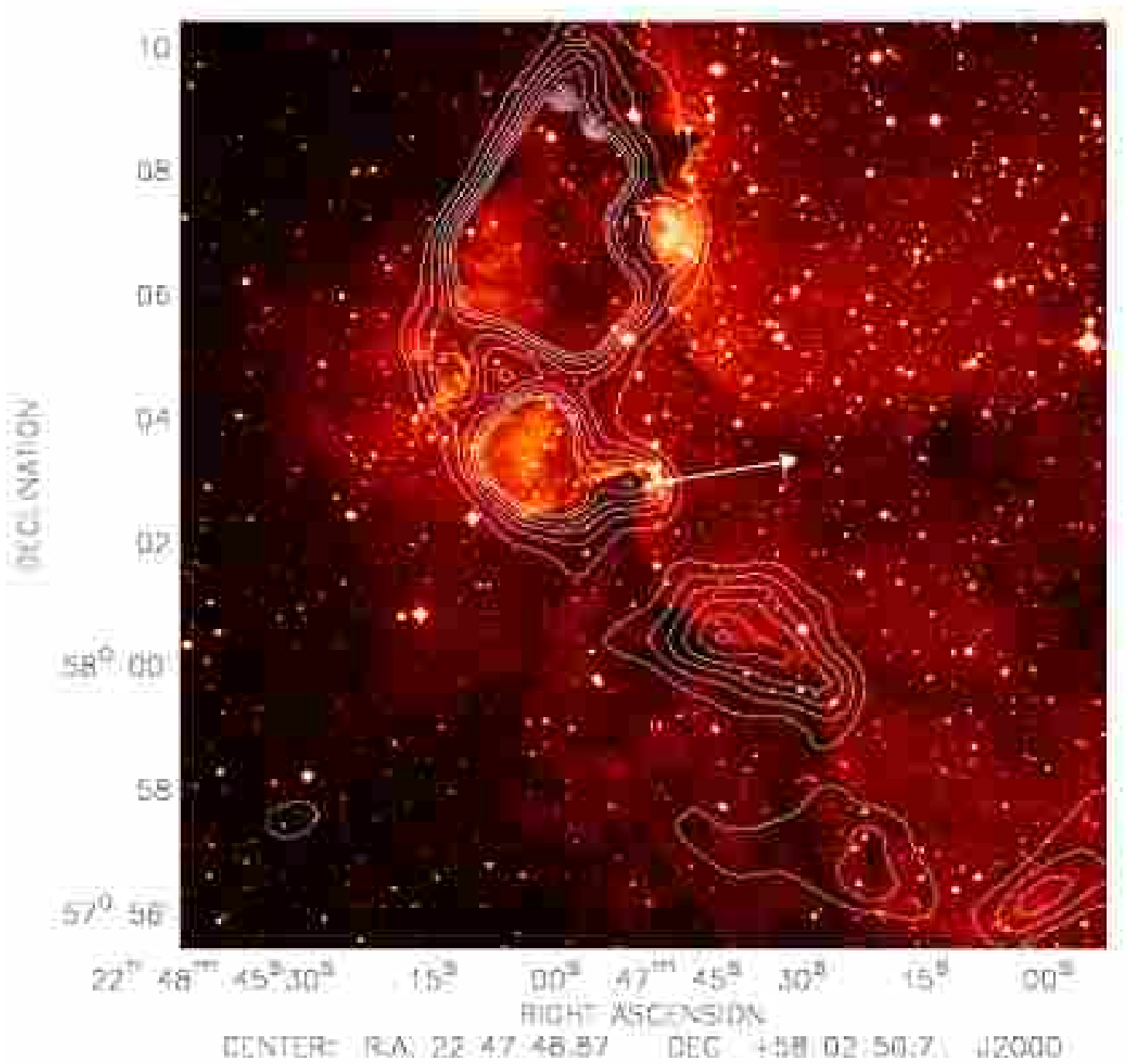}
\includegraphics*[scale=0.50]{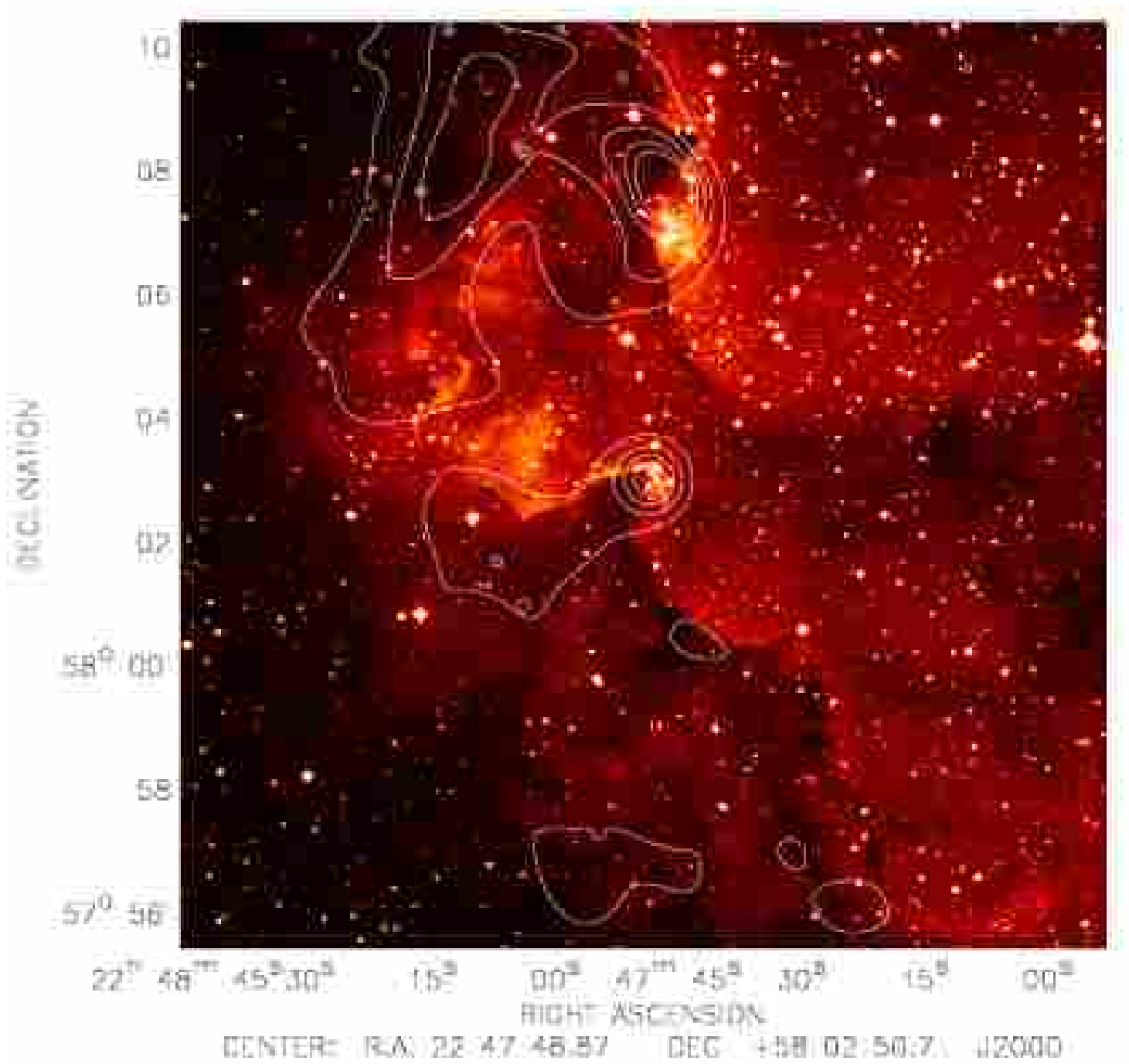}\\
\caption{\bf{(cont.)} \sf SFO 42 MSX contours start at 15$\sigma$ for clarity}
\end{figure*}
\end{center}

\begin{center}
\begin{figure*}
\includegraphics*[scale=0.50]{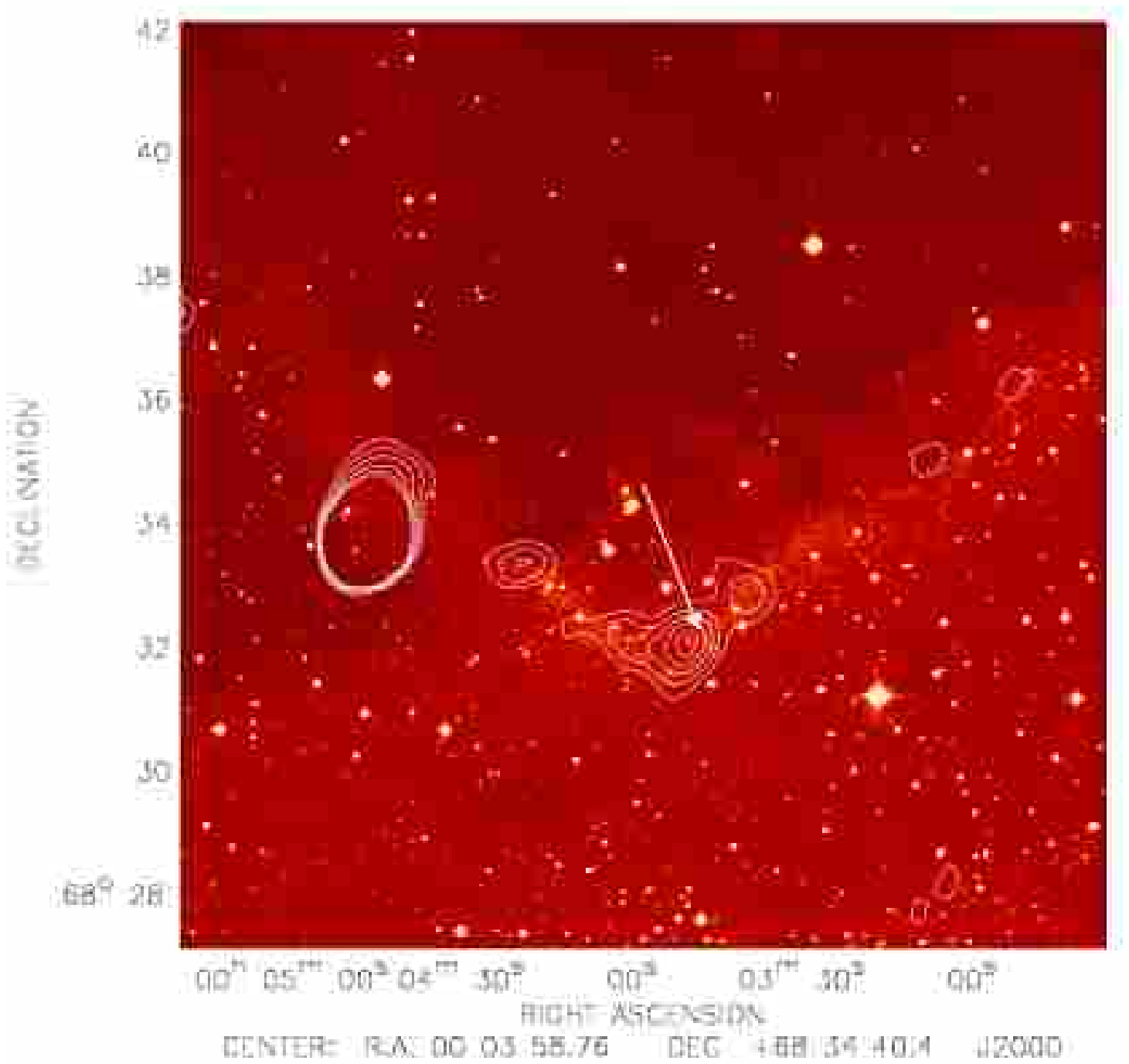}
\includegraphics*[scale=0.50]{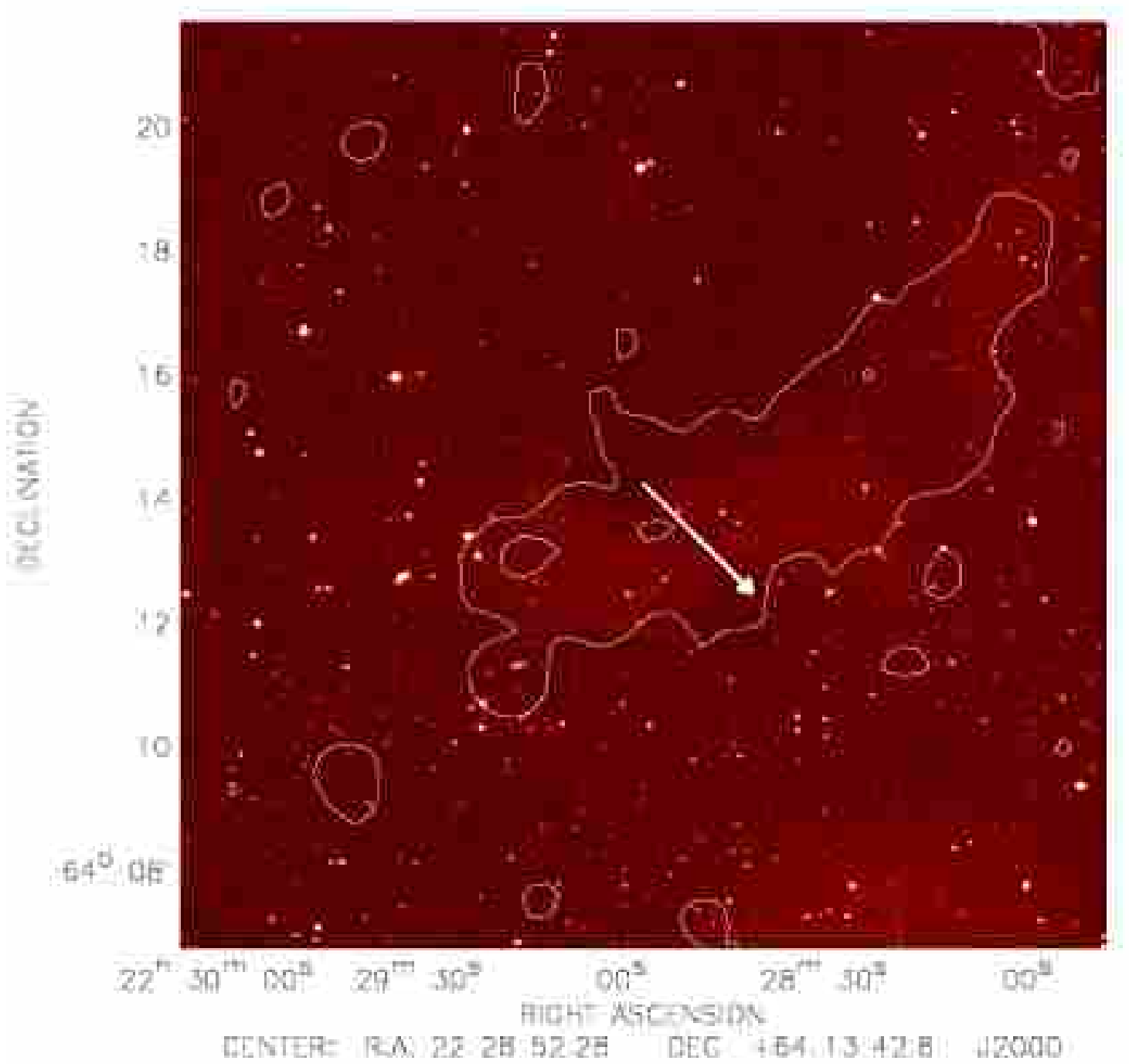}
\caption{SFO 2 and 44 NVSS contours overlaid on DSS images}
\label{fig:2and44}
\end{figure*}
\end{center}

\end{document}